\documentstyle[11pt]{article}
\begin{document}
\input{psfig}
\catcode`@=11
\def\seceqaa{\@addtoreset{equation}{section}
           \def\theequation{A\arabic{equation}}}
\def\seceqbb{\@addtoreset{equation}{section}
           \def\theequation{B\arabic{equation}}}
\def\seceqcc{\@addtoreset{equation}{section}
           \def\theequation{C\arabic{equation}}}
\catcode`@=12

\title{Derivation of  O$(q^4)$ 
Effective Pion-Nucleon Lagrangian Within 
Heavy Baryon Chiral Perturbation Theory} 
\author{A. Misra\thanks{e-mail: aalok@iitk.ac.in},\\
Department of physics, Indian Institute of Technology,\\
Kanpur 208 016, UP, India}
\maketitle
\vskip 0.5 true in

\begin{abstract}
We construct a complete list of O$(q^4)$ terms 
directly within Heavy Baryon Chiral Perturbation Theory (HBChPT) 
in the absence of external fields and assuming
isospin symmetry. In addition to a phase rule  recently developed 
\cite {1n}, a variety of algebraic identities and reparameterization
invariance, are used to
ensure linear independence of the terms and their low energy coupling
constants. We first construct O($q^4$) terms
for off-shell nucleons,  and then perform the on-shell reduction,
again within HBChPT. We discuss an application of the O$(q^4$) terms to the
evaluation of the O($q^4$) operator insertions in the ``contact
graph" of pion double charge exchange. 
\end{abstract}



\clearpage 

\section{Introduction}

Heavy Baryon Chiral Perturbation Theory (HBChPT) is a nonrelativistic
(with respect to the ``heavy" baryons) effective field theory
used for studying meson-baryon interactions at low energies, typically
below the mass of the first non-Goldstone resonance
(See \cite{jm,bkm,bkm1}). The
degrees of  freedom of SU(2) ($\equiv$ isospin)  HBChPT 
(which will be considered in this paper) are the (derivatives of) pion-triplet 
and the nucleon fields.

Recently, a method was developed to generate HBChPT Lagrangian 
(${\cal L}_{\rm HBChPT}$) for off-shell nucleons {\it directly within HBChPT},
which as stated in \cite{1n},
can prove useful when applying HBChPT  to nuclear processes in which
the nucleons  are bound, and hence off-shell. This method has the
advantage of not having to bother to start with the relativistic BChPT
Lagrangian and then carry out the nonrelativistic reduction. It
is thus shorter than the standard approach to HBChPT
as given in \cite{em}, showed explicitly up to O$(q^3)$ in \cite{1n}.
In the context of off-shell nucleons, the upshot of
the method developed is a phase rule (See (\ref{eq:phoffsh}),\cite {1n})
to implement charge conjugation  invariance (along
with Lorentz invariance, parity, hermiticity and isospin symmetry) 
directly within HBChPT.  The phase rule, along with additional
reductions from a variety of algebraic identities, was used to construct,
directly within HBChPT, a complete list  of off-shell O($q^3$) terms 
(in the absence of any external
fields and in the isospin-conserving approximation). We also showed
that the  on-shell limit of the list of terms obtained
matches the corresponding list in \cite{em} (in which the HBChPT Lagrangian
up to O$(q^3$) was constructed starting
from the relativistic BChPT Lagrangian). The extension of \cite{1n}
to O$(q^4,\phi^{2n})(\phi\equiv$pion field) was carried out
at the time of writing of \cite{1n}. We have since extended the work
to O$(q^4,\phi^{2n+1})$ to give the complete O$(q^4)$ list.

For a complete and precise calculation in the single-nucleon sector to one loop,
e.g., 1-loop corrections to pion production
off a single (on-shell) nucleon, because of convergence problems 
(assocated with the amplitude ``$D_2$" for
pion production off a single nucleon), one needs to 
go up to O$(q^4)$ (See \cite{bkm1,bkm2,mms}).
A complete list of the {\it divergent} O($q^4$) 
$\pi$-nucleon interaction  terms in the
presence of external fields was constructed in \cite{mms}, 
but again starting from the relativistic
BChPT Lagrangian. In this paper, we construct a complete list of 
off-shell O$(q^4)$ terms {\it working entirely within  the framework  of
HBChPT}.  
We then identify the finite terms at O$(q^4)$.
Even though, unlike \cite{mms}, we drop  the external fields and work
in the isospin-conserving approximation, it should be noted that the
techniques developed in \cite{1n} and this paper, can be easily extended
to include the external fields (in the covariant derivative and as
suitable field strengths) as well as include isospin
violation.

In addition to the phase rule and 
a whole new set of  algebraic identities (specific to
O$(q^4)$ terms), an additional source for reduction in 
the number of independent O$(q^4,\phi^{2n})\ (\phi$ is the pion field) 
low energy coupling constants (LECs) is
invariance of ${\cal L}_{\rm HBChPT}$ under infinitesimal variations
in the nucleon/baryon velocity, referred to as reparameterization
invariance (See Section 5).

Section 2 has the basics and sets up the notations. In Section 3, 
reductions obtained in addition  to (\ref{eq:phoffsh}) 
due to algebraic
identities such as Jacobi(-like) identities, etc. are discussed.
In Section 4, the complete lists
of O$(q^4)$ terms is given.
In Section 5, further reductions in the number of independent 
coupling constants due to reparameterization invariance, is 
discussed.  In section 6, we discuss an application of the O$(q^4)$ list in 
the multi-nucleon sector. In Section 7 we discuss
the derivation of the on-shell O$(q^4$) Lagrangian, again within HBChPT
using the techniques of \cite{1n}. Section 8 has the conclusion which 
includes remarks on the work done.

\section{The Basics}

The HBChPT Lagrangian is written in terms of the ``upper component"
H (and its hermitian adjoint ${\bar{\rm H}}$), 
exponentially parameterized matrix-valued meson fields $U,\ u\equiv\sqrt{U}$, 
baryon (``$v_\mu,\rm S_\nu$") and
pion-field-dependent (``${\rm D}_\mu, u_\nu,\chi_\pm$") building
blocks defined below:
\begin{equation}  
\label{eq:Hdef}
{\rm H}\equiv e^{i{\rm m}v\cdot x}{1\over2}(1+\rlap/v)\psi,
\end{equation}
where $\psi\equiv$ Dirac spinor
and $m\equiv$ the nucleon mass; 
\begin{eqnarray} 
\label{eq:Sdef}
& & 
v_\mu\equiv{\rm nucleon\ veclocity},\nonumber\\
&  & {\rm S}_\nu\equiv{i\over2}\gamma^5\sigma_{\nu\rho}v^\rho
\equiv{\rm Pauli-Lubanski\ spin\ operator};
\end{eqnarray}
\begin{equation}
\label{eq:Udef}
U={\rm exp}\biggl(i{\phi\over F_\pi}\biggr),\ {\rm where}\
\phi\equiv\vec\pi\cdot\vec\tau,
\end{equation}
where $\vec\tau\in$ nucleon isospin generators;
${\rm D}_\mu=\partial_\mu+\Gamma_\mu$ where 
$\Gamma_\mu\equiv {1\over 2}[u^\dagger,\partial_\mu u]$;
$u_{\mu}\equiv i(u^{\dagger}{\partial}_{\mu}u - 
u{\partial}_{\mu}u^{\dagger})$; $\chi_\pm\equiv u^\dagger\chi u^\dagger\pm
u\chi^\dagger u$, where $\chi\equiv \rm M_\pi^2$ for this paper.

Terms of the ${\cal L}_{\rm (H)BChPT}$ constructed from products of 
building blocks will automatically be chiral invariant. 
Symbolically, a term in ${\cal L}_{\rm HBChPT}$ 
can be written as just a product of the building blocks to
various powers (omitting $\rm H,{\bar{\rm H}}$ as will be done in the
rest of the paper except for Section 5):
\begin{equation} 
\label{eq:prodbb}
{\rm D}_\alpha^m 
u_\beta^n\chi_+^p\chi_-^q v_\sigma^l\rm S_\kappa^r\equiv(m,n,p,q)
\equiv{\rm O}(q^{m+n+2p+2q}).
\end{equation}

A systematic path integral derivation for ${\cal L}_{\rm HBChPT}$ based
on a paper by Mannel et al \cite {mnnl}, starting from
${\cal L}_{\rm BChPT}$ was first given by Bernard et al \cite {bkm2}.
As shown by them, after integrating
out h from the generating
functional, one arrives at ${\cal L}_{\rm HBChPT}$ :
\begin{equation}
\label{eq:lag}
{\cal L}_{\rm HBChPT} = {\bar{\rm H}}\biggl( {\cal A} +
\gamma^0 {\cal B}^\dagger\gamma^0 {\cal C}^{-1}{\cal B}
\biggr)\rm H,
\end{equation}
an expression in the upper components only i.e. for non-relativistic
nucleons. So the terms of ${\cal L}_{\rm HBChPT}$ in this paper are 
given as operators on the H-spinors. For off-shell nucleons,
$\gamma^0 {\cal B}^\dagger\gamma^0 {\cal C}^{-1}{\cal B}\in {\cal A}$, 
and hence, listing
${\cal A}$-type terms will suffice for this
paper.

For the sake of completeness, we will state the phase rule derived in
\cite {1n}. 
HBChPT terms (that are Lorentz scalar - isoscalars of even parity) made 
hermitian using a prescription for constructing hermitian (anti-)commutators
discussed in \cite {1n}, consisting of $q\ i\chi_-$'s,
$P[\ ,\ ]$'s 
and $j$ (which can take only the values 0 or 1) 
$\epsilon^{\mu\nu\rho\lambda}$'s for which 
the following phase rule is satisfied, are the only terms allowed:
\begin{equation} 
\label{eq:phoffsh}
(-1)^{q+P+j}=1.
\end{equation}
In \cite {1n}, (\ref{eq:phoffsh}) 
was used to generate complete lists up to O$(q^3)$ in
the absence of external vector and axial-vector fields. In
this paper, the same phase rule is used to construct complete lists of 
O$(q^4)$, which become relevant in e.g., 
multi $\pi$-production vertices (e.g. $\pi+N\to m\pi+N$)
embedded in nuclear processes, when working up to O$(q^3)$, e.g.,
for processes involving effectively two nucleons such as pion double charge
exchange (DCX) (See \cite{kj,mk} and Section 6). 

Let $A,B,C,D$ be operators chosen from the pion-field dependent
building blocks of (\ref{eq:prodbb}).  In what follows, and
especially in Section 3, use will be made of a notation of \cite{krause}:
$(A,B)\equiv [A,B]$ or $[A,B]_+$. One can
then show that apart from the  (0,0,0,2) and (0,0,2,0)
terms (using the notation of  (\ref{eq:prodbb})),
the following is the  
complete list of  O$(q^4,\phi^{2n})$ terms (using (\ref{eq:phoffsh})):
\begin{eqnarray}  
\label{eq:list}
& (i) & (A,(B,(C,D)))\equiv
(a)[A,[B,[C,D]_+]];\ (b)[A,[B,[C,D]]_+];\nonumber\\
& & (c)[A,[B,[C,D]]]_+;\ (d)[A,[B,[C,D]_+]_+]_+; \nonumber\\ 
& (ii) & ((A,B),(C,D))\equiv  
(a)[[A,B],[C,D]_+];\ (b)[[A,B]_+,[C,D]];\ \nonumber\\
& & (c)[[A,B],[C,D]]_+;\ (d)[[A,B]_+,[C,D]_+]_+;\ \nonumber\\
& (iii) & i(A,(B,(C,D)))\equiv
(a)i[A,[B,[C,D]_+]_+];\ (b)i[A,[B,[C,D]_+]]_+;\ \nonumber\\
& & (c)i[A,[B,[C,D]]_+]_+;\ (d)i[A,[B,[C,D]]];\ \nonumber\\
& (iv) & i((A,B),(C,D))\equiv  
(a)i[[A,B],[C,D]];\ (b)i[[A,B]_+,[C,D]_+];\ \nonumber\\
& & (c)i[[A,B],[C,D]_+]_+;\ (d)i[[A,B]_+,[C,D]]_+;\nonumber\\
& (v) & i(A,(B,C))\equiv (a)i[A,[B,C]]_+;\ (b)i[A,[B,C]_+];\ (c)A\leftrightarrow B;\nonumber\\
& & (d)i[[A,B],C]_+;\ (e) i[[A,B]_+,C];\ \nonumber\\
& (vi) &  (A,(B,C))\equiv (a)[A,[B,C]];\ (b)[A,[B,C]_+]_+;\ (c)A\leftrightarrow B\nonumber\\
& & (d)[[A,B],C];\ (e)[[A,B]_+,C]_+,
\end{eqnarray}
where it is understood that of all the possible
terms implied by $(i)(A,(B,(C,D)))$, $(i)((A,B),(C,D))$ and
$(i)(A,(B,C))$, only those that are allowed by (\ref{eq:phoffsh}) are
to be included. For (0,0,2,0)- and (0,0,0,2)- type terms, one needs to
include
\begin{equation}
\label{eq:chi+-}
(A,A)\equiv(\chi_+,\chi_+);\ (\chi_-,\chi_-).
\end{equation}
The list (\ref{eq:list}) holds good for O$(q^4,\phi^{2n+1})$ terms with the
difference that there is an additional  factor of $i$ multiplying
the terms in $(i), (ii)$ and $(vi)$,
and the $i$ in $(iii), (iv)$ and $(v)$, is absent. The reason for
including $i$ only in some combinations of terms has to do with
imposing charge conjugation invariance 
along with other symmetries {\it directly within HBChPT} (See \cite{1n}). 
The terms of (\ref{eq:list}) and its analog for O$(q^4,\phi^{2n+1})$
are not all independent since they can be related by a number of linear
relations: see next section (and \cite{1n} for O($q^3)$).

\section{Further Reduction due to Algebraic identities}

In this section, we discuss
further reduction in addition to the  ones obtained from (\ref{eq:phoffsh}).
The main result from \cite{1n} is that one need not consider
trace-dependent terms in SU(2) HBChPT if one assumes isospin conservation.
Thus, trace-dependent O$(q^4)$ terms can be eliminated in
preference for trace-independent terms. We discuss reduction due to
algebraic identities in the  various categories of (\ref{eq:list}).
Some of the algebraic reductions require one to consider  more than one
category at a time, e.g., for  O$(q^4,\phi^{2n})$ terms,
the Jacobi-like identities in (\ref{eq:oddanticoom}) require one to consider 
$(i), (i)(A\leftrightarrow B), (ii)$. 
After (\ref{eq:curvature})[curvature relation],
we discuss the algebraic reductions in 
O$(q^4,\phi^{2n})$ and O$(q^4,\phi^{2n+1})$ terms, separately. 

In the absence
of external vector  and axial-vector fields, one can show that:
\begin{equation}
\label{eq:curvature}
[{\rm D}_\mu,{\rm D}_\nu]={1\over 4}[u_\mu, u_\nu],
\end{equation}
which is referred to as the ``curvature relation." This relation will be
used extensively in conjunction with some Jacobi-like identities 
discussed below. It is because of  this identity that one requires to
consider some (4,0,0,0), (0,4,0,0) and (2,2,0,0) terms together,
e.g., in (\ref{eq:LCindoddanticoom}).

\subsection{O($q^4,\phi^{2n})$ Terms}

In this subsection, we consider reduction in the number
of  independent O$(q^4,\phi^{2n})$ terms due to various algebraic
identities. The following are the algebraic identities
responsible  for reduction in number of O$(q^4,\phi^{2n})$
terms:
(\ref{eq:ABoddanticoom1}), (\ref{eq:ABoddanticoom2}), (\ref{eq:oddanticoom}),
(\ref{eq:evenanticoom}), (\ref{eq:umured}), (\ref{eq:oddanticoom1}), 
(\ref{eq:uchi+0}) and (\ref{eq:evencoom}). 
For (\ref{eq:oddanticoom})
and (\ref{eq:evenanticoom}), there are two sets each of terms
(one $\epsilon^{\mu\nu\rho\lambda}$-dependent
and the other $\epsilon^{\mu\nu\rho\lambda}$-independent),
that get eliminated. One of the two sets ($\epsilon^{\mu\nu\rho\lambda}$-
independent) for (\ref{eq:oddanticoom}) has been discussed
in  detail in this subsection. 
The details for the other sets are given as appendices.

\underline{\underline{$p=q=0$ in (\ref{eq:prodbb})$\equiv(A,(B,(C,D))); ((A,B),(C,D))$}}

This includes $(i) - (iv)$ of (\ref{eq:list}). All
 terms in each of the first four types (of terms)
in (\ref{eq:list}) [$(i) - (iv)$] are linearly independent for 
unequal field operators A, B, C, D. 
However for (4,0,0,0), (0,4,0,0) and (2,2,0,0), L.C.-independent terms, one
needs to consider A=C, B=D in (i) in equation (\ref{eq:list}). Using
\begin{equation} 
\label{eq:ABoddanticoom1}
[A,[B,[A,B]_+]] = - [A,[B,[A,B]]_+] 
\end{equation}
only three of the four terms in (i) of equation (\ref{eq:list}), 
are linearly independent. Similarly, using
\begin{equation} 
\label{eq:ABoddanticoom2}
[[A,B],[A,B]_+] = - [[A,B]_+,[A,B]],
\end{equation}
only three of  the four terms in (ii) of equation (\ref{eq:list}), 
are linearly independent.

There are some reductions possible due to some Jacobi-like identities by 
considering :  $(i), (i)(A\leftrightarrow B), 
(ii)$ of (\ref{eq:list})($\equiv\epsilon^{\mu\nu\rho\lambda}$ -independent terms), and 
$(iii),(iii)(A\leftrightarrow B),(iv)$ of (\ref{eq:list}) ($\equiv
\epsilon^{\mu\nu\rho\lambda}$-dependent terms).
The reason why one can not hope to get reductions by considering
any other pairs of types of terms in $(i) -  (iv)$ (in (\ref{eq:list})), 
is because one can get (linear) algebraic relationships only between those
terms which are (both) independent of (have) an overall factor of $i$. 

\underline{$(i), (i)(A\leftrightarrow B), (ii)$ of (\ref{eq:list})}

One can show the following 8 Jacobi-like identities:
\begin{eqnarray}
\label{eq:oddanticoom}
& & [A,[B,[C,D]_+]] - [[A,B],[C,D]_+] = (i)(a)(A\leftrightarrow B)\nonumber\\
& & [A,[B,[C,D]_+]] - [[A,B]_+,[C,D]_+]_+ =-(i)(d)(A\leftrightarrow B)
\nonumber\\
& & [A,[B,[C,D]]_+] - [[A,B]_+,[C,D]] = -(i)(b)(A\leftrightarrow B)\nonumber\\
& & [A,[B,[C,D]]_+] - [[A,B],[C,D]]_+ = (i)(c)(A\leftrightarrow B)\nonumber\\
& & [A,[B,[C,D]]]_+ - [[A,B]_+,[C,D]] = -(i)(c)(A\leftrightarrow B)\nonumber\\
& & [A,[B,[C,D]_+]_+]_+ - [[A,B],[C,D]_+] = (i)(d)(A\leftrightarrow B)
\nonumber\\
\end{eqnarray}
Since we have 6 identities in 12 terms, we can take any 6 as linearly 
independent, say $(i)\biggl((a) - (d)\biggr)$ and 
$(ii)\biggl((a),(b)\biggr)$ of (\ref{eq:list}). 
Obviously,  to ensure linear independence, one can not choose 
these four terms such that
any three belong to the same (Jacobi-like)identity

(1) Using (\ref{eq:oddanticoom}) and (\ref{eq:curvature}), one needs to consider
the following (4,0,0,0), (0,4,0,0)  and (2,2,0,0) $\epsilon^{\mu\nu\rho\lambda}$-independent
terms together:
\begin{eqnarray}
\label{eq:LCindoddanticoom}
& & (v\cdot{\rm D},({\rm D}_\mu,(v\cdot{\rm D},{\rm D}^\mu))),\ 
({\rm D} _\mu,(v\cdot{\rm D},(v\cdot{\rm D},{\rm D}^\mu))),\ 
((v\cdot{\rm D},{\rm D}_\mu),(v\cdot{\rm D},{\rm D}^\mu))\nonumber\\
& & (v\cdot u,(u_\mu,(v\cdot u,u^\mu))),\ (u_\mu,(v\cdot u,(v\cdot u,u^\mu))),\
((v\cdot u,u_\mu),(v\cdot u, u^\mu)),\nonumber\\
& & (v\cdot{\rm D},({\rm D}_\mu,(v\cdot u,u^\mu))),\ 
({\rm D}_\mu,(v\cdot{\rm D},(u_\mu,v\cdot u))),\ 
((v\cdot{\rm D},{\rm D}_\mu),(v\cdot u,u^\mu)))\nonumber\\
& & 
(v\cdot u,(u_\mu,(v\cdot{\rm D},{\rm D}^\mu))),\ 
(u_\mu,(v\cdot u,(v\cdot{\rm D},{\rm D}^\mu))).\ 
\nonumber\\
& & 
\end{eqnarray}
One needs to do a  careful counting of the total number of identities
that one can write down using (\ref{eq:oddanticoom}) and
(\ref{eq:curvature}), and the total number of terms in those
identities. We do the same below.
 
Using (\ref{eq:ABoddanticoom1}),
(\ref{eq:ABoddanticoom2}), (\ref{eq:curvature}) and (\ref{eq:oddanticoom}),
one sees that one gets

(a) 14 identities in 21 terms:
\begin{eqnarray}
\label{eq:v.uv.Dset1}
& & [v\cdot{\rm D},[{\rm D}_\mu,[v\cdot{\rm D},{\rm D}^\mu]_+]]-{1\over4}
[[v\cdot u,u^\mu],[v\cdot{\rm D},{\rm D}_\mu]_+]
=[{\rm D}_\mu,[v\cdot{\rm D},[v\cdot{\rm D},{\rm D}^\mu]_+]]
\nonumber\\
& & [v\cdot{\rm D},[{\rm D}_\mu,[v\cdot{\rm D},{\rm D}^\mu]_+]]-
[[v\cdot {\rm D},{\rm D}^\mu]_+,[v\cdot{\rm D},{\rm D}_\mu]_+]_+
=-[{\rm D}_\mu,[v\cdot {\rm D},[v\cdot{\rm D},{\rm D}^\mu]_+]_+]_+
\nonumber\\
& & -[v\cdot{\rm D},[{\rm D}_\mu,[v\cdot{\rm D},{\rm D}^\mu]_+]]-{1\over{16}}
[[v\cdot u,u^\mu],[v\cdot u,u_\mu]]_+
={1\over4}[{\rm D}_\mu,[v\cdot{\rm D},[v\cdot u,u^\mu]]]_+
\nonumber\\
& & {1\over4}[v\cdot{\rm D},[{\rm D}_\mu,[v\cdot u,u^\mu]]]_++{1\over4}
[[v\cdot u,u^\mu],[v\cdot{\rm D},{\rm D}_\mu]_+]
=-{1\over4}[{\rm D}_\mu,[v\cdot{\rm D},[v\cdot u,u^\mu]]]_+
\nonumber\\
& & [v\cdot{\rm D},[{\rm D}_\mu,[v\cdot{\rm D},{\rm D}^\mu]_+]_+]_+-
[[v\cdot u,u^\mu],[v\cdot{\rm D},{\rm D}_\mu]_+]
=[{\rm D}_\mu,[v\cdot{\rm D},[v\cdot{\rm D},{\rm D}^\mu]_+]_+]_+
\nonumber\\
& & 
[v\cdot{\rm D},[{\rm D}_\mu,[v\cdot u,u^\mu]_+]]
-[[v\cdot u,u^\mu]_+,[v\cdot{\rm D},{\rm D}_\mu]_+]_+
=-[{\rm D}_\mu,[v\cdot{\rm D},[v\cdot u,u^\mu]_+]_+]_+
\nonumber\\
& & [v\cdot u,[u_\mu,[v\cdot{\rm D},{\rm D}^\mu]_+]]
-[[v\cdot u,u^\mu],[v\cdot{\rm D},{\rm D}_\mu]_+]
=[u_\mu,[v\cdot u,[v\cdot{\rm D},{\rm D}^\mu]_+]]\nonumber\\
& & {1\over4}[v\cdot u,[u_\mu,[v\cdot{\rm D},{\rm D}^\mu]_+]]
-[[v\cdot u,u^\mu],[v\cdot{\rm D},{\rm D}_\mu]_+]
=[u_\mu,[v\cdot u,[v\cdot{\rm D},{\rm D}^\mu]_+]_+]_+\nonumber\\
& & [v\cdot u,[u_\mu,[v\cdot{\rm D},{\rm D}^\mu]_+]_+]_+
-[[v\cdot u,u^\mu],[v\cdot{\rm D},{\rm D}_\mu]_+]
=[u_\mu,[v\cdot u,[v\cdot{\rm D},{\rm D}^\mu]_+]_+]_+\nonumber\\
& & [v\cdot u,[u_\mu,[v\cdot u,u^\mu]_+]]
-[[v\cdot u,u_\mu],[v\cdot u,u^\mu]_+]
=[u_\mu,[v\cdot u,[v\cdot u, u^\mu]_+]]\nonumber\\
& & [v\cdot u,[u_\mu,[v\cdot u,u^\mu]_+]]
-[[v\cdot u,u_\mu]_+,[v\cdot u,u^\mu]_+]_+
=[u_\mu,[v\cdot u,[v\cdot u, u^\mu]_+]_+]_+\nonumber\\
& & [v\cdot u,[u_\mu,[v\cdot u,u^\mu]_+]]
-[[v\cdot u,u_\mu],[v\cdot u,u^\mu]]_+
=[u_\mu,[v\cdot u,[v\cdot u, u^\mu]]]_+
\nonumber\\
& & [v\cdot u,[u_\mu,[v\cdot u,u^\mu]]]_+
-[[v\cdot u,u_\mu]_+,[v\cdot u,u^\mu]]
=[u_\mu,[v\cdot u,[v\cdot u, u^\mu]]]_+ 
\nonumber\\
& & [v\cdot u,[u_\mu,[v\cdot u,u^\mu]_+]_+]_+
-[[v\cdot u,u_\mu],[v\cdot u,u^\mu]_+]
=[u_\mu,[v\cdot u,[v\cdot u, u^\mu]_+]_+]_+,
\end{eqnarray}
and

(b) two identities in five terms:
\begin{eqnarray}
\label{eq:v.uv.Dset2}
& & [v\cdot{\rm D},[{\rm D}_\mu,[v\cdot u,u^\mu]_+]]
-{1\over4}[[v\cdot u,u^\mu],[v\cdot u,u_\mu]_+]
=[{\rm D}_\mu,[v\cdot{\rm D},[v\cdot u,u^\mu]_+]]\nonumber\\
& & [v\cdot{\rm D},[{\rm D}_\mu,[v\cdot u,u^\mu]_+]_+]_+-{1\over4}
[[v\cdot u,u^\mu],[v\cdot u,u_\mu]_+]
=[{\rm D}_\mu,[v\cdot{\rm D},[v\cdot u,u^\mu]_+]_+]_+.\nonumber\\
& & 
\end{eqnarray}

The reason for considering (\ref{eq:v.uv.Dset1}) and
(\ref{eq:v.uv.Dset2}) separately is because the terms contained in
them do not  mix. One can thus take  $i=2,3,4,19,20,36,37$ and
$i=28,44,45$ of Table 1 as the two sets of linearly independent terms.

(2) Similarly, one will need
to consider the following (4,0,0,0), (0,4,0,0) and (2,2,0,0) - type terms 
together:
\begin{eqnarray}
\label{eq:LCindoddanticoom1}
& & ({\rm D}_\nu,({\rm D}_\mu,({\rm D}^\nu,{\rm D}^\mu))),\ 
(({\rm D}_\nu,{\rm D}_\mu),({\rm D}^\nu,{\rm D}^\mu))\nonumber\\
& & (u_\nu,(u_\mu,(u^\nu,u^\mu))),
\ ((u_\nu,u_\mu),(u^\nu, u^\mu)),\nonumber\\
& & ({\rm D}_\nu,({\rm D}_\mu,(u^\nu,u^\mu))),\ 
(({\rm D}_\nu,{\rm D}_\mu),(u^\nu,u^\mu)))\nonumber\\
& & (u_\nu,(u_\mu,({\rm D}^\nu,{\rm D}^\mu))). 
\nonumber\\
& & 
\end{eqnarray}
As is shown in Appendix A, algebraic identities based
on (\ref{eq:oddanticoom}) and the
curvature relation (\ref{eq:curvature}), can be used to
select $i=13,14,25,26,28$ and $i=43,44,45$ of Table  1,
as the two sets of  linearly independent terms. 

For the (2,2,0,0)-type terms, using (\ref{eq:oddanticoom}), the
following result will be used for constructing
complete lists of O$(q^4)$ terms.
Using (\ref{eq:evenanticoom}), one gets 6+6=12 identities in
12+8=20 terms considered as following triplets:
\begin{eqnarray}
\label{eq:v.uv.D}
& & (v\cdot{\rm D},(({\rm D}_\mu,v\cdot u),u^\mu)),\
(u^\mu,(({\rm D}_\mu,v\cdot u),v\cdot{\rm D}),\
((v\cdot{\rm D},u^\mu),({\rm D}_\mu,v\cdot u)),\nonumber\\
& & ({\rm D}_\mu,((v\cdot{\rm D},u^\mu),v\cdot u)),\
(v\cdot u,((v\cdot{\rm D},u^\mu),{\rm D}_\mu)),
\end{eqnarray}
implying that one can take eight linearly independent
terms, say $i=71,..,78$.

\underline{$(iii),(iii)(A\leftrightarrow B)$ and $(iv)$ of (\ref{eq:list})}

One can show the following Jacobi-like identities to be true:
\begin{eqnarray}
\label{eq:evenanticoom}
 & & 
i[A,[B,[C,D]_+]_+] - i[[A,B]_+,[C,D]_+] = -(iii)(a)(A\leftrightarrow B)
\nonumber\\
& & i[A,[B,[C,D]_+]_+] - i[[A,B],[C,D]_+]_+ = (iii)(b)(A\leftrightarrow B)
\nonumber\\
& & i[A,[B,[C,D]_+]]_+ - i[[A,B]_+,[C,D]_+] = -(iii)(b)(A\leftrightarrow B)
\nonumber\\
& & i[A,[B,[C,D]]_+]_+ - i[[A,B],[C,D]] = (iii)(c)(A\leftrightarrow B)\nonumber\\
& & i[A,[B,[C,D]]_+]_+ - i[[A,B]_+,[C,D]_+] = -(iii)(d)(A\leftrightarrow B)
\nonumber\\
& & i[A,[B,[C,D]]] - i[[A,B],[C,D]] = (iii)(d)(A\leftrightarrow B)\nonumber\\
\end{eqnarray}
Again we have 6 identities in 12 terms, implying one can take 6 as 
linearly independent, say $(iii)\biggl((a) - (d)\biggr)$ 
and $(iv)\biggl((a),(b)\biggr)$ of (\ref{eq:list}).

(3) 
The identities (\ref{eq:evenanticoom}) along with the curvature relation,
require one  to consider the following category of  
$\epsilon^{\mu\nu\rho\lambda}$-dependent terms of the 
type (4,0,0,0), (0,4,0,0) and (2,2,0,0) terms together:
\begin{eqnarray} 
\label{eq:LCevenanticoom}
& & i\epsilon^{\mu\nu\rho\lambda}v_\rho\Biggl[
({\rm D}_\mu,({\rm D}_\nu,({\rm D}_\lambda,{\rm S}\cdot{\rm D}))),\ 
(u_\mu,(u_\nu,(u_\lambda,{\rm S}\cdot u))),\nonumber\\
& & 
({\rm S}\cdot{\rm D},({\rm D}_\mu,({\rm D}_\nu,[{\rm D}_\nu,{\rm D}_\lambda]))\
({\rm D}_\mu,({\rm S}\cdot{\rm D},[{\rm D}_\nu,{\rm D}_\lambda])),\nonumber\\
& & ({\rm S}\cdot u,(u_\mu,([u_\nu,u_\lambda])),
\ (u_\mu,({\rm S}\cdot u,[u_\nu,u_\lambda])),\nonumber\\
& & ([u_\mu,u_\nu],({\rm D}_\lambda,{\rm S}\cdot{\rm D})),
\ (u_\mu,(u_\nu,({\rm D}_\lambda,{\rm S}\cdot{\rm D}))),\nonumber\\
& & ({\rm D}_\mu,({\rm S}\cdot{\rm D},[u_\nu,u_\lambda])),\
({\rm S}\cdot{\rm D},({\rm D}_\mu,[u_\nu,u_\lambda])),\nonumber\\
& & (u_\mu,({\rm S}\cdot u,[{\rm D}_\nu,{\rm D}_\lambda])),\ 
({\rm S}\cdot u,(u_\mu,[{\rm D}_\nu,{\rm D}_\lambda])),\nonumber\\
& & ({\rm D}_\mu,({\rm D}_\nu,(u_\lambda,{\rm S}\cdot u))),
\ ([{\rm D}_\mu,{\rm D}_\nu],(u_\lambda,{\rm S}\cdot u))\Biggr]. 
\end{eqnarray}

One will need to use the following four equations  in addition:
\begin{equation}
\label{eq:antisymsym}
\epsilon^{\mu\nu\rho\lambda}
\biggl([[{\rm D}_\mu,{\rm D}_\nu]_+,[{\rm D}_\lambda,{\rm S}\cdot{\rm D}]]_+
= [[{\rm D}_\mu,{\rm D}_\nu]_+,[{\rm D}_\lambda,{\rm S}\cdot{\rm D}]_+]\biggr)
= 0,
\end{equation}
\begin{equation}
\label{eq:antisym}
\epsilon^{\mu\nu\rho\lambda}\biggl[
({\rm D}_\mu,({\rm D}_\nu,({\rm D}_\lambda,{\rm S}\cdot{\rm D}))
= - ({\rm D}_\nu,({\rm D}_\mu,({\rm D}_\lambda,{\rm S}\cdot{\rm D}))
\biggr],
\end{equation}
\begin{equation}
\label{eq:antisymsym1}
i\epsilon^{\mu\nu\rho\lambda}v_\rho\biggl([[u_\mu,u_\nu]_+,[{\rm D}_\lambda,{\rm S}\cdot{\rm D}]]_+
= [[u_\mu,u_\nu]_+,[\rm D_\lambda,\rm S\cdot\rm D]]_+=0\biggr),
\end{equation}
\begin{equation}
\label{eq:antisym1}
i\epsilon^{\mu\nu\rho\lambda} v_\rho\biggl(
(u_\mu,(u_\nu,({\rm D}_\lambda,{\rm S}\cdot{\rm D}))
= - (u_\nu,(u_\mu,({\rm D}_\lambda,{\rm S}\cdot{\rm D}))\biggr).
\end{equation}

Analogous to (\ref{eq:LCindoddanticoom}) and (\ref{eq:LCindoddanticoom1}),
one needs to do a  careful counting of the total number of identities
that one can write down using (\ref{eq:evenanticoom}),
(\ref{eq:curvature}) and (\ref{eq:antisymsym})-(\ref{eq:antisym1}), 
and the total number of terms in those
identities. We do the same in Appendix B.

A similar analysis can be carried out  for terms with $v\leftrightarrow\rm S$
in (\ref{eq:LCevenanticoom}).

(4) 
The identities (\ref{eq:evenanticoom}) along with the curvature relation,
require one  to consider the following category of  $\epsilon^{\mu\nu\rho\lambda}$-
dependent terms of the type (4,0,0,0), (0,4,0,0) and (2,2,0,0) terms together:
\begin{eqnarray} 
\label{eq:LCevenanticoom1}
& & i\epsilon^{\mu\nu\rho\lambda}v_\rho{\rm S}_\lambda\Biggl(
(u_\kappa,(u_\mu,(u^\kappa,u_\nu))),\ (u_\mu,(u_\kappa,(u^\kappa,u_\nu))),\
((u_\mu,u_\kappa),(u_\nu,u^\kappa)),\nonumber\\
& & (\rm D_\kappa,(\rm D_\mu,(\rm D^\kappa,\rm D_\nu))),\
((\rm D_\mu,(\rm D_\kappa,(\rm D^\kappa,\rm D_\nu))), \ 
((\rm D_\kappa,\rm D_\mu),(\rm D^\kappa,\rm D_\nu)),\nonumber\\
& & (({\rm D}_\kappa,({\rm D}_\mu,(u^\kappa,u_\nu))),\ 
(({\rm D}_\mu,({\rm D}_\kappa,(u^\kappa,u_\nu))),\ 
(({\rm D}_\kappa,{\rm D}_\mu),(u^\kappa,u_\nu)),\nonumber\\
& & (u_\mu,(u_\kappa,({\rm D}^\kappa,{\rm D}_\nu))),\
(u_\mu,(u_\kappa,({\rm D}^\kappa,{\rm D}_\nu)))\Biggr).
\end{eqnarray} 
As is shown in Appendix C, algebraic
identities based upon (\ref{eq:evenanticoom}) and the curvature relation
(\ref{eq:curvature}), can be used to select a set of linearly independent
terms from from (\ref{eq:LCevenanticoom1}). This is done in Appendix C.
  
A similar table can be constructed with
$(u_\kappa,{\rm D}^\kappa)\rightarrow
(v\cdot u, v\cdot\rm D)$ in (\ref{eq:LCevenanticoom1}).

For the (2,2,0,0)-type terms, using (\ref{eq:evenanticoom}), the
following results will be used for constructing
complete lists of O$(q^4)$ terms.

(a) The following set of 12+8
\footnote{The 8 is because 
$i\epsilon^{\mu\nu\rho\lambda}v_\rho(({\rm D}_\mu,{\rm S}\cdot u),
({\rm D}_\nu,u_\lambda))$ is common to both
$(i)$ and $(ii)$ in (\ref{eq:epsS.u1}).}
=20 terms need to be considered together as
the following triplets:
\begin{eqnarray}
\label{eq:epsS.u1}
& (i) & iv_\rho\epsilon^{\mu\nu\rho\lambda}\biggl(
({\rm D}_\mu,(({\rm D}_\nu,
u_\lambda),{\rm S}\cdot u)),\ ({\rm S}\cdot u,(({\rm D}_\nu,
u_\lambda),{\rm D}_\mu));\nonumber\\ 
& & (({\rm D}_\mu,{\rm S}\cdot u),({\rm D}_\nu, u_\lambda));\nonumber\\  
& (ii) & 
({\rm D}_\mu,(({\rm D}_\nu,{\rm S}\cdot u),u_\lambda)),\ 
(u_\lambda,(({\rm D}_\nu,{\rm S}\cdot u),{\rm D}_\mu)),\nonumber\\ 
& & (({\rm D}_\mu,{\rm S}\cdot u),({\rm D}_\nu, u_\lambda))\biggr).
\end{eqnarray}
Using (\ref{eq:evenanticoom}), 
one can eliminate all except eight, say:
$i=99,...,106$ of Table 1. A similar analysis
is carried out to select $i=77,..,84;91,..,98;107,..,114;$
$115,..,118,120,121,123$ $;122,124,..130;143,..,150;151,..,158;
159,..,166;$\\
$167,..,174$ of Table 1.

(b) Using (\ref{eq:evenanticoom}), one gets 6 identities in 12  
terms considered as following triplet:
\begin{eqnarray} 
\label{eq:dkapukap1}
&  &  
i\epsilon^{\mu\nu\rho\lambda} v_\rho{\rm S}_\lambda
\biggl(({\rm D}_\mu,(({\rm D}_\nu,u_\kappa),u^\kappa)),\ 
(({\rm D}_\mu,u^\kappa),({\rm D}_\nu,u^\kappa)),\nonumber\\
& & (u^\kappa,(({\rm D}_\mu,u^\kappa),{\rm D}_\nu))\biggr),
\end{eqnarray}
implying that one can take six linearly independent terms,
say $i=131,..,136$ of Table 1. A similar analysis
is carried out to select $i=71,..,76;85,..,90;$ $137,..142;175,..,180;$
$181,..,186$ of Table 1.

For (0,4,0,0)-type
terms, writing $u_\mu=u_\mu^a\tau^a$, one will need to consider the following reductions:
\begin{eqnarray}
\label{eq:umured}  
& (i) & [\tau^a,[\tau^b,[\tau^c,\tau^d]_+]]=[\tau^a,[\tau^b,[\tau^c,\tau^d]]_+]
=0;\ \nonumber\\ 
& & [\tau^a,[\tau^b,[\tau^c,\tau^d]]]_+,\ [\tau^a,[\tau^b,[\tau^c,\tau^d]_+]_+]_+ 
\neq0;\nonumber\\
& (ii) &  [[\tau^a,\tau^b],[\tau^c,\tau^d]_+]=0;\ 
[[\tau^a,\tau^b],[\tau^c,\tau^d]]_+,\ 
[[\tau^a,\tau^b]_+,[\tau^c,\tau^d]_+]_+\neq0;\nonumber\\
& (iii) & i[\tau^a,[\tau^b,[\tau^c,\tau^d]_+]]_+=0;\nonumber\\
& & i[\tau^a,[\tau^b,[\tau^c,\tau^d]]_+]_+,\ 
i[\tau^a,[\tau^b,[\tau^c,\tau^d]_+]_+],\
i[\tau^a,[\tau^b,[\tau^c,\tau^d]]]\neq0;\nonumber\\
& (iv) & i[[\tau^a,\tau^b]_+,[\tau^c,\tau^d]_+]=0;\ 
i[[\tau^a,\tau^b],[\tau^c,\tau^d]_+]_+,\
i[[\tau^a,\tau^b],[\tau^c,\tau^d]]\neq0.\nonumber\\
& & 
\end{eqnarray}

\underline{\underline{$p\neq0$ or $q\neq0$ in (\ref{eq:prodbb})$\equiv(A,(B,C))$}}

This includes $(v)$ and $(vi)$ of (\ref{eq:list}).

\underline{($v$) of (\ref{eq:list})} 

By using  the following three Jacobi-like identities which  
are generalized Jacobi 
identities as used in graded Lie algebra in supersymmetric theories 
\begin{eqnarray}
\label{eq:oddanticoom1}
& & i[A,[B,C]]_+ - i[[A,B],C]_+ = i[B,[A,C]_+]\nonumber\\
& & i[A,[B,C]]_+ - i[[A,B]_+,C] = -i[B,[A,C]]_+\nonumber\\
& & i[A,[B,C]_+] - i[[A,B]_+,C] = -i[B,[A,C]_+],
\end{eqnarray}
one sees that one needs to consider only three of the six terms that figure in 
the above three identities, say $i[A,[B,C]_+], i[A,[B,C]]_+$ and
$i[B,[A,C]_+]$ as linearly independent terms. These three 
identities are similar to the ones that occur in SUSY graded Lie algebra for
$A,C\equiv$ fermionic and $B\equiv$ bosonic fields, $A,B\equiv$ fermionic and 
$C\equiv$ bosonic fields,  
and $A,B,C\equiv$ fermionic fields, respectively. The identities
in (\ref{eq:oddanticoom1}) are used in, e.g., 
$\epsilon^{\mu\nu\rho\lambda}$-independent (1,1,0,1)-type terms. When
applying (\ref{eq:oddanticoom1}) to (2,0,1,0), because
of (\ref{eq:curvature}), one will need to consider the following
terms together:
\begin{eqnarray}
\label{eq:epsDuchi+}
& & i\epsilon^{\mu\nu\rho\lambda}v_\rho{\rm S}_\lambda\biggl(
({\rm D}_\mu,({\rm D}_\nu,\chi_+)),\
([{\rm D}_\mu,{\rm D}_\nu],\chi_+);\nonumber\\
& & (u_\mu,(u_\nu,\chi_+)),\ ([u_\mu,u_\nu],\chi_+)\biggr).
\end{eqnarray}
Further noting that $u_\mu$ is an isovector and $\chi_+$ is an
isoscalar, one sees that:
\begin{equation}
\label{eq:uchi+0}
[u_\mu,\chi_+]=0.
\end{equation}
Applying (\ref{eq:oddanticoom1}) and (\ref{eq:uchi+0}) to 
(\ref{eq:epsDuchi+}), one sees that one can take two linearly independent
terms, say 
$i\epsilon^{\mu\nu\rho\lambda}v_\rho{\rm S}_\lambda\biggl(
[{\rm D}_\mu,[{\rm D}_\nu,\chi_+]]_+,
[[u_\mu,u_\nu],\chi_+]_+\biggr)$.

\underline{$(vi)$ of (\ref{eq:list})} 

By considering the following three Jacobi(-like) identities : 
\begin{eqnarray} 
\label{eq:evencoom}
& & [A,[B,C]] - [[A,B],C] = [B,[A,C]]\nonumber\\
& & [A,[B,C]_+]_+ - [[A,B]_+,C]_+ = -[B,[A,C]]\nonumber\\
& & [A,[B,C]] - [[A,B]_+,C]_+ = -[B,[A,C]_+]_+,
\end{eqnarray}
one needs to consider only three of the six terms, say $[A,[B,C]], [A,[B,C]_+]_+$
and $[B,[A,C]]$.
The identities in (\ref{eq:evencoom}) are used in, e.g., 
$\epsilon^{\mu\nu\rho\lambda}$-dependent (1,1,0,1)-type terms.

\subsection{O$(q^4,\phi^{2n+1}$) Terms}

In this subsection, we consider the reduction in the 
number of O$(q^4,\phi^{2n+1})$ terms because of algebraic identities.
The discussion in this subsection will be much briefer than the preceding
(subsection).

\underline{$(i)-(iv)$ of (\ref{eq:list})$^\prime$}

(1) The identities (\ref{eq:oddanticoom})
are the same for O$(q^4,\phi^{2n+1})$
except for an overall factor of $i$. We will denote the analogue
of (\ref{eq:oddanticoom}) for O$(q^4,\phi^{2n+1})$ terms 
as (\ref{eq:oddanticoom})$^\prime$.\footnote{Similarly, the analogs of
(\ref{eq:list}), (\ref{eq:evenanticoom}), (\ref{eq:oddanticoom1}) 
and (\ref{eq:evencoom})  will be denoted
by (\ref{eq:list})$^\prime$, 
(\ref{eq:evenanticoom})$^\prime$, (\ref{eq:oddanticoom1})$^\prime$ 
and (\ref{eq:evencoom})$^\prime$.}
Using it together with (\ref{eq:curvature}), one sees that
one needs to consider the following set of terms together:
\begin{eqnarray}
\label{eq:oddset1}
& & i(v\cdot{\rm D},({\rm D}_\mu,({\rm D}^\mu,{\rm S}\cdot u))),\
i({\rm D}_\mu,(v\cdot{\rm D},({\rm D}^\mu,{\rm S}\cdot u))),\ 
i((v\cdot{\rm D},{\rm D}_\mu),({\rm D}^\mu,{\rm S}\cdot u)),\nonumber\\
& & i({\rm S}\cdot u,({\rm D}_\mu,({\rm D}^\mu,v\cdot{\rm D}))),\
i({\rm D}_\mu,({\rm S}\cdot u,({\rm D}^\mu,v\cdot{\rm D})));\nonumber\\
&  & i(v\cdot u,(u_\mu,({\rm S}\cdot u,{\rm D}^\mu))),\
i(u_\mu,(v\cdot u,({\rm S}\cdot u,{\rm D}^\mu))),\
i((v\cdot u,u_\mu),({\rm S}\cdot u,{\rm D}^\mu)),\nonumber\\
& & i({\rm D}_\mu,({\rm S}\cdot u,(u^\mu,v\cdot u)))\
i({\rm S}\cdot u,({\rm D}_\mu,(u^\mu,v\cdot u)));\nonumber\\
& & i({\rm D}_\mu,({\rm D}^\mu,(v\cdot{\rm D},{\rm S}\cdot u))),\
i({\rm D}^2,(v\cdot{\rm D},{\rm S}\cdot u)),\nonumber\\
& & i({\rm S}\cdot u,(v\cdot{\rm D},{\rm D}^2)),\ 
i(v\cdot{\rm D},({\rm S}\cdot u,{\rm D}^2)).
\end{eqnarray}
One can show that of the terms listed
in (\ref{eq:oddset1}), one can take 
$i=208,209,210,245,..,252,300$ of Table 2 as a set of linearly independent
terms.

Similarly, using (\ref{eq:oddanticoom})$^\prime$ and (\ref{eq:curvature}),
one can show that of
\begin{eqnarray}
\label{eq:oddset2}
& (a) & i(v\cdot{\rm D},({\rm D}_\mu,({\rm S}\cdot{\rm D},u^\mu))),\
i({\rm D}_\mu,(v\cdot{\rm D},({\rm S}\cdot{\rm D},u^\mu))),\
i((v\cdot{\rm D},{\rm D}_\mu),({\rm S}\cdot{\rm D},u^\mu)),\nonumber\\
& & i(u_\mu,({\rm S}\cdot{\rm D},({\rm D}^\mu,v\cdot{\rm D})))\
i({\rm S}\cdot{\rm D},(u_\mu,({\rm D}^\mu,v\cdot{\rm D})));\nonumber\\
& & i(v\cdot u,(u_\mu,(u^\mu,{\rm S}\cdot {\rm D}))),\
i(u_\mu,(v\cdot u,(u^\mu,{\rm S}\cdot {\rm D}))),\ 
i((v\cdot u,u_\mu),(u^\mu,{\rm S}\cdot{\rm D})),\nonumber\\
& & i({\rm S}\cdot {\rm D},(u_\mu,(u^\mu,v\cdot u))),\
i(u_\mu,({\rm S}\cdot {\rm D},(u^\mu,v\cdot u)))\nonumber\\
& & i(u^\mu,(u_\mu,(v\cdot u,{\rm D}^\mu))),\
i({\rm S}\cdot u,(u_\mu,v\cdot u,{\rm D}^\mu)),\nonumber\\
& & i((u_\mu,{\rm S}\cdot u),(v\cdot u,{\rm D}^\mu)),\
i({\rm D}_\mu,(v\cdot u,(u^\mu,{\rm S}\cdot u))),
\nonumber\\
& & i(v\cdot u,({\rm D}_\mu,(u^\mu,{\rm S}\cdot u)));\nonumber\\
& & i(u^\mu,(u_\mu,(v\cdot u,{\rm S}\cdot{\rm D}))),\ 
i(u^2,(v\cdot u,{\rm S}\cdot{\rm D})),\nonumber\\
& & i({\rm S}\cdot{\rm D},(v\cdot u,u^2)),
\ i(v\cdot u,({\rm S}\cdot{\rm D},u^2)).
\end{eqnarray}
$i=213,214,244,261,..,267,268,315$ 
of Table 2 form a set of linearly independent terms;
\begin{eqnarray}
\label{eq:oddset3}
& (b) & 
i(v\cdot{\rm D},({\rm S}\cdot{\rm D},({\rm D}^\mu,u_\mu))),\
i({\rm S}\cdot{\rm D},(v\cdot{\rm D},({\rm D}^\mu,u_\mu))),\
i((v\cdot{\rm D},{\rm S}\cdot{\rm D}),({\rm D}^\mu,u_\mu)),\nonumber\\
& & i(u_\mu,({\rm D}^\mu,(v\cdot{\rm D},{\rm S}\cdot{\rm D}))),\
i({\rm D}_\mu,(u^\mu,(v\cdot{\rm D}{\rm S}\cdot{\rm D})));\nonumber\\
& & 
i(v\cdot  u,({\rm S}\cdot u,(u^\mu,{\rm D}_\mu))),\
i({\rm S}\cdot u,(v\cdot u,({\rm D}^\mu,u_\mu))),\
i((v\cdot u,{\rm S}\cdot u),({\rm D}^\mu,u_\mu)),\nonumber\\
& & i({\rm D}_\mu,(u^\mu,(v\cdot u,{\rm S}\cdot u))),\
i(u_\mu,({\rm D}^\mu,(v\cdot u,{\rm S}\cdot u)))
\end{eqnarray}
$i=211,212,221,253,..260,301$
of Table 2 form a set of linearly independent terms;

\begin{eqnarray}
\label{eq:oddset4}
& (c) & 
i({\rm D}_\mu,({\rm S}\cdot{\rm D},(v\cdot{\rm D},u^\mu))),\
i({\rm S}\cdot{\rm D},({\rm D}_\mu,(v\cdot{\rm D},u^\mu))),\
i(({\rm S}\cdot{\rm D},{\rm D}_\mu),(v\cdot{\rm D},u^\mu)),\nonumber\\
& & i(u_\mu,(v\cdot{\rm D},({\rm D}^\mu,{\rm S}\cdot{\rm D}))),\
i(v\cdot{\rm D},(u_\mu,({\rm D}^\mu,{\rm S}\cdot{\rm D}))),\nonumber\\
& & i(u_\mu,({\rm S}\cdot u,(u^\mu,v\cdot {\rm D}))),\
i({\rm S}\cdot u,(u_\mu,(u^\mu,v\cdot {\rm D}))),\
i((u_\mu,{\rm S}\cdot u),(u^\mu,v\cdot {\rm D}))),\nonumber\\
& & i(v\cdot {\rm D},(u_\mu,(u^\mu,{\rm S}\cdot u)),\ 
i(u_\mu,(v\cdot {\rm D},(u^\mu,{\rm S}\cdot u));\nonumber\\
& & i(u_\mu,(u^\mu,({\rm S}\cdot u,v\cdot{\rm D}))),\
i(u^2,({\rm S}\cdot u,v\cdot{\rm D})),\nonumber\\
& & i(v\cdot{\rm D},({\rm S}\cdot u,u^2)),\ 
i({\rm S}\cdot u,(v\cdot{\rm D},u^2)).
\end{eqnarray}
$i=215,216,219,269,..,276,303$.
of Table 2 form a set of linearly independent terms;
\begin{eqnarray}
\label{eq:oddset5}
& (d) &
i({\rm D}_\mu,({\rm S}\cdot{\rm D},({\rm D}^\mu,v\cdot u))),\
i({\rm S}\cdot{\rm D},({\rm D}_\mu,({\rm D}^\mu,v\cdot u))),\
i(({\rm D}_\mu,{\rm S}\cdot{\rm D}),({\rm D}^\mu,v\cdot u))),\nonumber\\
& & i(v\cdot u,({\rm D}_\mu,({\rm D}^\mu,{\rm S}\cdot{\rm D})),\ 
i({\rm D}_\mu,(v\cdot u,({\rm D}^\mu,{\rm S}\cdot{\rm D}));\nonumber\\
& & i(u_\mu,({\rm S}\cdot u,(v\cdot u,{\rm D}^\mu))),\
i({\rm S}\cdot u,(u_\mu,(v\cdot u,{\rm D}^\mu))),\
i(({\rm S}\cdot u,u_\mu),(v\cdot u,{\rm D}^\mu)),\nonumber\\
& & i({\rm D}_\mu,(v\cdot u,(u^\mu,{\rm S}\cdot u))),\
i(v\cdot u,({\rm D}_\mu,(u^\mu,{\rm S}\cdot u)));\nonumber\\
& & i({\rm D}_\mu,({\rm D}^\mu,({\rm S}\cdot{\rm D},v\cdot u))),\
i({\rm D}^2,({\rm S}\cdot{\rm D},v\cdot u)),\nonumber\\
& & i(v\cdot u,({\rm S}\cdot{\rm D},{\rm D}^2)),\ 
i({\rm S}\cdot{\rm D},(v\cdot u,{\rm D}^2))
\end{eqnarray}
$i=217,218,220,227,302,307,..,312,316$
of Table 2 form a set of linearly independent terms;
\begin{eqnarray}
\label{eq:oddset6}
& (e) & 
i(v\cdot{\rm D},({\rm S}\cdot{\rm D},(v\cdot {\rm D},v\cdot u))),\
i({\rm S}\cdot{\rm D},(v\cdot{\rm D},(v\cdot {\rm D},v\cdot u))),\
i((v\cdot{\rm D},{\rm S}\cdot{\rm D}),(v\cdot {\rm D},v\cdot u))),\nonumber\\
& & i(v\cdot u,(v\cdot{\rm D},(v\cdot{\rm D},{\rm S}\cdot{\rm D})),\
i(v\cdot {\rm D},(v\cdot u,(v\cdot{\rm D},{\rm S}\cdot{\rm D}));\nonumber\\
& & 
i(v\cdot u,({\rm S}\cdot u,(v\cdot u,v\cdot {\rm D}))),\
i({\rm S}\cdot u,(v\cdot u,(v\cdot u,v\cdot {\rm D}))),\
i(({\rm S}\cdot u,v\cdot u),(v\cdot u,v\cdot {\rm D})),\nonumber\\
& & i(v\cdot u,(v\cdot{\rm D},(v\cdot u,{\rm S}\cdot u)),\
i(v\cdot {\rm D},(v\cdot u,(v\cdot u,{\rm S}\cdot u))\nonumber\\
&  & i(v\cdot u,({\rm S}\cdot{\rm D},(v\cdot{\rm D})^2)),\ 
i({\rm S}\cdot{\rm D},(v\cdot u,(v\cdot{\rm D})^2)),\nonumber\\
& & i((v\cdot{\rm D})^2,({\rm S}\cdot{\rm D},v\cdot u)),\ 
i(v\cdot{\rm D},(v\cdot{\rm D},({\rm S}\cdot{\rm D},v\cdot u)));\nonumber\\
& & i(v\cdot{\rm D},{\rm S}\cdot u,(v\cdot u)^2)),\
i({\rm S}\cdot u,(v\cdot{\rm D},(v\cdot u)^2)),\nonumber\\
& & i((v\cdot u)^2,({\rm S}\cdot u,v\cdot{\rm D})),\
i(v\cdot u,(v\cdot u,({\rm S}\cdot u,v\cdot{\rm D})).
\end{eqnarray}
$i=222,..,225,277,..,284,314$
of Table 2 form a set of linearly independent terms;
\begin{eqnarray}
\label{eq:oddset9}
& (f) & i(v\cdot{\rm D},(v\cdot{\rm D},(v\cdot{\rm D},{\rm S}\cdot u))),\
i((v\cdot{\rm D})^2,(v\cdot{\rm D},{\rm S}\cdot u)),\nonumber\\
& & i[{\rm S}\cdot u,(v\cdot{\rm D})^3]_+,
\ i(v\cdot{\rm D},({\rm S}\cdot u,(v\cdot{\rm D})^2))
\end{eqnarray}
$i=229,230$
of Table 2 form a set of linearly independent terms.
For $u\leftrightarrow\rm D$ in (\ref{eq:oddset9}),
$i=305,306$
of Table 2 form a set of linearly independent terms.

(2) Using (\ref{eq:evenanticoom})$^\prime$ and (\ref{eq:curvature}),
one sees that one has  to consider the following set of terms together:
\begin{eqnarray}
\label{eq:oddset11}
& & \epsilon^{\mu\nu\rho\lambda}\biggl(
({\rm D}_\mu,({\rm D}_\nu,({\rm D}_\rho,u_\lambda))),\
([{\rm D}_\mu,{\rm D}_\nu],({\rm D}_\rho,u_\lambda))),\nonumber\\
& & (u_\mu,({\rm D}_\nu,[{\rm D}_\rho,{\rm D}_\lambda])),\
({\rm D}_\mu,(u_\nu,[{\rm D}_\rho,{\rm D}_\lambda]));\nonumber\\
& & (u_\mu,(u_\nu,(u_\rho,{\rm D}_\lambda))),\
([u_\mu,u_\nu],(u_\rho,{\rm D}_\lambda),\nonumber\\
& & ({\rm D}_\mu,(u_\nu,[u_\rho, u_\lambda])),\
(u_\mu,({\rm D}_\nu,[u_\rho,u_\lambda]))\biggr).
\end{eqnarray}
One can show that of  the terms
listed in (\ref{eq:oddset11}), one can take $i=231,232,233,285,286$
of Table 2 as a set of linearly independent terms.

A similar  analysis is carried out to select $i=234,235,236,287,288$;
$237,238,239,290,291$ of Table 2.

Also, using (\ref{eq:evenanticoom})$^\prime$ and (\ref{eq:curvature}),
one can show that of the following set of terms:
\begin{eqnarray}
\label{eq:oddset14}
& & \epsilon^{\mu\nu\rho\lambda}v_\rho\biggl(
(v\cdot{\rm D},({\rm D}_\mu,({\rm D}_\nu,u_\lambda))),\
((v\cdot{\rm D},{\rm D}_\mu),({\rm D}_\nu,u_\lambda))),\
({\rm D}_\mu,(v\cdot{\rm D},({\rm D}_\nu,u_\lambda))),\nonumber\\
& & (u_\mu,({\rm D}_\nu,(v\cdot{\rm D},{\rm D}_\lambda))),\
({\rm D}_\mu,(u_\nu,(v\cdot{\rm D},{\rm D}_\lambda)));\nonumber\\
& & (v\cdot u,(u_\mu,(u_\nu,{\rm D}_\lambda))),\
(u_\mu,(v\cdot u,(u_\nu,{\rm D}_\lambda))),\
((u_\mu,v\cdot u),(u_\nu,{\rm D}_\lambda)),\nonumber\\
& & ({\rm D}_\mu,(u_\nu,(v\cdot u,u_\lambda))),\
(u_\mu,({\rm D}_\nu,(v\cdot u,u_\lambda)))\biggr),
\end{eqnarray}
one need consider 13 independent terms, say, 
$i=240,..,243,292,..,300$ of Table 2.

\underline{\underline{$p\neq0$ or $q\neq0$ in (\ref{eq:prodbb})$\equiv(A,(B,C))$}}

\underline{$(v)$ of (\ref{eq:list})$^\prime$}

Using (\ref{eq:curvature}) and (\ref{eq:oddanticoom1})$^\prime$, 
one sees that one will have to consider the following set of terms  together:
\begin{eqnarray}
\label{eq:oddset15}
& & 
({\rm S}\cdot{\rm D},(v\cdot{\rm D},\chi_-)),\
(v\cdot{\rm D},({\rm S}\cdot{\rm D},\chi_-)),\
((v\cdot{\rm D},{\rm S}\cdot{\rm D}),\chi_-);\nonumber\\
& & (v\cdot u,({\rm S}\cdot u,\chi_-)),\ 
({\rm S}\cdot u,(v\cdot u,\chi_-)),\ 
(({\rm S}\cdot u,v\cdot u),\chi_-).
\end{eqnarray}
Further noting that $u_\mu$ and $\chi_-$ are isovectors, we see
that
\begin{equation}
\label{eq:uchi-0}
[v\cdot u,[{\rm S}\cdot u,\chi_-]_+]=[{\rm S}\cdot u,[v\cdot u,\chi_-]_+]=
[[v\cdot u,{\rm S}\cdot u]_+,\chi_-]=0.
\end{equation}
Applying (\ref{eq:curvature}), (\ref{eq:oddanticoom1})$^\prime$
and (\ref{eq:uchi-0}) to  (\ref{eq:oddset15}),  we see 
that we get two linearly independent terms, say,
$[{\rm S}\cdot{\rm D},[v\cdot{\rm D},\chi_-]]_+,\
i[{\rm S}\cdot u,[v\cdot u,\chi_-]]_+$.

\underline{$(vi)$ of (\ref{eq:list})$^\prime$}

Using (\ref{eq:evencoom})$^\prime$, one sees that one will have
to consider the following set of terms together:
\begin{equation}
\label{eq:oddset16}
i({\rm S}\cdot{\rm D},(v\cdot u,\chi_+)),
\ i(v\cdot u,({\rm S}\cdot {\rm D},\chi_+)),\
i(({\rm S}\cdot{\rm D},v\cdot u),\chi_+);
v\leftrightarrow{\rm S}.
\end{equation}
Further, noting that ${\rm D}_\mu\equiv$isoscalar($\equiv\partial_\mu$)
+isovector($\equiv\Gamma_\mu$), $u_\nu$ is an isovector and $\chi_+$ is
an isoscalar, we see
\begin{equation}
\label{eq:Duchi+0}
[[{\rm S}\cdot{\rm D},v\cdot u],\chi_+]=[[v\cdot{\rm D},{\rm S}\cdot u],\chi_+]=
[v\cdot u,[{\rm S}\cdot{\rm D},\chi_+]]=[{\rm S}\cdot u,[v\cdot{\rm D},\chi_+]]=0.
\end{equation}
Applying (\ref{eq:curvature}),
(\ref{eq:evencoom})$^\prime$ and (\ref{eq:Duchi+0}) to (\ref{eq:oddset16}),
we see that we get one linearly independent
term,  say, $[[{\rm S}\cdot{\rm D},v\cdot u]_+,\chi_+]_+,$; similarly for
$v\leftrightarrow{\rm S}$.

Note that  because of  parity constraints and
the algebra of the $\rm S_\mu$s (See \cite{1n}), 
there are no Levi Civita-dependent 
(2,0,0,1)-, (0,2,0,1)- and (1,1,1,0)-type terms.

\section{The Lists of Independent Terms in ${\cal L}_{\rm HBChPT}$
(off-shell nucleons)}

In this section, using (\ref{eq:phoffsh}), and the algebraic reductions
of Section 3, we list  all possible
${\cal A}$-type terms of O$(q^4,\phi^{2n})$, 
and O$(q^4,\phi^{2n+1})$ in Tables
1 and 2, that are allowed by (\ref{eq:phoffsh})
and have not been eliminated in Section 3.
As noted in Section 2 (and \cite{1n}), for off-shell nucleons,
$\gamma^0{\cal B}^\dagger\gamma^0{\cal C}^{-1}{\cal B}\in {\cal A}$.
Hence, it is sufficient to list only ${\cal A}$-type terms (for off-shell
nucleons).

Let us summarize the basis of construction of list of linearly independent
terms using (\ref{eq:phoffsh}) and section 3. Using
the notation of \cite{krause}, one groups terms allowed by (\ref{eq:phoffsh})
as in (\ref{eq:list}). Then, following Section 3, we write down
all possible identities by considering different groups 
of terms together (e.g. $(iii)$ and $(iv)$ of (\ref{eq:list})).
Sometimes, one has to consider together
several types of  terms but belonging
to the same types of groups.
There are two different cases to be considered: (a) the
different term types involve
different permutations of the same building blocks, and (b) some
of the building blocks of the different term types are different. 
We will illustrate both cases by considering an example each. 

(a) For example, one has to consider the triplet of terms
\footnote{The reason for considering the terms as triplets is
because (\ref{eq:evenanticoom}) shows there are linear 
relations involving them.}
\begin{eqnarray}
\label{eq:epsS.u3}
& & i\epsilon^{\mu\nu\rho\lambda}v_\rho{\rm S}_\lambda
({\rm D}_\mu,(({\rm D}_\nu,u_\lambda),{\rm S}\cdot u))\in (iii)\ {\rm of}
\ (\ref{eq:list}),\nonumber\\
& & i\epsilon^{\mu\nu\rho\lambda}v_\rho{\rm S}_\lambda
({\rm S}\cdot u,(({\rm D}_\nu,u_\lambda),{\rm D}_\mu))
\in (iii) (``A\leftrightarrow B")\  {\rm of}\ (\ref{eq:list}),\nonumber\\
& & i\epsilon^{\mu\nu\rho\lambda}v_\rho{\rm S}_\lambda
(({\rm D}_\mu,{\rm S}\cdot u),(({\rm D}_\nu,u_\lambda)),\in 
(iv)\ {\rm  of}\ (\ref{eq:list}),
\end{eqnarray}
in conjunction with 
\begin{eqnarray}
\label{eq:epsS.u4}
& & i\epsilon^{\mu\nu\rho\lambda}v_\rho{\rm S}_\lambda
({\rm D}_\mu,(({\rm D}_\nu,{\rm S}\cdot u),u_\lambda))\in
(iii)\ {\rm  of}\ (\ref{eq:list}),\nonumber\\
& & i\epsilon^{\mu\nu\rho\lambda}v_\rho{\rm S}_\lambda
(u_\lambda,(({\rm D}_\nu,{\rm S}\cdot u),{\rm D}_\mu))\in
(iii) (``A\leftrightarrow B")\  {\rm of}\ (\ref{eq:list}),\nonumber\\
& & i\epsilon^{\mu\nu\rho\lambda}v_\rho{\rm S}_\lambda
(({\rm D}_\mu,u_\lambda),({\rm D}_\nu,{\rm S}\cdot u))\in
(iv)\ {\rm of} (\ref{eq:list}).
\end{eqnarray}
[Note that 
$i\epsilon^{\mu\nu\rho\lambda}v_\rho{\rm S}_\lambda
({\rm D}_\mu,(({\rm D}_\nu,u_\lambda),{\rm S}\cdot u))$ and
$i\epsilon^{\mu\nu\rho\lambda}v_\rho{\rm S}_\lambda
({\rm D}_\mu,(({\rm D}_\nu,{\rm S}\cdot u),u_\lambda))$ both $\in
(iii)$ of (\ref{eq:list}), and so forth.] The reason for doing
so is, as we notice in the example, 
$i\epsilon^{\mu\nu\rho\lambda}v_\rho{\rm S}_\lambda
(({\rm D}_\mu,u_\lambda),({\rm D}_\nu,{\rm S}\cdot u))\in (iv)$
of (\ref{eq:list}), is common to both the triplets.

(b) For example,
one needs to  consider the following
triplet of terms
\footnote{The reason for considering the terms as  triplets
is because (\ref{eq:oddanticoom1}) shows that there are linear 
relations between them.}
:
\begin{equation}
\label{eq:DDchi-1}
(v\cdot{\rm D},({\rm S}\cdot{\rm D},\chi_-)),\ ({\rm S}\cdot{\rm D},
(v\cdot{\rm D},\chi_-)),\
((v\cdot{\rm D},{\rm S}\cdot{\rm D},\chi_-)
\end{equation}
in conjunction with
\begin{equation}
\label{eq:uuchi-}
[v\cdot u,[{\rm S}\cdot u,\chi_-]]_+,\ [{\rm S}\cdot u,[v\cdot u,\chi_-]]_+,\
[[v\cdot u,{\rm S}\cdot u],\chi_-]_+.
\footnote{Note that because of (\ref{eq:uchi-0}), one
need not consider
$[v\cdot u,[{\rm S}\cdot u,\chi_-]_+],\ [{\rm S}\cdot u,[v\cdot u,\chi_-]_+],\
[[v\cdot u,{\rm S}\cdot u]_+,\chi_-]$.} 
\end{equation}
This is so because of (\ref{eq:curvature}).

Using the algebraic identities
of Section 3, if we end up with $m$ independent identities in $n(>m)$ terms,
then we  can take $(n-m)$ linearly independent terms. 
So, for the above examples,  (a) $m=6+6=12$ and $n=12+8=20$,
implying that one can take eight linearly independent terms;
(b) $m=7$ and $n=9$ implying that one can take two linearly independent
terms.  Care has to be
taken in the choice of those $(n-m)$ terms, namely, no subset of these terms
should satisfy any relations. Let us first consider the example in (a). 
A valid choice is the set of eight terms in Table 1: $i=99,..,106$. The
following set of eight terms, however, {\it are not} linearly independent:
$iv_\rho\epsilon^{\mu\nu\rho\lambda}\biggl(
[{\rm D}_\mu,[[{\rm D}_\nu,u_\lambda]_+,{\rm S}\cdot u]_+];$
$[{\rm S}\cdot u,[[{\rm D}_\nu,u_\lambda]_+,{\rm D}_\mu]_+];\
[[{\rm D}_\mu,{\rm S}\cdot u]_+,[{\rm D}_\nu,u_\lambda]_+];$
$[{\rm D}_\mu,[[{\rm D}_\nu,u_\lambda],{\rm S}\cdot u]]\biggr);$
$i=101,..,104$(of Table 1). The reason 
is that from the first identity in (\ref{eq:evenanticoom}), one sees 
that the first three terms are not linearly independent.
Note that the allowed set of eight terms  is {\it not a unique choice}.
An equivalent choice would be
$iv_\rho\epsilon^{\mu\nu\rho\lambda}
[[{\rm D}_\mu,{\rm S}\cdot u]_+,[{\rm D}_\nu,u_\lambda]_+];$ 
$i=99,..,101,103,..,106$ of Table 1.
Now, coming to the example considered in (b).
Two valid equivalent  choices are:
$[{\rm S}\cdot{\rm D},[v\cdot{\rm D},\chi_-]]_+,
\ [{\rm S}\cdot u,[v\cdot u,\chi_-]]_+$
or
$[{\rm S}\cdot{\rm D},[v\cdot{\rm D},\chi_-]_+],
\ [[{\rm S}\cdot u,v\cdot u],\chi_-]_+$;
and, e.g.,  
$ [{\rm S}\cdot u,[v\cdot u,\chi_-]]_+,\ [v\cdot u,[{\rm S}\cdot u,\chi_-]]_+$
are not linearly independent.

Even though the phase rule (\ref{eq:phoffsh}) and linear independence of
terms are
sufficient for listing terms in the O$(q^4$) HBChPT Lagrangian for off-shell
nucleons, however, if for a given choice of terms and 
group of terms in (\ref{eq:list}), we find similar group of terms
in \cite{mms}, then while  listing the $(n-m)$ terms,
preference is  given to including terms that also
figure in Table 1 of \cite{mms}. The reason for doing
the same is that this allows for an easy identification of 
the finite terms, given
that the divergent (counter) terms have been worked out in \cite{mms}.

In tables 1 and 2, the allowed 
4-tuples $(m,n,p,q)$  are listed along with the corresponding
terms. The main aim is to find the number of finite O($q^4)$
terms, given that the UV divergent terms have already been worked out in
\cite{mms}. For this purpose, the terms in tables 1 and 2
are labeled as F denoting the finite terms and D denoting the  
divergent terms. For the purpose of comparison with 
\cite{mms}, we have also indicated
which terms in table 1 of \cite{mms} (setting the external fields
to zero and assuming isospin symmetry) the D-type terms correspond
to.  The LECs of O$(q^4)$ terms in
\cite{mms} are denoted by $d_i, i=1$ to 199. 
Further, the $i=188$ term in Table 1 of \cite{mms} should have
${\rm S}_\rho$ instead of $v_\rho$.

\subsection{O$(q^4,\phi^{2n})$ Terms}
 
These terms are listed in  Table 1.
One  gets a total of 207 O$(q^4,\phi^{2n})$
terms, the LECs of two of which, as will be shown in Section 5,
are fixed relative to those of lower order terms. 
The last column of Table 1 will be explained
in Section 7.

\subsection{O($q^4,\phi^{2n+1})$ Terms}

These terms are listed in  Table 2.
One gets a total of 113 O$(q^4,\phi^{2n+1})$ terms.
The last column of Table 2 will ll be explained in Section 7.

Overall, one gets 230 finite and 90 divergent (counter) terms
at O$(q^4)$. 

\section{Further Reduction in LECs due to Reparameterization Invariance}

In this section, we discuss those O$(q^4)$ terms whose
low energy coupling constants (LECs) are fixed relative to
O$(q^{1,2,3})$ terms.

We first show that for off-shell nucleons, 
all such O$(q^4)\ A-$type terms belong to a 
certain class of (2,2,0,0)-type of terms, and the O$(q^2,\phi^{2n})$
term is ${\bar{\rm H}}(v\cdot u)^2\rm H$, 
and the O$(q^3,\phi^{2n})$ term is 
$i{\bar{\rm H}}u_\mu v\cdot u{\rm D}^\mu{\rm H}$+h.c. 
We then show 
that by using the formulation of Luke and Manohar \cite {lm} for what is called
reparameterization invariance (RI),
one arrives at the same conclusion, implying that the above reduction in the
number of independent LEC's is equivalent to reparameterization invariance. 
Then, for on-shell nucleons, we see that even though RI poses no
constraints on the $A-$type terms, but as its consequence,
the LECs of quite a few terms
arising from 
$(\gamma^0B^\dagger\gamma^0C^{-1}B)^{(4)}$,
are fixed relative to the LECs of lower order terms.

\subsection{Off-Shell Nucleons}

For this subsection, even though not explicitly written everywhere,
but a nonrelativisitc term will imply an $A-$type HBChPT term. This
is because for off-shell nucleons, 
$\gamma^0B^\dagger\gamma^0C^{-1}B\in A$. 

In general, a nonrelativistic term can be written as:
\begin{equation} 
\label{eq:HBChPTgen}
\biggl(\epsilon^{\lambda_1\lambda_2\lambda_3\lambda_4}\biggr){\bar{\rm H}}
\prod_{i=1}^{M_1}{\cal V}_{\mu_i}\prod_{j=1}^{M_2}{\cal A}_{\nu_j}
({\cal V}_\alpha{\cal V}^\alpha)^{m_1}({\cal A}_\beta{\cal A}^\beta)^{m_2}
({\cal V}_\mu{\cal A}^\mu)^{m_3}(\chi_+)^p(\chi_-)^q\rm H,
\end{equation}
and ${\cal V}_{\mu_i}\equiv v_{\mu_i}\ {\rm or}\ {\rm D}_{\mu_i}$ 
and ${\cal A}_{\nu_j}\equiv u_{\nu_j}\ {\rm or}\ \rm S_{\nu_j}$ 
and only tensor contractions are indicated in (\ref{eq:HBChPTgen}). The upper component H and
${\bar{\rm H}}$ will be dropped in the equations and tables after equation (\ref{eq:vdotD2}).

Using
\begin{equation} 
\label{eq:Dsq}
{\bar{\rm H}}{\rm D}^2\rm H\equiv {\bar{\psi}}P_+\biggl(
-(i\rlap/\rm D-\rm m)^2+{i\over8}\sigma^{\mu\nu}[u_\mu,u_\nu]\biggr)P_+\psi,
\end{equation}
($P_+\equiv{1\over2}(1+\rlap/v)$) 
and 
\begin{equation} 
\label{eq:vdotD1}
-2i{\rm m}{\bar{\rm H}}v\cdot{\rm D}{\rm H}=
{\bar{\psi}}P_+({\rm D}^2+{\rm m}^2)P_+\psi-{\bar{\rm H}}\rm D^2\rm H
\end{equation}
one gets:
\begin{equation}
\label{eq:vdotD2}
 -2i{\rm m} 
{\bar{\rm H}}v\cdot{\rm D}{\rm H}\equiv{\bar{\psi}}P_+\Biggl(
({\rm D}^2+{\rm m}^2)
-\biggl(-(i\rlap/{\rm D}-{\rm m})^2+{i\over8}\sigma^{\mu\nu}[u_\mu,u_\nu]\biggr)
\Biggr)P_+\psi\ .
\end{equation}
Using (\ref{eq:Dsq}) and (\ref{eq:vdotD2}) one can construct the 
Tables 3 and 5.  In these tables,
by considering all possible terms up to O$(q^4)$,
it is  shown that one needs to  consider only one  kind 
of (2,2,0,0) term in the Levi-Civita(L.C)-independent category 
(Table 4), whose LECs are going to be fixed
relative to the LECs of lower order terms (O$(q^2, q^3)$). 
In Tables 3 and 5
$(\chi_+)^p(\chi_-)^q$ have been dropped because they are the same in 
HBChPT and BChPT.

(a) Table 3 (L.C.-independent terms):

We consider $({\cal V}_\alpha{\cal V}^\alpha)^{m_1},
\ ({\cal A}_\beta{\cal A}^\beta)^{m_2}$
and $({\cal V}_\mu{\cal A}^\mu)^{m_3}$ separately and see if
they and their relativistic counterparts are of the same chiral order.
This is done in Table 3. 
To understand Table 3 better, let us consider the
example belonging to the $(v\cdot{\rm D})^{l_1}(\rm D_\nu\rm D^\nu)^{l_2}$-type
terms.  A term in Table 1 that belongs to
this category is ${\rm D}_\mu v\cdot{\rm D}{\rm D}^\mu v\cdot\rm D)+{\rm h.c.}$.
What Table 3 says is that it is possible
to find a relativistic counterpart
of this term which is of the same chiral order $\equiv 0(q^4)$. One
can explicitly check that 
-${1\over{4{\rm m}\ ^2}}\rm D_\mu \{\} \rm D^\mu
-{1\over4}\{\}^2+{1\over{4{\rm m}\ ^2}}\{\}^3+{\rm h.c.}$,
on nonrelativistic reduction gives
${\rm D}_\mu v\cdot{\rm D}{\rm D}^\mu v\cdot{\rm D}+{\rm h.c.}$;
$\{\}\equiv({\rm D}^2+{\rm m}\ ^2)
-\biggl(-(i\rlap/{\rm D}-{\rm m})^2
+{i\over8}\sigma^{\mu\nu}[u_\mu,u_\nu]\biggr)$.

From inspection of Table 3, it becomes clear that for
Levi-Civita-independent terms, except for $(u_\mu\rm D^\mu)^{j_2}$
$\in({\cal V}_\mu{\cal A}^\mu)^{m_3}$, the nonrelativistic terms
are of the same chiral order as their relativistic counterparts.
This exception
is because the ${1\over{\rm m}}$-reduction of
${\bar{\psi}}(u_\mu\rm D^\mu)^{j_2}\psi$ gives, in addition to the
expected
term ${\bar{\rm H}}(u_\mu\rm D^\mu)^{j_2}\rm H$, also lower order terms.
For example, for $j_2=2$, one gets $i{\bar{\rm H}}v\cdot u u_\mu\rm D^\mu
\rm H$
and ${\bar{\rm H}}(v\cdot u)^2\rm H$, which are of O($q^3$) and O($q^2$).
For $j_2\geq 2$, it becomes impossible
to find a linear combination of relativistic counterparts that will
exactly cancel the lower order terms because there will be terms of at least
two different lower orders (which is the case for $j_2=2$) to be taken
care of.  For $j_2=1$, it is possible to construct 
a linear combination of relativistic counterparts that gives
only the non-relativistic terms one is interested in, canceling
the lower order terms, e.g. for
${\bar{\rm H}}iu^\mu v\cdot u\rm D_\mu\rm H+{\rm h.c.}$
($\equiv i=21$ O$(q^3,\phi^{2n})$ term in \cite {1n}). So,
there is no reduction  in O$(q^3)$ terms.

Up to O$(q^4)$, it is sufficient to consider only
${\bar{\rm H}}(u_\mu\rm D^\mu)^2\rm H$. Its relativistic counterpart  
${\bar{\psi}} (u_\mu\rm D^\mu)^2\psi$, on ${1\over{\rm m}}$-reduction
gives  
${\bar{\rm H}}(v\cdot u)^2\rm H$ as the O$(q^2)$ and
$-i{\rm m}{\bar{\rm H}}[v\cdot u (u_\mu{\rm D}^\mu)+(u_\mu\rm D^\mu)v\cdot
u]\rm H$ as the O$(q^3)$ non-relativistic terms.
One can modify the relativistic  counterpart in two ways to eliminate one
of the two terms of different lower orders. One can eliminate the O$(q^3)$
non-relativistic term, by considering
${\bar{\psi}}[(u_\mu\rm D^\mu)^2+i{\rm m}\rlap/u(u_\mu{\rm D}^\mu)+$
$i{\rm m}(u_\mu\rm D^\mu)\rlap/u]\psi$. However, this way, one also
gets the O$(q^2)$ term ${\bar{\rm H}}(v\cdot u)^2\rm H$, implying that
the relativistic counterpart is of O$(q^2)$. Alternatively, one can
eliminate
the O$(q^2)$ non-relativistic term by considering   
${\bar{\psi}}[(u_\mu{\rm D}^\mu)^2+i{\rm m}\rlap/u (u_\mu{\rm
D}^\mu)]\psi$. This way one also gets the O$(q^3)$ term
$i{\bar{\rm H}}u\cdot{\rm D}v\cdot u{\rm H}$, implying
that the relativistic counterpart is of O$(q^3)$.
Hence, the LEC's of ${\bar{\rm H}}(u^\mu{\rm D}^\mu)^2\rm H$ is fixed
relative to 
${\bar{\rm H}}(v\cdot u)^2\rm H$ and $i{\bar{\rm H}}u_\mu v\cdot u{\rm D}^\mu$.

Using the algebraic reductions of Section 3, one sees that
there are two
O$(q^4,\phi^{2n})$ (L.C.-independent) terms whose LECs are fixed
relative to the O$(q^2)$ term ${\bar{\rm H}}(v\cdot u)^2\rm H$ and
the O$(q^3)$ term $i{\bar{\rm H}}u_\mu v\cdot u{\rm D}^\mu\rm H+h.c.$-
given in Table 4.

We will now show (in Table 5) that up to O($q^4$), one gets no
such reduction in the number of 
independent O($q^4)$ LECs for L.C.-dependent terms.

(b) Table 5 (L.C.-dependent terms):

Using the conclusion arrived upon from Table 3, it is sufficient to consider:
\begin{equation} 
\label{eq:LCdep1}
\epsilon^{\lambda_1\lambda_2\lambda_3\lambda_4}\ 
\prod_{i=1}^{M_1}{\cal V}_{\mu_i}\prod_{j=1}^{M_2}{\cal A}_{\nu_j}
\times\biggl(1,\ ({\cal A}_\mu{\cal V}^\mu)^{m_3}\biggr),
\end{equation}
i.e. one need not consider
$\epsilon^{\lambda_1\lambda_2\lambda_3\lambda_4}\prod_{i=1}^{M_1}{\cal V}_{\mu_i}
\prod_{j=1}^{M_2}{\cal A}_{\nu_j}({\cal V}_\alpha
{\cal V}^\alpha)^{m_1}({\cal A}_\beta{\cal A}^\beta)^{m_2}$. Using  the
algebra of $S_\mu$s and the antisymmetry of 
$\epsilon^{\lambda_1\lambda_2\lambda_3\lambda_4}$,
it is sufficient to consider L.C.-dependent terms with one or no $S_\mu$ and
one or no $v_\mu$. In Table 5, one sees
whether (\ref{eq:LCdep1}) up to O$(q^4$) and their relativistic
counterparts are of of the same chiral order.

To understand Table 5  better, let us consider the example
of\\ 
$\epsilon^{\mu\nu\rho\lambda}{\rm S}_{\nu_l}
\prod_{i=1}{\rm D}_{\mu_i}\prod_{j\neq l} u_{\nu_j}(v\cdot u)$ in Table 5.
A term in Table 1 that belongs to this class of terms is
$i\epsilon^{\mu\nu\rho\lambda}{\rm S}_\rho
[{\rm D}_\mu,[[{\rm D}_\nu,u_\lambda]_+,v\cdot u]]_+$. What Table 5 tells
is that it is possible to construct
the relativistic counterpart of the same chiral order (i.e O$(q^4)$). One
can explicitly check that 
${\epsilon^{\mu\nu\rho\lambda}\over{\rm m}}
{\bar{\psi}}\gamma^5\gamma_\rho
[{\rm D}_\mu,[[{\rm D}_\nu,u_\lambda]_+,[{\rm D}_\kappa,u^\kappa]_+]]_+\psi$
$+4i{\rm m}{\bar{\psi}}\sigma^{\nu\lambda}[[{\rm D}_\nu,
[{\rm D}_\kappa,u^\kappa]_+],u_\lambda]_+\psi\equiv O(q^4)$
after nonrelativistic reduction gives 
$i\epsilon^{\mu\nu\rho\lambda}{\rm S}_\rho
[{\rm D}_\mu,[[{\rm D}_\nu,u_\lambda]_+,v\cdot u]]_+\equiv O(q^4)$, 
as one of its O$(q^4)$ terms.
\footnote{Strictly speaking,
one will also get contributions from the 
``cross terms" $\gamma^0{\cal B}^\dagger\gamma {\cal C}^{-1}{\cal B}$:
$[[[v\cdot{\rm D},v\cdot u],{\rm S}\cdot u]_+,v\cdot\rm D]$,
$[[[{\rm S}\cdot{\rm D},v\cdot u],v\cdot u]_+,v\cdot\rm D]$,
$[[[v\cdot{\rm D},v\cdot u],u^\mu]_+,u_\mu]_+$,
$[[[v\cdot{\rm D},v\cdot u],v\cdot u]_+,v\cdot u]_+$,
$\epsilon^{\mu\nu\rho\lambda}
v_\rho{\rm S}_\lambda[[[v\cdot{\rm D},v\cdot u],u_\mu]_+,u_\nu]_+$,
$i[[[{\rm D}_\mu,v\cdot u],v\cdot u]_+,u^\mu]_+$,
$i[[[v\cdot{\rm D},v\cdot u],v\cdot u]_+,v\cdot u]_+$,
$\epsilon^{\mu\nu\rho\lambda}v_\rho{\rm S}_\lambda
[[[{\rm D}_\mu,v\cdot u],v\cdot u]_+,u_\nu]_+$. One can show that 
all these terms can be obtained independently from
the nonrelativistic reduction of other 
relativistic terms.} 
From Table 5, one sees that for Levi-Civita-dependent terms, 
the nonrelativistic
terms are of the same chiral order as their relativistic counterparts.

The LECs of the O$(q^4,\phi^{2n})$ terms
of Table  4, are also fixed relative to the LEC's of lower
order terms because of the ``cross terms" 
$\gamma^0{\cal B}^\dagger\gamma^0{\cal C}^{-1}{\cal B}$
in (\ref{eq:lag}).
One can show that the $B^{(3)}$ obtained from the nonrelativistic reduction
of $i{\bar{\psi}}[[{\rm D}_\mu,u^\mu]_+,\rlap/u]_+\psi,$
${\bar{\psi}}[[{\rm D}_\mu,\rlap/u]_+,u^\mu]_+\psi,
i{\bar{\psi}}[{\rm D}^\mu,[\rlap/u,u_\mu]_+]_+\psi$ together
with the $B^{(1)}$ obtained from the nonrelativistic reduction
of $i{\bar{\psi}}\rlap/\rm D\psi$, using 
$\gamma^0 {\cal B}^\dagger\gamma^0 {\cal C}^{-1}{\cal B}$
give 
${\bar{\rm H}}[{\rm D}_\mu,[[{\rm D}_\nu,u^\nu]_+,u^\mu]_+]_+\rm H$,
${\bar{\rm H}}[{\rm D}_\mu,[[{\rm D}_\nu,u^\mu]_+,u^\nu]_+]_+\rm H$ and \\
${\bar{\rm H}}[{\rm D}_\mu,[{\rm D}_\nu,[u^\mu,u^\nu]_+]_+]_+\rm H$. 
This implies that the LEC of 
${\bar{\rm H}}[{\rm D}_\mu,[[{\rm D}_\nu,u^\nu]_+,u^\mu]_+]_+\rm H$,
${\bar{\rm H}}[{\rm D}_\mu,[[{\rm D}_\nu,u^\mu]_+,u^\nu]_+]_+\rm H$
and ${\bar{\rm H}}[{\rm D}_\mu,[{\rm D}_\nu,[u^\mu,u^\nu]_+]_+]_+\rm H$ 
will also be fixed relative to the Dirac term.

So, the conclusion one arrives at from the discussion so far, is that
for O$(q^4)$ terms, it is only a certain class of (2,2,0,0) 
terms whose LECs are fixed relative to the O$(q^m,\phi^{2n}), m=1,2,3$ terms.
It should be noted that all three terms of Table 4 
are such that the derivatives act on the baryon field, 
in addition to acting on the meson field.

Now we show the connection between the above  and  reparameterization
invariance (RI).

Using the formalism of Luke and Manohar \cite {lm},
${\cal L}_{\rm HBChPT}$ can be written in terms of 
manifestly reparameterization invariant nucleon field
${\cal H}_v$:
\begin{equation}
\label{eq:RIH} 
{\cal H}_v\equiv\Lambda({p\over{\rm m}},v){\rm H}_v=(1+
i{{\rlap/{\rm D}}\over{2\rm m}}){\rm H}_v+{\cal O}({1\over{\rm m^2}}),
\end{equation}
where $\Lambda({p\over{\rm m}},v)$ represents the Lorentz-boost 
matrix relating ${\rm H}_{{p\over{\rm m}}}$ to ${\rm H}_v$ (${\rm H}_v$ has
been hitherto denoted by ${\rm H}$; the label $v$ was assumed),
(p is the total 4-momentum of the nucleon), 
and the unimodular reparameterization invariant velocity operator 
``${\cal U}_\mu$" $\equiv{{\cal V}_\mu\over{|{\cal V}|}}$,
where ${\cal V}_\mu\equiv v_\mu+i{{\rm D}_\mu\over\rm m}$, and
$|{\cal V}|\equiv\sqrt{{\cal V}^2}$ $=\sqrt{1-2i{{v\cdot\rm D}\over\rm m}
-{{\rm D}^2\over{\rm m^2}}}$. 
\begin{equation}
\label{eq:RIU}
 {\cal U}_\mu = {\cal V}_\mu + 
iv_\mu{{v\cdot{\rm D}}\over{\rm m}}+O({1\over{\rm m^2}}).
\end{equation}

We will consider one example  from Tables 4. 
The first term in Table 4 can be written as: 
\begin{equation}
\label{eq:RIex}
 {\bar{\cal H}}_v
[{\cal U}_\mu,[u_\nu,[{\cal U}^\nu,u^\mu]_+]_+]_+{\cal H}_v.
\end{equation}

While using (\ref{eq:RIH}) (and (\ref{eq:RIU})) to rewrite
(\ref{eq:RIex}) in terms of H, $\rm D_\mu$ (and $u_\nu$), one  can
replace ${\cal H}_v$ by H because the $i\rlap/{\rm D}$ occurs
as  $v\cdot\rm D$ or  $\rm D^2$ or $\sigma^{\mu\nu}[u_\mu,u_\nu]$
(using the curvature relation), which can be obtained from other
reparameterization invariant terms containing ${\cal U}^2-1$ and 
$\sigma^{\mu\nu}[u_\mu,u_\nu]$.
\footnote{Strictly speaking, this is not 
a complete argument for dropping the $i\rlap/{\rm D}$
in ${\cal H}_v$ because ${\cal U}^2-1$ generates both $iv\cdot\rm D$
and $\rm D^2$. For  off-shell nucleons, LM's formalism does not show
how  to obtain $iv\cdot\rm D$ and $\rm D^2$ independently from
different manifestly reparameterization invariant  terms. 
But using (\ref{eq:Dsq}) and (\ref{eq:vdotD2}),
for off-shell nucleons, both $iv\cdot\rm D$ and $\rm D^2$ can
be obtained from independent HBChPT terms (written  in terms of H).} 
After doing so, one gets:
\begin{eqnarray} 
\label{eq:RIex1}
& & -{1\over{\rm m^2}}{\bar{\rm H}}\biggl([{\rm D}_\mu,[u_\nu,[{\rm D}^\nu,
u^\mu]_+]_+]_+-2i{\rm m}[{\rm D}^\nu,[v\cdot u, u_\nu]_+]_+\nonumber\\
& &  -2i{\rm m}
[u_\mu,[{\rm D}^\mu,v\cdot  u]_+]_+
-8{\rm m}^2(v\cdot u)^2\biggr)\rm H.
\end{eqnarray}
Now, $i[{\rm D}^\nu,[v\cdot u, u_\nu]_+]_+\equiv iv\cdot u u\cdot{\rm D} + 
iu_\mu v\cdot u{\rm D}^\mu$ + h.c. $\equiv 2iu_\mu v\cdot u{\rm D}^\mu $ + h.c.
($\equiv i=21$ term in \cite {1n})+$i[{\rm D}^\nu,[u_\nu,v\cdot u]]$. One need
not  consider  $i[{\rm D}_\mu,[u^\mu,v\cdot u]]$  as it can be obtained
independently.
Similarly, $i{\bar{\rm H}}[u_\mu,[{\rm D}^\mu,v\cdot u]_+]_+{\rm H}
=2i{\bar{\rm H}}u_\mu v\cdot u{\rm D}^\mu{\rm H}+{\rm h.c.}+$
$i{\bar{\rm H}}[u_\mu,[{\rm D}^\mu,v\cdot u]]\rm H$;
$i{\bar{\rm H}}[u_\mu,[{\rm D}^\mu,v\cdot u]]\rm H$ can be obtained
independently.
 So, one sees that the LEC of 
${\bar{\rm H}}[{\rm D}_\mu,[u_\nu,[{\rm D}^\nu,u^\mu]_+]_+]_+\rm H$ is fixed
relative to the LECs of ${\bar{\rm H}}(v\cdot u)^2\rm H$ and 
$i{\bar{\rm H}}u_\mu v\cdot u{\rm D}^\mu{\rm H}$. One can similarly 
show the same for the other terms in Table 4. 

\subsection{On-Shell Nucleons}

For on-shell nucleons, by the application of (\ref{eq:ruleonsh}), all 
three terms in Table 4 get eliminated as $A$-type terms. However, the first
as well as 43 other terms, arising from
$(\gamma^0B^\dagger\gamma^0C^{-1}B)^{(4)}$, have their LECs
fixed  relative to O$(q^{1,2}$) terms. The following is
the list of the 44 terms:  
\begin{eqnarray}
\label{eq:RIcross}
& & 
i\epsilon^{\mu\nu\rho\lambda}
v_\rho {\rm D}_\lambda u_\mu u_\nu {\rm S}\cdot{\rm D}
+{\rm h.c.} ;\ i{\rm S}\cdot u[u^\mu,v\cdot u]{\rm D}_\mu
+{\rm h.c.} ;\nonumber\\
& & iu^\mu[{\rm S}\cdot u,v\cdot u]{\rm D}_\mu
+{\rm h.c.} ;\ iu^\mu[u_\mu,v\cdot u]{\rm S}\cdot{\rm D}
+{\rm h.c.} ;\nonumber\\
& & i{\rm D}^\mu{\rm S}\cdot u u_\mu v\cdot u
+{\rm h.c.};\
i{\rm D}^\mu u_\mu{\rm S}\cdot u v\cdot u 
+{\rm h.c.};\nonumber\\
& & 
i[{\rm S}\cdot{\rm D},(v\cdot u)^3]_+;\
i\epsilon^{\mu\nu\rho\lambda}v_\rho{\rm S}_\lambda
{\rm D}_\mu(v\cdot u)^2{\rm D}_\nu;\nonumber\\
& & {\rm D}_\mu u^2{\rm D}^\mu;\
i\epsilon^{\mu\nu\rho\lambda}v_\rho{\rm S}_\lambda
{\rm D}_\mu u^2{\rm D}_\nu;\nonumber\\
& & 
{\rm D}_\mu\chi_+{\rm D}^\mu;\
i\epsilon^{\mu\nu\rho\lambda}v_\rho{\rm S}_\lambda
{\rm D}_\mu\chi_+{\rm D}_\nu;\nonumber\\
& & iv\cdot u\chi_+{\rm S}\cdot{\rm D}+{\rm h.c.};\
i[v\cdot u,u_\mu]u^\mu{\rm S}\cdot {\rm D} 
+{\rm h.c.};\
\nonumber\\
& & i[v\cdot u,{\rm S}\cdot u]u_\mu{\rm D}^\mu+{\rm h.c.};\
i\chi_-u^\mu{\rm D}_\mu+{\rm h.c.};\nonumber\\
& & \epsilon^{\mu\nu\rho\lambda}v_\rho{\rm S}_\lambda \chi_-
u_\mu{\rm D}_\nu+{\rm h.c.};
\nonumber\\
& &  i[u_\mu,v\cdot u]{\rm S}\cdot u{\rm D}^\mu+{\rm h.c.};\
u_\mu{\rm D}_\nu u^\mu{\rm D}^\nu+{\rm h.c.};\
\nonumber\\
& & 
i\epsilon^{\mu\nu\rho\lambda}v_\rho{\rm S}_\lambda
{\rm D}_\kappa u_\mu{\rm D}^\kappa u_\nu
+{\rm h.c.};\
i\epsilon^{\mu\nu\rho\lambda} v_\rho
{\rm S}\cdot u{\rm D}_\mu u_\lambda{\rm D}_\nu+
{\rm h.c.};\nonumber\\
& &  i{\rm D}_\mu u^\mu v\cdot u{\rm S}\cdot u
+{\rm h.c.};\nonumber\\
& & i{\rm S}\cdot{\rm D} u^2v\cdot u+{\rm
h.c.};\
{\rm D}^\mu u^\nu u_\mu{\rm D}_\nu+{\rm h.c.};\nonumber\\
& & i\epsilon^{\mu\nu\rho\lambda}v_\rho{\rm S}_\lambda{\rm D}^\kappa
[u_\mu,u_\kappa] {\rm D}_\nu;\
i{\rm D}^\mu[v\cdot u,u_\mu]_+{\rm S}\cdot u+{\rm h.c.};\nonumber\\
& & i{\rm D}_\mu[{\rm S}\cdot u,u_\mu]v\cdot u
+{\rm h.c.};\
i[[u^2,v\cdot u]_+,{\rm S}\cdot{\rm D}]_+;\
i[u^\mu v\cdot u u_\mu,{\rm S}\cdot{\rm D}]_+;\nonumber\\
& & 
i[{\rm D}^\mu,[v\cdot u,[u^\mu,{\rm S}\cdot u]]_+];\
i[{\rm S}\cdot{\rm D},[v\cdot u,\chi_+]_+]_+;\
\nonumber\\
& & i\epsilon^{\mu\nu\rho\lambda}v_\rho{\rm S}_\lambda
[{\rm D}_\nu,[v\cdot u,[{\rm D}_\mu,v\cdot u]]];\
[{\rm D}_\mu,[u_\nu,[{\rm D}_\mu,u^\nu]]]_+;
\nonumber\\
& & i\epsilon^{\mu\nu\rho\lambda}v_\rho{\rm S}_\lambda
[{\rm D}_\nu,[u_\kappa,[{\rm D}_\mu,u^\kappa]]];\
[{\rm D}^\nu,[u^\mu,[{\rm D}_\mu,u_\nu]_+]_+]_+;
\nonumber\\
& & 
i\epsilon^{\mu\nu\rho\lambda}v_\rho{\rm S}_\lambda
[{\rm D}_\nu,[u^\kappa,[u_\mu,{\rm D}_\kappa]_+]_+];\
[{\rm D}^\nu,[u^\mu,[u_\nu,{\rm D}_\mu]]]_+;
\nonumber\\
& & i\epsilon^{\mu\nu\rho\lambda}v_\rho{\rm S}_\lambda
[{\rm D}_\nu,[u^\kappa,[u_\mu,{\rm D}_\kappa]]];\
i\epsilon^{\mu\nu\rho\lambda}v_\rho{\rm S}_\lambda
[{\rm D}_\nu,[v\cdot u,[{\rm D}_\mu,v\cdot u]]_+]_+;
\nonumber\\
& & 
i\epsilon^{\mu\nu\rho\lambda}v_\rho{\rm S}_\lambda
[{\rm D}_\nu,[u_\mu,[v\cdot{\rm D},v\cdot u]]_+]_+;\
i[{\rm D}_\mu, [\chi_-,u^\mu]]_+;\nonumber\\
& & 
\epsilon^{\mu\nu\rho\lambda}v_\rho{\rm S}_\lambda
[{\rm D}_\nu,[\chi_-,u_\mu]];\
i\epsilon^{\mu\nu\rho\lambda}v_\rho{\rm S}_\lambda
[{\rm D}_\nu,[v\cdot u,[v\cdot{\rm D},u_\mu]]];\nonumber\\
& & 
i\epsilon^{\mu\nu\rho\lambda}v_\rho{\rm S}_\lambda
[{\rm D}_\nu,[u_\mu,[v\cdot{\rm D},v\cdot u]]].
\end{eqnarray}
Their coefficients can be
determined from (\ref{eq:B1C2B1onsh}), (\ref{eq:B1C1C1B1onsh})
and (\ref{eq:B3B1onsh}) of Section 7.

Hence, in conclusion, 
for off-shell nucleons, one gets 207 O$(q^4,\phi^{2n})$ and 
113 O$(q^4,\phi^{2n+1})$
terms. Of these, reparameterization/Lorentz invariance fixes the LECs of 
of three terms of the L.C.-dependent (2,2,0,0) category given in
Table 4; no such reduction in the number of independent LECs is
obtained in the L.C.-dependent terms up to O($q^4$).
As a consequence, of the 207 O$(q^4,\phi^{2n})$ terms (obtained
in the Section 5), there are  (207 - 3)+113 = 317
linearly independent O$(q^4)$ LECs. For on-shell nucleons,
the LECs of 44 terms are fixed relative to lower order terms.

\section{Application}

We now discuss an application of the O($q^4$) terms
of Section 4 to evaluation of the contribution of the O($q^4$) 
operator insertions to the ``contact graph" of pion DCX  
(See Figs. 1 and 2) at threshold
and assuming static  nucleons. In \cite{mk} we included
Fig 1, whose vertices, given in
Fig 2, were taken to LO, which is O($q$). In \cite{bkm2}, the vertices
were corrected (but only on-shell) to O($q^3$).
We will show that one gets no contribution from the 
O$(q^4$) terms (at threshold, and
in the static limit  of the nucleons). We will also
discuss generalization of this result.
The notations of \cite{mk} will be used
in this section.

Let $p_{1,2}^\mu$  and $p^\mu_{3,4}$ be the 4-momenta of  the
incoming and outgoing nucleons
(respectively), and $q_{1,2}^\mu$ the 4-momenta of the incoming and outgoing pions
(respectively).  The velocity parameters       
of the two participating nucleons are both chosen to have the static limit
values, with only a non-vanishing time component , i.e.
$v^\mu_1=v^\mu_2=(1,\vec 0)$. Also the nucleons will be
treated as if they were on-shell  (sometimes referred to as impulse
approximation). In the HBChPT  formalism,
$p^0$ denotes only the contribution to the time component of the
total nucleon momenta ($\equiv mv+p$),
$\it in\ addition$ to the rest mass energy $m$ (for  the choice of the
nucleon velocity to possess only a non-zero time component).
In the present  case $p^0=E_B$, which as stated above, we drop.
Then if we go to the c.m. frame of the nucleons:
\begin{equation}
\label{eq:cm}
p_1^\mu=(0,\vec p);\ p_2^\mu=(0,-\vec p);\
p_3^\mu=(0,\vec p^\prime);\ p_4^\mu=(0,-\vec p^\prime);\
q_1^\mu=q_2^\mu= (M_\pi,\vec 0).
\end{equation}
(Note: The external pions are at zero kinetic energy [threshold].)
Equation (\ref{eq:cm}) implies:
\begin{equation}
\label{eq:conseqcm}
v\cdot {\cal P}={\rm S}\cdot q_{1,2}= {\cal P}\cdot q_{1,2}=0,
\end{equation}
where ${\cal P}\equiv p_1$ or $p_3$ or $p_1-p_3$
In addition, one can show that:
\begin{equation}
\label{eq:anticoomGammau}
[\Gamma^{(2)}_\mu,u_\nu^{(1)}]_+|_{\pi^0=0}=0,
\end{equation}
where the superscripts refer to the powers of $\phi$.

One can show that all 113 terms
of Table 2 (terms in table 1 will not contribute to the required vertices)
will consist of one or more of  the 
following terms (written symbolically): 
\begin{eqnarray}
\label{eq:coordspace}
& & {\rm S}\cdot\partial\pi^+;\ v\cdot\partial\pi^-;\ v\cdot\partial
\biggl((\pi^+)^2\pi^-\biggr);\nonumber\\ 
& & [\Gamma^{(2)}_\mu,u_\nu^{(1)}]_+|_{\pi^0}.
\end{eqnarray}
In addition, using (\ref{eq:cm}), in momentum space, some terms
in Table 2 also consist of:
\begin{equation}
\label{eq:moomspace}
v\cdot q_{1,2}{\rm M}_\pi^2+q_1\leftrightarrow -q_2.
\end{equation}
Using (\ref{eq:conseqcm}) and (\ref{eq:anticoomGammau}) in (\ref{eq:coordspace})
and (\ref{eq:moomspace}), one
sees that one receives no contribution from any of the 113 terms of
Table 2.

Thus, O$(q^4)$ terms give no contribution to the contact DCX graph
at threshold and in the impulse approximation (for the nucleons).
One can in fact generalize this result. For this, one notes that because
of parity constraints,
L.C.-independent terms contributing to an O$(\phi^{2n+1}$)-vertex
have to consist of an ${\rm S}_\mu$, and similarly, L.C-dependent
terms contributing to an O($\phi^{2n+1}$)-vertex
have to be ${\rm S}_\mu$-independent. Further, no
L.C. -dependent term will contribute to the contact graph. One
way to see this is to note that for the ${\bar p}(\pi^+)^2\pi^- n$-vertex,
in momentum space, L.C-dependent
terms will consist of one of the following terms:
\begin{eqnarray}
\label{eq:eps0}
& & \epsilon_{\mu\nu\rho\lambda}
v^\mu p_1^\nu q_1^\rho q_2^\lambda;\nonumber\\
& & \epsilon_{\mu\nu\rho\lambda}
v^\mu p_3^\nu q_1^\rho q_2^\lambda;\nonumber\\
& & \epsilon_{\mu\nu\rho\lambda}
v^ \mu p_1^\nu p_3^\rho q_1^\lambda;\nonumber\\
& & \epsilon_{\mu\nu\rho\lambda}
v^ \mu p_1^\nu p_3^\rho q_2^\lambda;\nonumber\\
& & \epsilon_{\mu\nu\rho\lambda}
p_1^\mu p_3^\nu q_1^\rho q_2^\lambda.
\end{eqnarray}
Using (\ref{eq:cm}),
one sees that all five terms in (\ref{eq:eps0}), vanish. One
can similarly arrive at a similar conclusion
for the ${\bar p}\pi^+n$-vertex.  Now, using (\ref{eq:conseqcm}),
one sees that the contribution of O($q^N$) term to the 
${\bar p}(\pi^+)^2\pi^-n$-vertex
of the contact DCX graph can be written as:
\begin{equation}
\label{eq:orderNcont1}
({\rm S}\cdot{\cal P})(v\cdot q_1)^{l_1}(v\cdot q_2)^{l_2}
(q_1\cdot q_2)^{l_3}{\rm M}_\pi^{2l_4}+q_1\leftrightarrow-q_2,
\end{equation}
where $N=1+l_1+l_2+2l_3+2l_4$.
This implies that only those terms that
correspond to $l_1+l_2\equiv$ even, i.e., terms in the 
Lagrangian of odd chiral orders,
will contribute to the contact DCX graph vertices.

\section{On-shell reduction}

In this section, we discuss the derivation of the on-shell O($q^4$)
${\cal L}_{\rm HBChPT}$, directly within HBChPT using
the techniques of \cite{1n}.

The main result obtained in \cite{1n} in the context of complete
on-shell reduction within HBChPT was the following rule:
\begin{eqnarray}
\label{eq:ruleonsh}
& & A-{\rm type\  terms\ of\ the\ form\
{\bar{\rm H}}{\rm S}\cdot{\rm D}{\cal O}\rm H\ +\ h.c.} \nonumber\\
& & {\rm or\ {\bar{\rm H}}v\cdot{\rm D}{\cal O}\rm H\ +\ h.c.} \nonumber\\
& & {\rm or\ {\bar{\rm H}}{\cal O}^\mu\rm D_\mu\rm H\ +\ h.c.\ can\ be\
eliminated}\nonumber\\
& & {\rm except\ for}\
{\cal O}_\mu\equiv\Biggl(i^{m_1+l_5+l_7+1},\ {\rm or}\
\epsilon^{\nu\lambda\kappa\rho}\times\Omega\Biggr)\times u_\mu\Lambda\nonumber\\
& & {\rm with}\ l_1\geq1,
\Omega\equiv 1(i)\ {\rm for}\ (-)^{m_1+l_5+l_7+1}=-1(1),\nonumber\\
& & {\rm or}\nonumber\\
& & {\cal O}_\mu\equiv\Biggl(i^{m_1+l_5+l_7},\ {\rm or}\
\epsilon^{\nu\lambda\kappa\rho}\times\Omega^\prime\Biggr)
\times {\rm D}_\mu\Lambda,\nonumber\\
& & {\rm with}\ l_1\geq1, \Omega^\prime\equiv 1(i)\ {\rm for}\ 
(-)^{m_1+l_5+l_7}=-1(1).
\end{eqnarray}
In (\ref{eq:ruleonsh})
\begin{equation}
\label{eq:exception8}
\Lambda\equiv\prod_{i=1}^{M_1}{\cal V}_{\nu_i}
\prod_{j=1}^{M_2} u_{\rho_j}
(v\cdot u)^{l_1}u^{2l_2}\chi_+^{l_3}\chi_-^{l_4}
([v\cdot{\rm D},\ ])^{l_5}
({\rm D}_\beta{\rm D}^\beta)^{l_6}(u_\alpha{\rm D}^\alpha)^{l_7},
\end{equation}
where ${\cal V}_{\nu_i}\equiv v_{\nu_i}\ \rm or\ \rm D_{\nu_i}$.
where ${\cal V}_{\nu_i}\equiv v_{\nu_i}\ \rm or\ \rm D_{\nu_i}$.
The number of ${\rm D}_{\nu_i}$s in (\ref{eq:exception8})
equals $m_1(\leq M_1)$.
Assuming that Lorentz invariance, isospin symmetry, parity and hermiticity
have been implemented, the choice of the factors
of $i$ in (\ref{eq:ruleonsh})
automatically incorporates the phase rule (\ref{eq:phoffsh}).
In (\ref{eq:ruleonsh}),
it is only the contractions of the building blocks that
has  been indicated.
It is  understood that all (anti-)commutators  in the HBChPT Lagrangian
are to be expanded out until one hits the first ${\rm D}_\mu$, so that the
$A$-type HBChPT term can be put in the form
${\bar{\rm H}}{\cal O}^\mu\rm D_\mu\rm H$
+h.c.

Using (\ref{eq:ruleonsh}), we perform on-shell reduction of terms in Tables 1 
and 2 . Terms that get eliminated are marked by an ``E" and those  
that are not are marked by an ``ON" in Tables  1 and 2 . Terms in Table 2 
with $i=211,229,244,253,261,268,269,276,277,278,305$ are marked by 
``ON$^\prime$." This is because it is not
the whole term, but only the on-shell ``component" of these terms that
do not get eliminated for on-shell nucleons. These terms can generically
be written as $i[[u_\mu,u_\nu]_+,[u_\rho,{\rm D}_\lambda]_+]_+$,in which it is
understood that the Lorentz indices  are contracted either within
themselves or by $v_\alpha$ and/or $\rm S_\beta$. The on-shell ``component"
of these terms can be shown to be equal to 
$i[u_\rho,[{\rm D}_\lambda,[u_\mu,u_\nu]_+]]$.

The complete on-shell O($q^4$) HBChPT Lagrangian can be shown
to be given by:
\begin{eqnarray}
\label{eq:onshO41}
& & {\cal L}_{\rm HBChPT}^{(4)} = A^{(4)}+
{1\over{2\rm m}}\gamma^0B^{(2)}\ ^\dagger\gamma^0 B^{(2)}\nonumber\\
& & +{1\over{2\rm m}}\biggl[\gamma^0 B^{(3)}\ ^\dagger\gamma^0 B^{(1)}
+\gamma^0B^{(1)}\ ^\dagger\gamma^0B^{(3)}\biggr]
\nonumber\\
& & -{1\over{4\rm m^2}}\biggl[\gamma^0 B^{(2)}\ ^\dagger\gamma^0 C^{(1)}B^{(1)}
+\gamma^0 B^{(1)}\ ^\dagger\gamma^0 C^{(1)}B^{(2)}\biggr]\nonumber\\
& & -{1\over{4\rm m^2}}\gamma^0 B^{(1)}\ ^\dagger\gamma^0 C^{(2)}B^{(1)}
+{1\over{8\rm m^3}}
\gamma^0 B^{(1)}\ ^\dagger\gamma^0(C^{(1)})^2B^{(1)}.
\end{eqnarray}
One can use the results of \cite{1n} for writing down
on-shell $B$ directly within HBChPT,
but for the case at hand, we find it equally convenient to use
the relativistic counterparts of $A^{(2)}$ and $A^{(3)}$
given in \cite{1n} for constructing on-shell $B^{(2)}$ and $B^{(3)}$.
Of course the $\alpha_3$ term (of (\ref{eq:B2onsh}))
has to be put in addition as $\gamma^5$ does
not contribute to $A$.  

Using
\begin{equation}
\label{eq:B2onsh}
B^{(2)}_{\rm OS}=\alpha_1\gamma^5[v\cdot u,{\rm S}\cdot u]
+\alpha_2\gamma^5[v\cdot u,{\rm S}\cdot u]_+
+\alpha_3\gamma^5\chi_- +i\alpha_4[v\cdot{\rm D},v\cdot u],
\end{equation}
(OS$\equiv$on-shell) one gets:
\begin{eqnarray}
\label{eq:B2B2onsh}
& & 
{1\over{2\rm m}}\gamma^0B^{(2)}\ ^\dagger\gamma^0 B^{(2)}=\nonumber\\
&  & \alpha_1^2[v\cdot u,{\rm S}\cdot u]^2
-\alpha_2^2[v\cdot u,{\rm S}\cdot u]_+^2
+\alpha_3^2\chi_-^2-\alpha_4^2[v\cdot{\rm D},v\cdot u]^2\nonumber\\
& & +\alpha_1\alpha_2\biggl(
-{1\over2}[[v\cdot u,u^\mu],[v\cdot u,u_\mu]_+]
+{i\over2}\epsilon^{\mu\nu\rho\lambda}v_\rho{\rm S}_\lambda
[[v\cdot u,u_\mu],[v\cdot u,u_\nu]_+]_+\biggr)\nonumber\\
& & -\alpha_1\alpha_3[[v\cdot u,{\rm S}\cdot u],\chi_-]
+i\alpha_1\alpha_4[[v\cdot u,{\rm S}\cdot u],[v\cdot{\rm D},v\cdot u]]_+
\nonumber\\
& &  -i\alpha_2\alpha_4[[v\cdot u,{\rm S}\cdot u]_+,[v\cdot{\rm D},
v\cdot u]]
+i\alpha_3\alpha_4[\chi_-,[v\cdot{\rm D},v\cdot u]]_+.
\end{eqnarray}
The set $\{\alpha_i\}$ can be related to the set $\{a_i\}$ of \cite{em}.

Using (\ref{eq:B2onsh}) and $C^{(1)}_{\rm OS}$, 
and eliminating all terms proportional to the nonrelativistic
eom by  field redefinition of H, one gets:
\begin{eqnarray}
\label{eq:B2C1B1onsh}
& & -{1\over{4\rm m^2}}\gamma^0 B^{(2)}\ ^\dagger\gamma^0 C^{(1)}B^{(1)}+{\rm h.c.}
\nonumber\\
& & 
=-{1\over{4\rm m^2}}\Biggl[2\alpha_1\biggl(-{1\over{16}}[v\cdot u,u^\mu]^2+ig_A^0[v\cdot u,
u_\mu]{\rm D}^\mu{\rm S}\cdot u+{i\over8}
\epsilon^{\mu\nu\rho\lambda}v_\rho{\rm S}_\lambda[v\cdot u,u_\mu]
[v\cdot u, u_\nu]\nonumber\\
& & 
+{g_A^0\over8}
\epsilon^{\mu\nu\rho\lambda}v_\rho[v\cdot u,u_\mu]{\rm D}_\nu u_\lambda
-{g_A^0\over2}\biggl[{i\over2}
[v\cdot u,{\rm S}\cdot u]{\rm D}_\mu u^\mu
-{i\over2}[v\cdot u,{\rm S}\cdot u][v\cdot{\rm D},v\cdot u]\nonumber\\
& & 
+{g_A^0\over8}[v\cdot u, u_\mu]v\cdot u u_\mu+{ig_A^0\over4}
\epsilon^{\mu\nu\rho\lambda}v_\rho{\rm S}_\lambda[v\cdot u,u_\mu]v\cdot u u_\nu
\biggr] \nonumber\\
& & 
-{ig_A^0\over2}[v\cdot u,{\rm S}\cdot u][v\cdot{\rm D},v\cdot u]
+{{g^0_A}^2\over2}\biggl[-{1\over4}[v\cdot u, u_\mu]
v\cdot u u^\mu+{i\over2}\epsilon^{\mu\nu\rho\lambda}v_\rho{\rm S}_\lambda
[v\cdot u,u_\mu]v\cdot u u_\nu\biggr]
\nonumber\\
& & 
-2g_A^0\biggl[-{i\over4}[v\cdot u,u_\mu]u^\mu{\rm S}\cdot{\rm D}
+{1\over8}\epsilon^{\mu\nu\rho\lambda}v_\rho[v\cdot u, u_\mu]u_\nu{\rm D}_\lambda\nonumber\\
& & -{i\over4}[v\cdot u,{\rm S}\cdot u]u_\mu{\rm D}^\mu
+{g_A^0\over{16}}[v\cdot u, u_\mu]v\cdot u u^\mu
\nonumber\\
& & -{g_A^0 \over8}\epsilon^{\mu\nu\rho\lambda}v_\rho{\rm S}_\lambda[v\cdot u, u_\mu]
v\cdot u u_\nu+{1\over4}[v\cdot u,u_\mu]{\rm S}\cdot u{\rm D}^\mu\biggr]
\nonumber\\
& & +{{g^0_A}^2\over8}[v\cdot u,u_\mu]u^\mu v\cdot u
-{i{g^0_A}^2\over4}
\epsilon^{\mu\nu\rho\lambda}v_\rho{\rm S}_\lambda[v\cdot u,u_\mu]u_\nu v\cdot u\biggr)
\nonumber\\
& &
-\alpha_2\biggl( 
-{1\over{16}}[v\cdot u,u^\mu]_+[v\cdot u,u_\mu]+ig_A^0[v\cdot u,
u_\mu]_+{\rm D}^\mu{\rm S}\cdot u+{i\over8}
\epsilon^{\mu\nu\rho\lambda}v_\rho{\rm S}_\lambda[v\cdot u,u_\mu]_+
[v\cdot u, u_\nu]\nonumber\\
& & 
+{g_A^0\over8}
\epsilon^{\mu\nu\rho\lambda}v_\rho[v\cdot u,u_\mu]_+{\rm D}_\nu u_\lambda
-{g_A^0\over2}\biggl[{i\over2}
[v\cdot u,{\rm S}\cdot u]_+{\rm D}_\mu u^\mu
-{i\over2}[v\cdot u,{\rm S}\cdot u]_+[v\cdot{\rm D},v\cdot u]\nonumber\\
& & 
+{g_A^0\over8}[v\cdot u, u_\mu]_+v\cdot u u_\mu+{ig_A^0\over4}
\epsilon^{\mu\nu\rho\lambda}v_\rho{\rm S}_\lambda[v\cdot u,u_\mu]_+v\cdot u u_\nu
\biggr] \nonumber\\
& & 
-{ig_A^0\over2}[v\cdot u,{\rm S}\cdot u]_+[v\cdot{\rm D},v\cdot u]
+{{g^0_A}^2\over2}\biggl[-{1\over4}[v\cdot u, u_\mu]_+
v\cdot u u^\mu+{i\over2}\epsilon^{\mu\nu\rho\lambda}v_\rho{\rm S}_\lambda
[v\cdot u,u_\mu]_+v\cdot u u_\nu\biggr]
\nonumber\\
& & 
-2g_A^0\biggl[-{i\over4}[v\cdot u,u_\mu]_+u^\mu{\rm S}\cdot{\rm D}
+{1\over8}\epsilon^{\mu\nu\rho\lambda}v_\rho[v\cdot u, u_\mu]_+u_\nu{\rm D}_\lambda\nonumber\\
& & -{i\over4}[v\cdot u,{\rm S}\cdot u]_+u_\mu{\rm D}^\mu
+{g_A^0\over{16}}[v\cdot u, u_\mu]_+v\cdot u u^\mu
\nonumber\\
& & -{g_A^0 \over8}\epsilon^{\mu\nu\rho\lambda}v_\rho{\rm S}_\lambda[v\cdot u, u_\mu]_+
v\cdot u u_\nu+{1\over4}[v\cdot u,u_\mu]_+{\rm S}\cdot u{\rm D}^\mu\biggr]
\nonumber\\
& & +{{g^0_A}^2\over8}[v\cdot u,u_\mu]_+u^\mu v\cdot u
-{i{g^0_A}^2\over4}
\epsilon^{\mu\nu\rho\lambda}v_\rho{\rm S}_\lambda[v\cdot u,u_\mu]_+u_\nu v\cdot u\biggr)
\nonumber\\
& & +\alpha_3\biggl(2\chi_-v\cdot{\rm D}{\rm S}\cdot{\rm D}
-i{g_A^0\over2}\chi_-[v\cdot{\rm D},v\cdot u] +{{g_A^0}^2\over2}\chi_-v\cdot u
{\rm S}\cdot u\nonumber\\
& & 
+g_A^0\chi_-\biggl[{1\over4}v\cdot u{\rm S}\cdot u+{i\over4}u_\mu{\rm D}^\mu
+{1\over2}\epsilon^{\mu\nu\rho\lambda}v_\rho{\rm S}_\lambda u_\mu{\rm D}_\nu\biggr]
\biggr)\nonumber\\
& & 
+\alpha_4\biggl({i\over2}[v\cdot{\rm D},v\cdot u][v\cdot u,{\rm S}\cdot u]
-2g_A^0\biggl[{1\over4}[v\cdot{\rm D},v\cdot u]^2-{1\over4}[v\cdot{\rm D},v\cdot u]{\rm D}_\mu u^\mu
\nonumber\\
& & +{ig_A^0\over4} [v\cdot{\rm D},v\cdot u]v\cdot u{\rm S}\cdot u+{i\over2}
\epsilon^{\mu\nu\rho\lambda}v_\rho{\rm S}_\lambda[v\cdot{\rm D},v\cdot u]{\rm D}_\mu 
u_\nu\biggr]\nonumber\\
& & +{g_A^0\over2}[v\cdot{\rm D},v\cdot u]^2+{i{g_A^0}^2\over2}
[v\cdot{\rm D},v\cdot u]v\cdot u{\rm S}\cdot u\nonumber\\
& & +2g_A^0\biggl[{ig_a^0\over4}[v\cdot{\rm D},v\cdot u]v\cdot u{\rm S}\cdot u
-{1\over4}[v\cdot{\rm D},v\cdot u]u_\mu{\rm D}^\mu\nonumber\\
& & +{i\over2}v_\rho{\rm S}_\lambda[v\cdot{\rm D},v\cdot u]
u_\mu{\rm D}_\nu\biggr]
-{i{g^0_A}^2\over2}[v\cdot{\rm D},v\cdot u]{\rm S}\cdot u
v\cdot u\biggr)
\nonumber\\
& & +{\rm h.c.}\Biggr]
\end{eqnarray}
Similarly, using:
\begin{equation}
\label{eq:A2onsh}
C^{(2)}_{\rm OS}=-i\alpha_1\epsilon^{\mu\nu\rho\lambda}v_\rho
{\rm S}_\lambda-\alpha_2(v\cdot u)^2-\alpha_5 u^2-\alpha_6\chi_+,
\end{equation}
and
\begin{equation}
\label{eq:B1}
B^{(1)}=-2i\gamma^5{\rm S}\cdot{\rm D}-{g_A^0\over2}\gamma^5v\cdot u,
\end{equation}
and eliminating all terms proportional to the nonrelativistic
eom by  field redefinition of H, one sees that:
\begin{eqnarray}
\label{eq:B1C2B1onsh}
& & -{1\over{4\rm m^2}}\gamma^0 B^{(1)}\ ^\dagger\gamma^0 C^{(2)}B^{(1)}
\nonumber\\
& & =-{1\over{4\rm m^2}}\Biggl[\alpha_1\Biggl(i\epsilon^{\mu\nu\rho\lambda}
v_\rho{\rm D}_\lambda u_\mu u_\nu
{\rm S}\cdot{\rm D}-i{g_A^0\over 4}{\rm S}\cdot u[u^\mu,v\cdot u]{\rm D}_\mu
-{g^0_A\over 8}\epsilon^{\mu\nu\rho\lambda}
v_\rho u_\mu[u_\lambda,v\cdot u]{\rm D}_\nu \nonumber\\
& & +{ig_A^0\over 4}
(-u^\mu[{\rm S}\cdot u,v\cdot u]{\rm D}_\mu
+u^\mu[u_\mu,v\cdot u]{\rm S}\cdot{\rm D})\nonumber\\
& & -{{g_A^0}^2\over 4}v\cdot u
(-{1\over 4}[u_\mu,v\cdot u]u^\mu+{i\over 2}
\epsilon^{\mu\nu\rho\lambda}v_\rho{\rm S}_\lambda
[u_\mu,v\cdot u]u_\nu)\nonumber\\
& & +\biggl[{g_A^0\over4}\epsilon^{\mu\nu\rho\lambda}v_\rho{\rm D}_\lambda u_\mu u_\nu v\cdot u
-{g_A^0\over 2}\biggl(-i{\rm D}^\mu{\rm S}\cdot u u_\mu v\cdot u\nonumber\\
& & -g_A^0[-{1\over4}u^2(v\cdot u)^2+{i\over2}
\epsilon^{\mu\nu\rho\lambda}v_\rho{\rm S}_\lambda
u_\mu u_\nu(v\cdot u)^2]\nonumber\\
& & +g_A^0(-{1\over 4}[u_\mu, v\cdot u] u^\mu v\cdot u+{i\over2} 
\epsilon^{\mu\nu\rho\lambda}v_\rho{\rm S}_\lambda
u_\mu v\cdot u u_\nu v\cdot u)+i{\rm D}^\mu u_\mu{\rm S}\cdot u v\cdot u\biggr)
+{\rm h.c.}\biggr]\nonumber\\
& & +{{g_A^0}^2\over4}
\epsilon^{\mu\nu\rho\lambda}v_\rho{\rm S}_\lambda 
v\cdot u u_\mu u_\nu v\cdot u\Biggr)\nonumber\\
& & +\alpha_2\Biggl(
{g_A^0}^2\biggl[{1\over4}(v\cdot u)^2-{1\over4}u_\mu (v\cdot u)^2 u_\mu
+{i\over2}\epsilon^{\mu\nu\rho\lambda}v_\rho{\rm S}_\lambda
u_\mu(v\cdot  u)^2u_\nu\biggr]\nonumber\\
& & +ig_A^0[{\rm S}\cdot{\rm D},(v\cdot u)^3]_++{{g_A^0}^2\over4}
(v\cdot u)^4+{\rm D}_\mu(v\cdot u)^2{\rm D}^\mu-2i
\epsilon^{\mu\nu\rho\lambda}v_\rho{\rm S}_\lambda
{\rm D}_\mu(v\cdot u)^2{\rm D}_\nu\Biggr)\nonumber\\
& & +\alpha_5\Biggl({g_A^0}^2\biggl[{1\over4}v\cdot u u^2v\cdot u-{1\over4}u_\mu u^2u^\mu
+{i\over2}\epsilon^{\mu\nu\rho\lambda}v_\rho{\rm S}_\lambda 
u_\mu u^2u_\nu\biggr]\nonumber\\
& & +{\rm D}_\mu u^2{\rm D}^\mu
-2i\epsilon^{\mu\nu\rho\lambda}v_\rho{\rm S}_\lambda 
{\rm D}_\mu u^2{\rm D}_\nu
+(ig_A^0{\rm S}\cdot{\rm D}u^2v\cdot u+{\rm h.c.})
+{{g_A^0}^2\over4}v\cdot u u^2v\cdot u\Biggr)
\nonumber\\
& & +\alpha_6\Biggl({g_A^0}^2\biggl[{1\over 4}v\cdot u\chi_+v\cdot u
-{1\over4}u_\mu\chi_+u^\mu+{i\over2}
\epsilon^{\mu\nu\rho\lambda}v_\rho{\rm S}_\lambda u_\mu\chi_+u_\nu\biggr]
\nonumber\\
& & +{\rm D}_\mu\chi_+{\rm D}^\mu
-2i\epsilon^{\mu\nu\rho\lambda}v_\rho{\rm S}_\lambda 
{\rm D}_\mu\chi_+{\rm D}_\nu\nonumber\\
& & +(ig_A^0v\cdot u\chi_+{\rm S}\cdot{\rm D}+{\rm h.c.})
+{{g_A^0}^2\over 4}v\cdot u\chi_+v\cdot u\Biggr)\Biggr]
\end{eqnarray}
Further, the O$(q^4)$ terms, independent of any
undetermined LECs are given by (after elimination of the terms 
proportional to the nonrelativistic eom, 
by  field-redefinition of H)
\begin{eqnarray}
\label{eq:B1C1C1B1onsh}
& & {1\over{8\rm m^3}}
\gamma^0 B^{(1)}\ ^\dagger\gamma^0(C^{(1)})^2B^{(1)}\nonumber\\
& & ={1\over{8\rm m^3}}\Biggl[3{g_A^0}^2\biggl[{1\over16}(v\cdot u)^4+{1\over16}u^4
\nonumber\\
& & +{1\over16}\biggl(-u^\mu u^\nu u_\mu u_\nu+v\cdot u u^\mu v\cdot u u_\mu
+u_\mu v\cdot u u^\mu v\cdot u-v\cdot u u^2v\cdot u-u^\mu(v\cdot u)^2u_\mu+u^\mu u^2u_\mu\biggr)
\nonumber\\
& & +{i\over8}\epsilon^{\mu\nu\rho\lambda}v_\rho u_\mu u_\nu[u_\lambda,
{\rm S}\cdot u]-{1\over16}[(v\cdot u)^2,u^2]_+-{i\over8}
\epsilon^{\mu\nu\rho\lambda}v_\rho{\rm S}_\lambda
[(u^2-(v\cdot u)^2),u_\mu u_\nu]_+
\biggr]\nonumber\\
& & -{g_A^0}^2\biggl[ig_A^0\biggl({1\over 4}(v\cdot u)^2-{1\over4}u^2
\biggr)[v\cdot{\rm D},{\rm S}\cdot u]
+{g_A^0\over8}\epsilon^{\mu\nu\rho\lambda}v_\rho
u_\mu u_\nu[v\cdot{\rm D},u_\lambda] \nonumber\\ 
& & +i{g_A^0\over4}\biggl(-[{\rm S}\cdot u,u_\mu][v\cdot{\rm D},u^\mu]
+[{\rm S}\cdot u,v\cdot u][v\cdot{\rm D},v\cdot u]\biggr)\nonumber\\ 
& &  +{ig_A^0\over2}{\rm S}\cdot u\biggl([v\cdot{\rm D},v\cdot u]v\cdot u
-[v\cdot{\rm D},u^\mu]u_\mu\biggr)\nonumber\\
& & +{g_A^0\over4}\biggl(-u^\mu[v\cdot{\rm D},{\rm S}\cdot u]v\cdot u
+v\cdot u[v\cdot{\rm D},{\rm S}\cdot u]v\cdot u\nonumber\\
& & -v\cdot u[v\cdot{\rm D},v\cdot u]{\rm S}\cdot u
+u^\mu[v\cdot{\rm D},u_\mu]{\rm S}\cdot u\biggr)
\nonumber\\
& & +{1\over4}v\cdot u[v\cdot{\rm D},[v\cdot{\rm D},v\cdot u]]-{u^\mu\over4}
[v\cdot{\rm D},[v\cdot{\rm D},u_\mu]]
+{i\over2}\epsilon^{\mu\nu\rho\lambda}v_\rho{\rm S}_\lambda
u_\mu[v\cdot{\rm D},[v\cdot{\rm D},u_\nu]]\biggr]\nonumber\\
& & -{1\over16}[u_\mu,v\cdot u]^2 
+{ig_A^0\over4}[{\rm S}\cdot u,{\rm D}_\mu [v\cdot u,u^\mu]]_+\nonumber\\
& & +2i\epsilon^{\mu\nu\rho\lambda}v_\rho
\biggl({{\rm S}_\lambda\over16}[u_\mu,v\cdot u][v\cdot u,u_\nu]
-{ig_A^0\over16}[u_\lambda,{\rm D}_\mu[v\cdot u,u_\nu]]_+\biggr)
\nonumber\\
& & -{ig_A\over4}\biggl(-[u^\mu,{\rm S}\cdot {\rm D}[v\cdot u,u_\mu]]_+
-[v\cdot u,v\cdot u {\rm D}[v\cdot u,{\rm S}\cdot u]]_+
+[u^\mu,{\rm D}_\mu,[v\cdot u,{\rm S}\cdot u]]_+\biggr)\nonumber\\
& & +ig_A^0\biggl({1\over4}[{\rm S}\cdot u,v\cdot u]
[v\cdot{\rm D},v\cdot u]-{ig_A^0\over4}[v\cdot u,v\cdot{\rm D}]^2
-{{g_A^0}^2\over4}{\rm S}\cdot u v\cdot u[v\cdot{\rm D},v\cdot u]\nonumber\\
& & -{ig_A^0\over4}u_\mu{\rm D}^\mu [v\cdot{\rm D},v\cdot u]
-{g_A^0\over2}\epsilon^{\mu\nu\rho\lambda}v_\rho{\rm S}_\lambda
u_\mu{\rm D}_\nu[v\cdot{\rm D},v\cdot u]-{{g_A^0}^2\over4}
{\rm S}\cdot u[v\cdot{\rm D},v\cdot u]v\cdot  u\nonumber\\
& & -{ig^0_A\over4}{\rm D}_\mu[v\cdot{\rm D},v\cdot u]u^\mu
-{g_A^0\over2}
\epsilon^{\mu\nu\rho\lambda}v_\rho{\rm S}_\lambda
{\rm D}_\mu[v\cdot{\rm D},v\cdot u]u_\nu+{\rm h.c.}\biggr)\nonumber\\
& & -{{g_A^0}^2\over4}\biggl(-[v\cdot u,v\cdot{\rm D}]^2
+ig_A^0[{\rm S}\cdot u,v\cdot u[v\cdot{\rm D},v\cdot u]]_+\biggr)\nonumber\\
& & -ig_A^0\biggl({1\over4}[u_\mu,v\cdot u]{\rm S}
\cdot u{\rm D}^\mu
+{ig_A^0\over4}v\cdot u{\rm  D}_\mu v\cdot u{\rm D}^\mu
-{ig_A^0\over4}u_\mu{\rm D}_\nu u^\mu{\rm D}^\nu\nonumber\\
& &-{g_A^0\over4}
\epsilon^{\mu\nu\rho\lambda}v_\rho{\rm S}_\lambda 
{\rm D}_\kappa u_\mu{\rm D}^\kappa u_\nu
+{\rm h.c.}\biggr)\nonumber\\
& & -{g_A^0}^3\biggl({1\over4}[{\rm S}\cdot u,[v\cdot{\rm D},\{(v\cdot u)^2-u^2\}]]
-{i\over16}
\epsilon^{\mu\nu\rho\lambda}v_\rho[u_\lambda,[v\cdot{\rm D},[u_\mu,u_\nu]]]
\nonumber\\
& & +{1\over4}[u^\mu,[v\cdot{\rm D},[{\rm S}\cdot u,u_\mu]]]_+
-{1\over4}[v\cdot u,[v\cdot{\rm D},[{\rm S}\cdot u,v\cdot u]]]_+\biggr)\nonumber\\
& & +{g_A^0\over4}\epsilon^{\mu\nu\rho\lambda} v_\rho\biggl(ig_A^0
{\rm S}\cdot u{\rm D}_\mu u_\lambda{\rm D}_\nu+{1\over4}[u_\mu,v\cdot u]
u_\lambda{\rm D}_\nu+{\rm h.c.}\biggr)
\nonumber\\
& & +{g_A^0}^2\biggl[{{g_A^0}^2\over4}(v\cdot u)^2[{1\over4}(v\cdot u)^2
-{1\over4}u^2+{i\over2}
\epsilon^{\mu\nu\rho\lambda}v_\rho{\rm S}_\lambda u_\mu u_\nu]\nonumber\\
& & +{ig_A^0\over4}\biggl([{\rm S}\cdot u,v\cdot{\rm D}](v\cdot u)^2
+{ig_A^0\over4}(v\cdot u)^4 -{ig_A^0\over4}u_\mu(v\cdot u)^2uu^\mu\nonumber\\
& & -{g_A^0\over2}\epsilon^{\mu\nu\rho\lambda}v_\rho{\rm S}_\lambda
u_\mu(v\cdot u)^2u_\nu+{\rm S}\cdot u v\cdot u[v\cdot{\rm D},v\cdot u]
\biggr)\nonumber\\
& & -{g_A^0\over2}{\rm S}\cdot u{\rm D}_\mu u^\mu
v\cdot u-{1\over16}[u^\mu,v\cdot u]u^\mu v\cdot u-{ig_A^0\over2}
{\rm D}_\mu u^\mu [v\cdot{\rm D},v\cdot u]\nonumber\\
& & -{ig_A^0\over2}{\rm D}_\mu u^\mu v\cdot u{\rm S}\cdot u
+{\rm h.c.}\biggr]\nonumber\\
& & -{{g_A^0}^3\over4}\biggl(iv\cdot u{\rm S} 
\cdot u[v\cdot{\rm D},v\cdot u]+{\rm h.c.}
-{g^0_A\over2} (v\cdot u)^4+{g_A^0\over2}
v\cdot u u^\mu v\cdot u u_\mu\nonumber\\
& & 
-ig_A^0\epsilon^{\mu\nu\rho\lambda}v_\rho{\rm S}_\lambda
v\cdot u u_\mu v\cdot u u_\nu\biggr)
\nonumber\\
& & +{{g_A^0}^2\over4}\biggl[-{g_A^0}^2\biggl({1\over4}
v\cdot u[(v\cdot u)^2-u^2]v\cdot u
-{1\over4}u_\mu[(v\cdot u)^2-u^2]u^\mu\nonumber\\
& & +{i\over2}
\epsilon^{\mu\nu\rho\lambda}v_\rho{\rm S}_\lambda u_\mu[(v\cdot u)^2-u^2]u_\nu\biggr)
-{\rm D}_\mu[(v\cdot u)^2-u^2]{\rm D}^\mu+2i
\epsilon^{\mu\nu\rho\lambda}v_\rho{\rm S}_\lambda
{\rm D}_\mu[(v\cdot u)^2-u^2]{\rm D}_\nu
\biggr]\nonumber\\
& & -({{ig_A^0}^3\over4}{\rm S}\cdot{\rm D}
[(v\cdot u)^2-u^2]v\cdot u+{\rm h.c.})-{{g_A^0}^4\over8}
v\cdot u[(v\cdot u)^2-u^2]v\cdot u-{ig_A^0\over2}
\epsilon^{\mu\nu\rho\lambda}v_\rho{\rm D}_\lambda u_\mu u_\nu v\cdot u
\nonumber\\
& & +{ig_A^0\over2}\biggl[(-{i\over2}{\rm D}^\mu u^\nu u_\mu 
{\rm D}_\nu+{\rm h.c.})-
\epsilon^{\mu\nu\rho\lambda}v_\rho{\rm S}_\lambda{\rm D}^\kappa[u_\mu,u_\kappa]
{\rm D}_\nu
+({1\over2}g_A^0{\rm D}^\mu[v\cdot u,u_\mu]_+{\rm S}\cdot u+{\rm h.c.})
\nonumber\\
& & -{ig_A^0\over4}\epsilon^{\mu\nu\rho\lambda}v_\rho 
u_\lambda[v\cdot u,u_\mu]{\rm D}_\nu
+{g_A^0\over2}\biggl(-u^\mu[v\cdot u,{\rm S}\cdot u]{\rm D}_\mu
-{ig_A^0\over4}v\cdot u[v\cdot u,u^\mu]u_\mu\nonumber\\
& & -{g_A ^0\over2}\epsilon^{\mu\nu\rho\lambda}v_\rho{\rm S}_\lambda
v\cdot u[v\cdot u,u_\mu]u_\nu
+u^\mu[v\cdot u ,u_\mu]{\rm S}\cdot{\rm D}\biggr)\biggr]\nonumber\\
& &+{g_A^0\over2}\biggl[
-{1\over4}\epsilon^{\mu\nu\rho\lambda}v_\rho
{\rm D}_\lambda u_\mu u_\nu v\cdot u+{i\over2}\biggl(
-{\rm D}_\mu[{\rm S}\cdot u,u_\mu]v\cdot u\nonumber\\
& & +ig_A^0[-{1\over4}u^2(v\cdot u)^2+{i\over2}
\epsilon^{\mu\nu\rho\lambda}v_\rho{\rm S}_\lambda u_\mu u_\nu(v\cdot u)^2]
\nonumber\\
& & -g_A^0[-{1\over4}u_\mu v\cdot u u^\mu v\cdot u+{i\over2}
\epsilon^{\mu\nu\rho\lambda}v_\rho{\rm S}_\lambda u_\mu v\cdot u u_\nu v\cdot u]
\biggr)+{\rm h.c.}\biggr]\nonumber\\
& &-{i{g_A^0}^3\over8}
\epsilon^{\mu\nu\rho\lambda}v_\rho{\rm S}_\lambda
v\cdot u u_\mu u_\nu v\cdot u. 
\end{eqnarray} 

Using: 
\begin{eqnarray}
\label{eq:B3onsh}
& & 
B^{(3)}_{\rm OS}=\beta_1\gamma^5[u^2,v\cdot u]_+ 
+\beta_2\gamma^5u^\mu v\cdot u u_\mu
\nonumber\\
& & 
+i\beta_3\epsilon^{\mu\nu\rho\lambda}\gamma^5{\rm S}_\mu[[u_\nu,u_\rho],u_\lambda]_+
+\beta_4\gamma^5[v\cdot u,\chi_+]_+
\nonumber\\
& & +\beta_5\gamma^5(v\cdot u)^3
+i\beta_6\epsilon^{\mu\nu\rho\lambda}\gamma^5v_\rho
{\rm S}_\lambda[v\cdot u,[u_\mu,u_\nu]]_+
\nonumber\\
& & +i\beta_7\epsilon^{\mu\nu\rho\lambda}\gamma^5v_\rho
{\rm S}_\lambda[u_\mu,[v\cdot u,u_\nu]]_+
+i\beta_8\gamma^5[v\cdot u,[{\rm S}\cdot{\rm D},v\cdot u]]
\nonumber\\
& & +i\beta_9\gamma^5[u^\mu,[{\rm S}\cdot{\rm D},u_\mu]]
+i\beta_{10}\gamma^5\biggl([v\cdot u,[v\cdot{\rm D},{\rm S}\cdot u]]_+
-[{\rm S}\cdot u,[v\cdot{\rm D},v\cdot u]]_+\biggr)\nonumber\\
& & 
+i\beta_{11}\gamma^5[u^\mu,[{\rm S}\cdot u,{\rm D}_\mu]_+]_+
+i\beta_{12}\gamma^5[u^\mu,[{\rm S}\cdot u,{\rm D}_\mu]]
\nonumber\\
& & 
+i\beta_{13}\gamma^5\biggl([v\cdot u,[{\rm S}\cdot{\rm D},v\cdot u]]_+
-[{\rm S}\cdot u,[v\cdot{\rm D},v\cdot u]]_+\biggr)
\nonumber\\
& & 
+\beta_{14}\gamma^5[\chi_-,{\rm S}\cdot u]
+i\beta_{15}\gamma ^5[v\cdot u,[v\cdot{\rm D},{\rm S}\cdot u]]
\nonumber\\
& & +i\beta_{16}\gamma^5[{\rm S}\cdot u,[v\cdot{\rm D},v\cdot u]],
\end{eqnarray}
${1\over{2\rm m}}\biggl[\gamma^0 B^{(3)}\ ^\dagger\gamma^0 B^{(1)}
+\gamma^0B^{(1)}\ ^\dagger\gamma^0B^{(3)}\biggr]$
and eliminating all terms proportional to the nonrelativistic
eom by  field redefinition of H, one gets:
\begin{eqnarray}
\label{eq:B3B1onsh}
& & {1\over{2\rm m}}\Biggl(\beta_1\biggl[-2i[[u^2,v\cdot u]_+,{\rm S}\cdot{\rm D}]_+
-{g^0_A\over2}[[u^2,v\cdot u]_+,v\cdot u]_+\biggr]\nonumber\\
& & +\beta_2\biggl[-2i[u^\mu v\cdot u u_\mu,{\rm S}\cdot{\rm D}]_+
+g_A^0[u^\mu v\cdot u u_\mu,v\cdot u]_+\biggr]\nonumber\\
& & 
+\beta_3\biggl[i{g_A^0\over2}
\epsilon^{\mu\nu\rho\lambda}[[[u_\nu,u_\rho],u_\lambda]_+,{\rm S}\cdot u]_+\nonumber\\
& & 
+2i[{\rm D}_\mu,\biggl([{\rm S}\cdot u,[u_\mu, v\cdot u]]_+
-[v\cdot u,[u^\mu,{\rm S}\cdot u]]_+]
+[u^\mu,[[v\cdot u,{\rm S}\cdot u]]_+\biggr)]\biggr]\nonumber\\
& & 
+\beta_4\biggl[-2i[{\rm S}\cdot{\rm D},[v\cdot u,\chi_+]_+]_+\nonumber\\
& & -{g_A^0\over2}[v\cdot u,[v\cdot u,\chi_+]_+]_+\biggr]
+\beta_5\biggl[-2i[{\rm S}\cdot{\rm D},(v\cdot u)^3]_+
-{g_A^0\over2}(v\cdot u)^4\biggr]\nonumber\\
& & +\beta_6\biggl[{1\over2}[-\epsilon^{\mu\nu\rho\lambda}v_\rho[{\rm D}_\lambda,[v\cdot u,[u_\mu,u_\nu]]_+]_+
\nonumber\\
& & 
-i{g_A^0\over2}\epsilon^{\mu\nu\rho\lambda}v_\rho{\rm S}_\lambda
[v\cdot u,[v\cdot u,[u_\mu,u_\nu]]_+]_+ 
-2i[{\rm D}_\mu,[v\cdot u,[{\rm S}\cdot u,u^\mu]]_+]\biggr]\nonumber\\
& & +\beta_7\biggl[-\epsilon^{\mu\nu\rho\lambda}v_\rho[{\rm D}_\lambda,[u_\mu,[v\cdot u,u_\nu]]_+]_+
-i[{\rm D}_\mu,[u^\mu,[{\rm S}\cdot u,v\cdot u]]_+] \nonumber\\
& & +{g_A^0\over2}\biggl(-{1\over2}[[u_\mu,v\cdot u],v\cdot u]_+,u^\mu]
+i\epsilon^{\mu\nu\rho\lambda}v_\rho{\rm S}_\lambda[[[u_\mu,v\cdot u],v\cdot u]_+u_\nu]_+\biggr)
\nonumber\\
& & -{ig_A^0\over2}\epsilon^{\mu\nu\rho\lambda}
v_\rho{\rm S}_\lambda[v\cdot u,[u_\mu,[v\cdot u,u_\nu]]_+]_+\biggr]
\nonumber\\
& & \beta_8\biggl[-{1\over2}[{\rm D}^\mu,[v\cdot u,[{\rm D}_\mu,v\cdot u]]]_+
+{ig_A^0\over2}[{\rm S}\cdot u,[v\cdot u,[v\cdot{\rm D},v\cdot u]]]_+\nonumber\\
& & -i\epsilon^{\mu\nu\rho\lambda}v_\rho{\rm S}_\lambda
[{\rm D}_\nu,[v\cdot u,[{\rm D}_\mu,v\cdot u]]] 
-i{g_A^0\over2}[v\cdot u,[v\cdot u,[{\rm S}\cdot {\rm D},v\cdot u]]]_+\biggr]\nonumber\\
& & \beta_9\biggl[-{1\over2}
[{\rm D}^\mu,[u_\nu,[{\rm D}_\mu,u^\nu]]]_+
+{ig_A^0\over2}[{\rm S}\cdot u,[u_\mu,[v\cdot{\rm D},u^\mu]]]_+
-i\epsilon^{\mu\nu\rho\lambda}v_\rho{\rm S}_\lambda
[{\rm D}_\nu,[u_\kappa,[{\rm D}_\mu,u^\kappa]]]\nonumber\\
& & -i{g_A^0\over2}[v\cdot u,[u_\mu,[{\rm S}\cdot{\rm D},u^\mu]]]_+
+\beta_{10}\biggl[{1\over2}[{\rm D}^\mu,[v\cdot u,[v\cdot{\rm D},u_\mu]]_+]\nonumber\\
& & 
-{ig_A^0\over2}[v\cdot u,[{\rm S}\cdot u,[v\cdot{\rm D},v\cdot u]]_+]
-{1\over2}[{\rm D}_\mu,[u^\mu,[v\cdot{\rm D},v\cdot u]]_+]\nonumber\\
& & 
+i\epsilon^{\mu\nu\rho\lambda}v_\rho{\rm  S}_\lambda
\biggl([{\rm D}_\nu,[v\cdot u,[v\cdot{\rm D},u_\mu]]_+]_+
-[{\rm D}_\nu,[u_\mu,[v\cdot{\rm D},v\cdot u]]_+]_+\biggr)
\nonumber\\
& & 
-{1\over2}[{\rm D}_\mu,[u^\mu,[v\cdot{\rm D},v\cdot u]]_+]
-{i\over2}
\epsilon^{\mu\nu\rho\lambda}v_\rho{\rm S}_\lambda
[{\rm D}_\nu,[u_\mu,[v\cdot{\rm D},v\cdot u]]_+]_+\nonumber\\
& & +i{g_A^0\over2}\biggl([v\cdot u,[v\cdot u[v\cdot{\rm D},{\rm S}\cdot u]]_+]
+[v\cdot u,[{\rm S}\cdot u,[v\cdot{\rm D},v\cdot u]]_+]\biggr)\biggr]\nonumber\\
& & 
+\beta_{11}\biggl[-{1\over2}[{\rm D}_\nu,[u^\mu,[{\rm D}_\mu,u^\nu]_+]_+]_+
-i\epsilon^{\mu\nu\rho\lambda}v_\rho{\rm S}_\lambda 
[{\rm D}_\nu,[u^\kappa,[u_\mu,{\rm D}_\kappa]_+]_+]
\nonumber\\
& & 
+{ig_A^0\over2}[{\rm S}\cdot u,[u^\mu,[v\cdot u,{\rm D}_\mu]_+]_+]_+
+{ig_A^0\over2}[{\rm S}\cdot u,[u^\mu,[v\cdot u,{\rm D}_\mu]]]_+\nonumber\\
& & 
-i{g^0_A\over2}[v\cdot u,[u^\mu,[{\rm S}\cdot u,{\rm D}_\mu]_+]_+]_+\biggr]
\nonumber\\
& & +\beta_{12}\biggl[-{1\over2}[{\rm D}^\nu,[u^\mu,[u_\nu,{\rm D}_\mu]]]_+
-i\epsilon^{\mu\nu\rho\lambda}v_\rho{\rm S}_\lambda
[{\rm D}_\nu,[u^\kappa,[u_\mu,{\rm D}_\kappa]]]\nonumber\\
& & 
-{ig_A^0\over2}[v\cdot u,[u^\mu,[{\rm S}\cdot u,{\rm D}^\mu]]]_+\biggr]\nonumber\\
& & 
+\beta_{13}\biggl[{1\over2}[{\rm D}^\mu,[v\cdot u,[{\rm D}_\mu,v\cdot u]]_+]
+{i\over2}\epsilon^{\mu\nu\rho\lambda}v_\rho{\rm S}_\lambda
[{\rm D}_\nu,[v\cdot u,[{\rm D}_\mu,v\cdot u]]_+]_+\nonumber\\
& & 
-{1\over2}[{\rm D}_\mu,[u^\mu,[v\cdot{\rm D},v\cdot u]]_+]
-{i\over2}
\epsilon^{\mu\nu\rho\lambda}v_\rho{\rm S}_\lambda
[{\rm D}_\nu,[u_\mu,[v\cdot{\rm D},v\cdot u]]_+]_+\nonumber\\
& & 
+{ig_A^0\over2} [v\cdot u,[v\cdot u,[{\rm S}\cdot{\rm D},v\cdot u]]_+]
-{ig_A^0\over2}[v\cdot u,[{\rm S}\cdot u,[v\cdot{\rm D},v\cdot u]]_+]\biggr]
\nonumber\\
& & +\beta_{14}\biggl[i[{\rm D}_\mu, [\chi_-,u^\mu]]_+
+g_A^0[{\rm S}\cdot u,[\chi_-,v\cdot u]]_+
-\epsilon^{\mu\nu\rho\lambda}v_\rho{\rm S}_\lambda
[{\rm D}_\nu,[\chi_-,u_\mu]]\nonumber\\
& & 
-{g_A^0\over2}[v\cdot u,[\chi_-,{\rm S}\cdot u]]_+\biggr]
+\beta_{15}\biggl[-{1\over2}[{\rm D}_\mu,[v\cdot u,[v\cdot{\rm D},u^\mu]]]_+\nonumber\\
& & 
+{ig_A^0\over2}[{\rm S}\cdot u,[v\cdot  u,[v\cdot{\rm D},v\cdot u]]]_+
-i\epsilon^{\mu\nu\rho\lambda}v_\rho{\rm S}_\lambda
[{\rm D}_\nu,[v\cdot u,[v\cdot{\rm D},u_\mu]]]\nonumber\\
& & 
-{ig_A^0\over2}[v\cdot u,[v\cdot u,[v\cdot{\rm D},{\rm S}\cdot u]]]_+\biggr]
+\beta_{16}\biggl[-{1\over2}[{\rm D}_\mu,[u^\mu,[v\cdot{\rm D},v\cdot u]]]_+\nonumber\\
& & 
-g_A^0[{\rm S}\cdot u,[v\cdot  u,[v\cdot{\rm D},v\cdot u]]]_+
-i\epsilon^{\mu\nu\rho\lambda}v_\rho{\rm S}_\lambda
[{\rm D}_\nu,[u_\mu,[v\cdot{\rm D},v\cdot u]]]\nonumber\\
& & 
-{ig_A^0\over2} [v\cdot u,[{\rm S}\cdot u,[v\cdot{\rm D},v\cdot u]]]_+
\biggr]\Biggr).\nonumber\\
& & 
\end{eqnarray}
The set $\{\beta_i\}$ can be related to the set $\{b_i\}$ of \cite{em}.

So, the final  result is:
\begin{eqnarray}
\label{eq:O4onshfinres}
& & {{\cal L}_{\rm HBChPT}^{(4)}}({\rm OS})=A^{(4)}
({\rm ON\  or\ ON}^\prime)+(\ref{eq:B2B2onsh})+(\ref{eq:B2C1B1onsh})
+(\ref{eq:B1C2B1onsh})
+(\ref{eq:B1C1C1B1onsh})+(\ref{eq:B3B1onsh}).\nonumber\\
& & 
\end{eqnarray}

\section{Conclusion}

A complete list of O$(q^4)$ terms for off-shell nucleons was 
obtained {\it working within HBChPT} 
using a phase rule obtained in \cite {1n}, along
with reductions from algebraic identities and reparameterization invariance.
We also obtain the on-shell O($q^4$) terms, again within the framework
of HBChPT.  For this paper, we set the external fields to zero and assume
isospin symmetry.
For off-shell nucleons, one gets a total of  207 ${\cal O}(q^4,\phi^{2n})$ terms (given in Table 1) 
the LECs of three of which (given in Table 4) are fixed relative to some
O$(q^{1,2,3})$ terms, and 113 O$(q^4,\phi^{2n+1})$ terms (given in Table 2).
Of the total of 320 terms, 230 are finite. For on-shell nucleons,
the LECs of 44 terms (given in (\ref{eq:RIcross})) are fixed relative to those of lower order terms.
As an application of the off-shell O($q^4$) list, we showed
that none of the O$(q^4$) terms contribute to the contact
graph of pion DCX at threshold and 
assuming static nucleons. In fact we argued that at threshold and
for static nulceons, only terms of even chiral orders will contribute
to the contact graph (of pion DCX).
For future work, one could use (\ref{eq:phoffsh}) for the construction
of  O$(q^4)$ terms
including external fields and assuming
isospin symmetry violation, within HBChPT. 
We also mention that  the authors of \cite{mms} and M.Mojzis
are working on constructing the full O$(q^4$) HBChPT Lagrangian, but
starting from the BChPT Lagrangian.
\section*{Acknowledgement}

The author would like to thank D.S.Koltun (University of Rochester)
for a critical reading of the manuscript.

\appendix

\section{}
\setcounter{equation}{0}
\seceqaa

In this appendix, we show how to obtain a set of linearly independent
terms from (\ref{eq:LCindoddanticoom1}).

Using (\ref{eq:curvature}) and (\ref{eq:oddanticoom}), one obtains
(a) the following four identities in nine terms:
\begin{eqnarray}
\label{eq:set1}
& & 
[{\rm D}_\mu,[{\rm D}_\nu,[{\rm D}^\mu,{\rm D}^\nu]_+]]
-[[{\rm D}_\mu,{\rm D}_\nu],[{\rm D}^\mu,{\rm D}^\nu]_+]_+
=-[{\rm D}_\mu,[{\rm D}_\nu,[{\rm D}^\mu,{\rm D}^\nu]_+]_+]_+\nonumber\\
& & 
-[{\rm D}_\mu,[{\rm D}_\nu,[{\rm D}^\mu,{\rm D}^\nu]_+]]
-[[u_\mu,u_\nu],[u^\mu,u^\nu]]_+
=-{1\over4}[{\rm D}_\mu,[{\rm D}_\nu,[u^\mu,u^\nu]]]_+\nonumber\\
& & 
[u_\mu,[u_\nu,[u^\mu,u^\nu]_+]]-[[u_\mu,u_\nu]_+,[u^\mu,u^\nu]_+]_+
=-[u_\mu,[u_\nu,[u^\mu,u^\nu]_+]_+]_+\nonumber\\
& & -[u_\mu,[u_\nu,[u^\mu,u^\nu]_+]]-[[u_\mu,u_\nu],[u^\mu,u^\nu]]_+
=-[u_\mu,[u_\nu,[u^\mu,u^\nu]]]_+,
\end{eqnarray}
and

(b) the following two identities in five terms:
\begin{eqnarray}
\label{eq:set2}
& & [{\rm D}]_\mu,[{\rm D}_\nu,[u^\mu,u^\nu]_+]]
-[[{\rm D}_\mu,{\rm D}_\nu]_+,[u^\mu,u^\nu]_+]_+
=-[u_\mu,[u_\nu,[u^\mu,u^\nu]_+]_+]_+\nonumber\\
& & [u_\mu,[u_\nu,[{\rm D}^\mu,{\rm D}^\nu]_+]]
-[[u_\mu,u_\nu]_+,[{\rm D}^\mu,{\rm D}^\nu]_+]_+
=-[u_\mu,[u_\nu,[{\rm D}^\mu,{\rm D}^\nu]_+]_+]_+.\nonumber\\
& & 
\end{eqnarray}

The reason for considering (\ref{eq:set1}) and (\ref{eq:set2})
separately is because the terms in them do not mix. One thus can take
$i=12,13,26,27,29$ and $i=46,47,48$ of Table  1 
as the two sets of  linearly
independent terms.

\section{}
\setcounter{equation}{0}
\seceqbb

In this appendix, we show how to obtain a set of linearly independent
terms from (\ref{eq:LCevenanticoom}).

Using (\ref{eq:curvature}), (\ref{eq:evenanticoom}),(\ref{eq:antisymsym})-
(\ref{eq:antisym1}), one gets the following 15 identities
in the following 22 terms:
\begin{eqnarray}
\label{eq:set3}
& & i\epsilon^{\mu\nu\rho\lambda}v_\rho\biggl(
[{\rm D}_\mu,[{\rm D}_\nu,[{\rm D}_\lambda,{\rm S}\cdot{\rm D}]_+]_+]
-{1\over4}[[u_\mu,u_\nu],[{\rm D}_\lambda,{\rm S}\cdot{\rm D}]_+]_+
=-[{\rm D}_\mu,[{\rm D}_\nu,[{\rm D}_\lambda,{\rm S}\cdot{\rm D}]_+]]_+
\biggr)
\nonumber\\
& & i\epsilon^{\mu\nu\rho\lambda}v_\rho\biggl(
{1\over4}[{\rm D}_\mu,[{\rm D}_\nu,[u_\lambda,{\rm S}\cdot u]]_+]_+
-{1\over{16}}[[u_\mu,u_\nu],[u_\lambda,{\rm S}\cdot u]]
=-{1\over4}[{\rm D}_\mu,[{\rm D}_\nu,[u_\lambda,{\rm S}\cdot u]]_+]_+
\biggr) \nonumber\\
& & i\epsilon^{\mu\nu\rho\lambda}v_\rho\biggl(
{1\over4}[{\rm D}_\mu,[{\rm D}_\nu,[u_\lambda,{\rm S}\cdot u]]]
-{1\over{16}}[[u_\mu,u_\nu],[u_\lambda,{\rm S}\cdot u]]
=-{1\over4}[{\rm D}_\mu,[{\rm D}_\nu,[u_\lambda,{\rm S}\cdot u]]]
\biggr) \nonumber\\
& & i\epsilon^{\mu\nu\rho\lambda}v_\rho\biggl(
[u_\mu,[u_\nu,[u_\lambda,{\rm S}\cdot u]_+]_+]
-[[u_\mu,u_\nu],[u_\lambda,{\rm S}\cdot u]_+]_+
=-[u_\mu,[u_\nu,[u_\lambda,{\rm S}\cdot u]_+]]_+ 
\biggr)\nonumber\\
& & i\epsilon^{\mu\nu\rho\lambda}v_\rho\biggl(
[u_\mu,[u_\nu,[u_\lambda,{\rm S}\cdot u]]_+]_+
-[[u_\mu,u_\nu],[u_\lambda,{\rm S}\cdot u]]
=-[u_\mu,[u_\nu,[u_\lambda,{\rm S}\cdot u]]_+]_+\biggr)\nonumber\\
& & i\epsilon^{\mu\nu\rho\lambda}v_\rho\biggl(
[u_\mu,[u_\nu,[u_\lambda,{\rm S}\cdot u]]]
-[[u_\mu,u_\nu],[u_\lambda,{\rm S}\cdot u]]
=-[u_\mu,[u_\nu,[u_\lambda,{\rm S}\cdot u]]]\biggr)\nonumber\\
& & i\epsilon^{\mu\nu\rho\lambda}v_\rho\biggl(
{1\over4}[{\rm S}\cdot{\rm D},[{\rm D}_\mu,[u_\nu,u_\lambda]]_+]_+
-{1\over{16}}[[{\rm S}\cdot u,u_\mu],[u_\nu,u_\lambda]]={1\over4}
[{\rm D}_\mu,[{\rm S}\cdot{\rm D},[u_\nu,u_\lambda]]_+]_+
\biggr)
\nonumber\\
& & i\epsilon^{\mu\nu\rho\lambda}v_\rho\biggl(
{1\over4}[{\rm S}\cdot{\rm D},[{\rm D}_\mu,[u_\nu,u_\lambda]]_+]_+
-{1\over{16}}[[{\rm S}\cdot {\rm D},{\rm D}_\mu]_+,[u_\nu,u_\lambda]]_+
=-{1\over4}[{\rm D}_\mu,[{\rm S}\cdot{\rm D},[u_\nu,u_\lambda]]]
\biggr)
\nonumber\\
& & i\epsilon^{\mu\nu\rho\lambda}v_\rho\biggl(
{1\over4}[{\rm S}\cdot{\rm D},[{\rm D}_\mu,[u_\nu,u_\lambda]]]
-{1\over{16}}[[{\rm S}\cdot u,u_\mu],[u_\nu,u_\lambda]]={1\over4}
[{\rm D}_\mu,[{\rm S}\cdot{\rm D},[u_\nu,u_\lambda]]]
\biggr)\nonumber\\
& & i\epsilon^{\mu\nu\rho\lambda}v_\rho\biggl(
[{\rm S}\cdot u,[u_\mu,[u_\nu,u_\lambda]]_+]_+
-[[{\rm S}\cdot u,u_\mu],[u_\nu,u_\lambda]]
=[u_\mu,[{\rm S}\cdot u,[u_\nu,u_\lambda]]_+]_+
\biggr)\nonumber\\
& & i\epsilon^{\mu\nu\rho\lambda}v_\rho\biggl(
[{\rm S}\cdot u,[u_\mu,[u_\nu,u_\lambda]]_+]_+
-[[{\rm S}\cdot u,u_\mu]_+,[u_\nu,u_\lambda]_+]
=-[u_\mu,[{\rm S}\cdot u,[u_\nu,u_\lambda]]]
\biggr)\nonumber\\
& & i\epsilon^{\mu\nu\rho\lambda}v_\rho\biggl(
[{\rm S}\cdot u,[u_\mu,[u_\nu,u_\lambda]]]
-[[{\rm S}\cdot u,u_\mu],[u_\nu,u_\lambda]]
=[u_\mu,[{\rm S}\cdot u,[u_\nu,u_\lambda]]]
\biggr)\nonumber\\
& & i\epsilon^{\mu\nu\rho\lambda}v_\rho\biggl(
[u_\mu,[u_\nu,[{\rm D}_\lambda,{\rm S}\cdot{\rm D}]_+]_+]
-[[u_\mu,u_\nu],[{\rm D}_\lambda,{\rm S}\cdot{\rm D}]_+]_+
=-[u_\mu,[u_\nu,[{\rm D}_\lambda,{\rm S}\cdot{\rm D}]_+]]_+
\biggr)\nonumber\\
& & i\epsilon^{\mu\nu\rho\lambda}v_\rho\biggl(
[u_\mu,[u_\nu,[{\rm D}_\lambda,{\rm S}\cdot{\rm D}]_+]_+]
-[[u_\mu,u_\nu],[{\rm D}_\lambda,{\rm S}\cdot{\rm D}]_+]_+
=-[u_\mu,[u_\nu,[{\rm D}_\lambda,{\rm S}\cdot{\rm D}]_+]]_+
\biggr)\nonumber\\
& & i\epsilon^{\mu\nu\rho\lambda}v_\rho\biggl(
[{\rm D}_\mu,[{\rm D}_\nu,[u_\lambda,{\rm S}\cdot u]_+]_+]
-[[u_\mu,u_\nu],[u_\lambda,{\rm S}\cdot{\rm D}]_+]_+
=-[{\rm D}_\mu,[{\rm D}_\nu,[u_\lambda,{\rm S}\cdot u]_+]]_+
\biggr)\nonumber\\
& & i\epsilon^{\mu\nu\rho\lambda}v_\rho\biggl(
[u_\mu,[{\rm S}\cdot u,[u_\nu,u_\lambda]]_+]_+
-{1\over4}[[u_\mu,{\rm S}\cdot u]_+,[u_\nu,u_\lambda]]_+
=-{1\over4}[{\rm S}\cdot u,[u_\mu,[u_\nu,u_\lambda]]]\biggr).\nonumber\\
& & 
\end{eqnarray}
One  can thus take $i=14,15,30,31,49,50,51$ of Table 1
as linearly independent terms.

\section{} 
\setcounter{equation}{0}
\seceqcc

In this appendix, we show how to obtain a set of linearly independent
terms from (\ref{eq:LCevenanticoom1}).

Using (\ref{eq:curvature}) and (\ref{eq:LCevenanticoom1}), one gets
24 identities in 30 terms:
\begin{eqnarray}
\label{eq:set4}
& & i\epsilon^{\mu\nu\rho\lambda}v_\rho{\rm S}_\lambda\biggl(
[{\rm D}_\kappa,[{\rm D}_\mu,[{\rm D}^\kappa,{\rm D}_\nu]_+]_+]
-[[{\rm D}_\kappa,{\rm D}_\mu]_+,[{\rm D}^\kappa,{\rm D}_\nu]_+]
=-[{\rm D}_\mu,[{\rm D}_\kappa,[{\rm D}^\kappa,{\rm D}_\nu]_+]_+]
\biggr)
\nonumber\\
& & i\epsilon^{\mu\nu\rho\lambda}v_\rho{\rm S}_\lambda\biggl(
[{\rm D}_\kappa,[{\rm D}_\mu,[{\rm D}^\kappa,{\rm D}_\nu]_+]_+]
-{1\over4}[[u_\kappa,u_\mu],[{\rm D}^\kappa,{\rm D}_\nu]_+]_+
=-[{\rm D}_\mu,[{\rm D}_\kappa,[{\rm D}^\kappa,{\rm D}_\nu]_+]]_+
\biggr)
\nonumber\\
& & i\epsilon^{\mu\nu\rho\lambda}v_\rho{\rm S}_\lambda\biggl(
[{\rm D}_\kappa,[{\rm D}_\mu,[{\rm D}^\kappa,{\rm D}_\nu]_+]]_+
-[[{\rm D}_\kappa,{\rm D}_\mu]_+,[{\rm D}^\kappa,{\rm D}_\nu]_+]
=-[{\rm D}_\mu,[{\rm D}_\kappa,[{\rm D}^\kappa,{\rm D}_\nu]_+]]_+
\biggr)
\nonumber\\
& & i\epsilon^{\mu\nu\rho\lambda}v_\rho{\rm S}_\lambda\biggl(
{1\over4}[{\rm D}_\kappa,[{\rm D}_\mu,[u^\kappa,u_\nu]]_+]_+
-{1\over{16}}[[u_\kappa,u_\mu],[u^\kappa,u_\nu]]
=-{1\over4}[{\rm D}_\mu,[{\rm D}_\kappa,[u^\kappa,u_\nu]]_+]_+
\biggr)
\nonumber\\
& & i\epsilon^{\mu\nu\rho\lambda}v_\rho{\rm S}_\lambda\biggl(
{1\over4}[{\rm D}_\kappa,[{\rm D}_\mu,[u^\kappa,u_\nu]]_+]_+
-{1\over4}[[u_\kappa,u_\mu],[{\rm D}^\kappa,u_\nu]_+]_+
=-{1\over4}[{\rm D}_\mu,[{\rm D}_\kappa,[u^\kappa,u_\nu]]]
\biggr)
\nonumber\\
& & i\epsilon^{\mu\nu\rho\lambda}v_\rho{\rm S}_\lambda\biggl(
[u_\kappa,[u_\mu,[u^\kappa,u_\nu]_+]_+]-[[u^\kappa,u_\mu]_+,[u_\kappa,u_\nu]_+]
=-[u_\mu,[u_\kappa,[u^\kappa,u_\nu]_+]_+]
\biggr)
\nonumber\\
& & i\epsilon^{\mu\nu\rho\lambda}v_\rho{\rm S}_\lambda\biggl(
[u_\kappa,[u_\mu,[u^\kappa,u_\nu]_+]_+]-[[u^\kappa,u_\mu],[u_\kappa,u_\nu]_+]_+
=-[u_\mu,[u_\kappa,[u^\kappa,u_\nu]_+]]_+
\biggr)
\nonumber\\
& & i\epsilon^{\mu\nu\rho\lambda}v_\rho{\rm S}_\lambda\biggl(
[u_\kappa,[u_\mu,[u^\kappa,u_\nu]_+]]_+-[[u^\kappa,u_\mu]_+,[u_\kappa,u_\nu]_+]
=-[u_\mu,[u_\kappa,[u^\kappa,u_\nu]_+]]_+
\biggr)
\nonumber\\
& & i\epsilon^{\mu\nu\rho\lambda}v_\rho{\rm S}_\lambda\biggl(
[u_\kappa,[u_\mu,[u^\kappa,u_\nu]]_+]_+-[[u^\kappa,u_\mu],[u_\kappa,u_\nu]]
=-[u_\mu,[u_\kappa,[u^\kappa,u_\nu]]_+]_+
\biggr)
\nonumber\\
& & i\epsilon^{\mu\nu\rho\lambda}v_\rho{\rm S}_\lambda\biggl(
[u_\kappa,[u_\mu,[u^\kappa,u_\nu]]_+]_+-[[u^\kappa,u_\mu]_+,[u_\kappa,u_\nu]]_+
=-[u_\mu,[u_\kappa,[u^\kappa,u_\nu]]]
\biggr)
\nonumber\\
& & i\epsilon^{\mu\nu\rho\lambda}v_\rho{\rm S}_\lambda\biggl(
[u_\kappa,[u_\mu,[u^\kappa,u_\nu]]]-[[u^\kappa,u_\mu],[u_\kappa,u_\nu]]
=-[u_\mu,[u_\kappa,[u^\kappa,u_\nu]]]
\biggr)
\nonumber\\
& & i\epsilon^{\mu\nu\rho\lambda}v_\rho{\rm S}_\lambda\biggl(
[u_\kappa,[u_\mu,[{\rm D}^\kappa,{\rm D}_\nu]_+]_+]
-[[u_\kappa,u_\mu]_+,[{\rm D}^\kappa,{\rm D}_\nu]_+]
=-[u_\mu,[u_\kappa,[{\rm D}^\kappa,{\rm D}_\nu]_+]_+]
\biggr)
\nonumber\\
& & i\epsilon^{\mu\nu\rho\lambda}v_\rho{\rm S}_\lambda\biggl(
[u_\kappa,[u_\mu,[{\rm D}^\kappa,{\rm D}_\nu]_+]]_+
-[[u_\kappa,u_\mu],[{\rm D}^\kappa,{\rm D}_\nu]_+]_+
=-[u_\mu,[u_\kappa,[{\rm D}^\kappa,{\rm D}_\nu]_+]]_+
\biggr)
\nonumber\\
& & i\epsilon^{\mu\nu\rho\lambda}v_\rho{\rm S}_\lambda\biggl(
[u_\kappa,[u_\mu,[{\rm D}^\kappa,{\rm D}_\nu]_+]]_+
-[[u_\kappa,u_\mu]_+,[{\rm D}^\kappa,{\rm D}_\nu]_+]
=-[u_\mu,[u_\kappa,[{\rm D}^\kappa,{\rm D}_\nu]_+]]_+
\nonumber\\
& & i\epsilon^{\mu\nu\rho\lambda}v_\rho{\rm S}_\lambda\biggl(
[{\rm D}_\kappa,[{\rm D}_\mu,[u^\kappa,u_\nu]_+]_+]
-[[{\rm D}_\kappa,{\rm D}_\mu]_+,[u^\kappa,u_\nu]_+]
=-[{\rm D}_\mu,[{\rm D}_\kappa,[u^\kappa,u_\nu]_+]_+]
\biggr)\nonumber\\
& & i\epsilon^{\mu\nu\rho\lambda}v_\rho{\rm S}_\lambda\biggl(
[{\rm D}_\kappa,[{\rm D}_\mu,[u^\kappa,u_\nu]_+]_+]
-{1\over4}[[u_\kappa,u_\mu],[u^\kappa,u_\nu]_+]_+
=-[{\rm D}_\mu,[{\rm D}_\kappa,[u^\kappa,u_\nu]_+]]_+
\biggr)\nonumber\\
& & i\epsilon^{\mu\nu\rho\lambda}v_\rho{\rm S}_\lambda\biggl(
[{\rm D}_\kappa,[{\rm D}_\mu,[u^\kappa,u_\nu]_+]]_+
-[[{\rm D}_\kappa,{\rm D}_\mu]_+,[u^\kappa,u_\nu]_+]
=-[{\rm D}_\mu,[{\rm D}_\kappa,[u^\kappa,u_\nu]_+]]_+
\biggr).\nonumber\\
& & 
\end{eqnarray}
One can thus take $i=7,22,23,34,55,56$ as linearly independent terms.

\begin{table} [htbp]
\centering
\caption{The Allowed O$(q^4,\phi^{2n})$
Terms}
\begin{tabular} {|c|c|c|c|c|} \hline
$i$ & $(m,n,p,q)$ & Terms &  F($\equiv$Finite) & E\\
& & &  D($\equiv$Divergent)[$d_i$] & ON \\ \hline
1 & (4,0,0,0)   & 
$(v\cdot\rm D)^4$ & D[$d_{197}$] & E  \\ \hline
2 & & ${\rm D}_\mu(v\cdot\rm D)^2{\rm D}^\mu$ & D[$d_{198}$] &  E  \\ \hline
3 & &  $[{\rm D}^2,(v\cdot\rm D)^2]_+$ & F & E   \\ \hline 
4 & & $v\cdot{\rm D}{\rm D}_\mu v\cdot{\rm D}{\rm D}^\mu+{\rm h.c.}$ & F 
& E  \\ \hline 
5 & & ${\rm D}^4$ & F &   E \\ \hline                                         
6 & & $i\epsilon^{\mu\nu\rho\lambda}
v_\rho{\rm S}_\lambda[{\rm D}^2,[{\rm D}_\mu,{\rm D}_\nu]]_+$ & F & E \\ 
\hline                                         
7 & & $i\epsilon^{\mu\nu\rho\lambda}
v_\rho\rm S_\lambda{\rm D}_\kappa[{\rm D}_\mu,{\rm D}_\nu]{\rm D}^\kappa$ & 
D[$d_{178}$] & E  \\ \hline                                         
8 & & $i\epsilon^{\mu\nu\rho\lambda}
v_\rho{\rm S}_\lambda{\rm D}_\mu{\rm D}^2{\rm D}_\nu$ & F  & E \\
\hline                                         
9 & & $i\epsilon^{\mu\nu\rho\lambda}
v_\rho{\rm S}_\lambda[(v\cdot{\rm D})^2,[{\rm D}_\mu,{\rm D}_\nu]]_+$ & F
& E \\ \hline                                         
10 & & $i\epsilon^{\mu\nu\rho\lambda}
v_\rho\rm S_\lambda
v\cdot {\rm D}[{\rm D}_\mu,{\rm D}_\nu]v\cdot{\rm D}$ & D[$d_{177}$]
& E \\ \hline                                         
11 & & $i\epsilon^{\mu\nu\rho\lambda}
v_\rho\rm S_\lambda{\rm D}_\mu(v\cdot{\rm D})^2{\rm D}_\nu$ & F  & E \\
\hline
12 & & $[\rm D_\mu,[\rm D_\nu,[\rm D^\mu,\rm D^\nu]_+]]$ & F & E \\ \hline
13 & & $[\rm D_\mu,[\rm D_\nu,[\rm D^\mu,\rm D^\nu]_+]_+]_+$ & F & E \\ \hline
14 & & $\epsilon^{\mu\nu\rho\lambda}v_\rho
[\rm D_\mu,[\rm D_\nu,[\rm D_\lambda,\rm S\cdot\rm D]_+]]_+$ & F & E \\
\hline
15 & & $\epsilon^{\mu\nu\rho\lambda}v_\rho
[\rm D_\mu,[\rm D_\nu,[\rm D_\lambda,\rm S\cdot\rm D]_+]_+]$ & F & E \\
\hline
16& & $\epsilon^{\mu\nu\rho\lambda}S_\rho 
[{\rm D}_\mu,[{\rm D}_\nu,[{\rm D}_\lambda,v\cdot\rm D]_+]_+]$ & F & E \\
\hline
17& & $\epsilon^{\mu\nu\rho\lambda}S_\rho 
[{\rm D}_\mu,[{\rm D}_\nu,[{\rm D}_\lambda,v\cdot\rm D]_+]]_+$ & F & E \\
\hline
18 & (0,4,0,0) &   $(v\cdot u)^4$ & D[$d_5$] & ON  \\ \hline
19 & & $[u^2,(v\cdot u)^2]_+$ & D[$d_3$] &  ON \\ \hline
20 & &  $v\cdot u u_\mu v\cdot u u^\mu+{\rm h.c.}$ &  E
& ON \\
\hline
21 & & $u^4$ & D[$d_2$] & ON \\ \hline
22 & & $i\epsilon^{\mu\nu\rho\lambda}
v_\rho{\rm S}_\lambda[u^2,[u_\mu,u_\nu]]_+$ & D[$d_{31}$] & ON \\
\hline
23 & & $i\epsilon^{\mu\nu\rho\lambda}
v_\rho\rm S_\lambda u_\kappa[u_\mu,u_\nu]u^\kappa$ &  D[$d_{32}$] & ON \\
\hline
24 & & 
$i\epsilon^{\mu\nu\rho\lambda}v_\rho{\rm S}_\lambda
[(v\cdot u)^2,[u_\mu,u_\nu]]_+$ & D[$d_{34}$] & ON \\
\hline
25 & & $i\epsilon^{\mu\nu\rho\lambda}
v_\rho\rm S_\lambda
v\cdot u[u_\mu,u_\nu]v\cdot u$ & D[$d_{35}$] & ON \\ \hline
26& & $[u_\mu,u_\nu]^2$ & F & ON \\ \hline
27& & $[u_\mu,u_\nu]_+^2$ & D[$d_1$]  & ON \\ \hline
28 & & $[[v\cdot u,u^\mu],[v\cdot u,u_\mu]_+]$ & F & ON \\ 
\hline
29 & & $[u_\mu,[u_\nu.[u^\mu,u^\nu]_+]]$ & F & ON \\
\hline
30& & $i\epsilon^{\mu\nu\rho\lambda}v_\rho
[{\rm S}\cdot u,[u_\lambda,[u_\mu,u_\nu]]_+]_+$
& D[$d_{24}$] & ON \\
\hline
31& & $i\epsilon^{\mu\nu\rho\lambda}v_\rho
[u_\lambda,[{\rm S}\cdot u,[u_\mu,u_\nu]]_+]_+$
& D[$d_{25}$] & ON \\
\hline
32& & $i\epsilon^{\mu\nu\rho\lambda}{\rm S}_\rho
[v\cdot u,[u_\lambda,[u_\mu,u_\nu]]_+]_+$
& E & ON \\ \hline
33& & $i\epsilon^{\mu\nu\rho\lambda}{\rm S}_\rho
[u_\lambda,[v\cdot u,[u_\mu,u_\nu]]_+]_+$
& E & ON \\ \hline
\end{tabular}
\end{table}

\addtocounter{table}{-1}
\begin{table} [htbp]
\centering
\caption{continued}
\begin{tabular} {|c|c|c|c|c|} \hline
$i$ & $(m,n,p,q)$ & Terms &  F($\equiv$Finite) & E\\
& & &  D($\equiv$Divergent)[$d_i$] & ON \\ \hline
34& & $i\epsilon^{\mu\nu\rho\lambda}v_\rho{\rm S}_\lambda
[[u_\mu,u_\kappa],[u_\nu,u^\kappa]_+]_+$ 
& D[$d_{67}$] & ON
\\ \hline
35& & $i\epsilon^{\mu\nu\rho\lambda}v_\rho{\rm S}_\lambda
[[u_\mu,v\cdot u],[u_\nu,v\cdot u]_+]_+$ 
& D[$d_{36}$] & ON \\ \hline
36 & & $[u_\mu,v\cdot u]_+^2$ & D[$d_4$] & ON \\
\hline
37 & & $u_\mu(v\cdot u)^2 u^\mu$ & F & ON \\
\hline
38 & (2,2,0,0) &  $v\cdot{\rm D} (v\cdot u)^2 v\cdot{\rm D}$ &  
D[$d_{146}$] & E \\ \hline
39 & & $[v\cdot{\rm D},(v\cdot u)]^2$ & 
D[$d_{128}$] &  ON \\ \hline
40
 & &  $v\cdot{\rm D}u^2v\cdot{\rm D}$ & D[$d_{149}$] & E \\ \hline
41 & & $[v\cdot{\rm D},u_\mu]^2$ & D[$d_{133}$] & ON \\
\hline
42 & & ${\rm D}_\mu (v\cdot u)^2 {\rm D}^\mu$ & D[$d_{147}$] & ON \\ \hline
43 & & $[{\rm D}_\mu,(v\cdot u)]^2$ & D[$d_{135}$] & ON \\
\hline
44 & & $v\cdot{\rm D}[v\cdot u,u^\mu]_+\rm D_\mu+h.c.$ & D[$d_{148}]$] & E \\
\hline
45 & & $[{\rm D}_\mu,[v\cdot{\rm D},[v\cdot u,u^\mu]_+]]$ & F & E \\
\hline
46 & & $[{\rm D}_\mu,[{\rm D}_\nu,[u^\mu,u^\nu]_+]]$ & F & E \\ \hline
47 & & $[u_\mu,[u_\nu,[{\rm D}^\mu,{\rm D}^\nu]_+]]$ & F & E \\
\hline
48& & $[{\rm D}_\mu,[{\rm D}_\nu,[u^\mu,u^\nu]_+]_+]_+$ & F & E \\
\hline
49 & & $i\epsilon^{\mu\nu\rho\lambda}v_\rho
[{\rm S}\cdot{\rm D},[{\rm D}_\mu,[u_\nu,u_\lambda]]_+]_+$
& F & E \\ \hline
50 & & $i\epsilon^{\mu\nu\rho\lambda}v_\rho
[{\rm D}_\mu,[{\rm D}_\nu,[u_\lambda,{\rm S}\cdot u]_+]]_+$
& F & E \\ \hline
51 & & $i\epsilon^{\mu\nu\rho\lambda}v_\rho
[{\rm D}_\mu,[{\rm D}_\nu,[u_\lambda,{\rm S}\cdot u]_+]_+]$
& F & E \\ \hline
52& & $i\epsilon^{\mu\nu\rho\lambda}{\rm S}_\rho
[v\cdot{\rm D},[{\rm D}_\mu,[u_\nu,u_\lambda]]_+]_+$
& F & E \\ \hline
53 & & $i\epsilon^{\mu\nu\rho\lambda}{\rm S}_\rho
[{\rm D}_\mu,[{\rm D}_\nu,[u_\lambda,v\cdot u]_+]]_+$
& F & E \\ \hline
54& & $i\epsilon^{\mu\nu\rho\lambda}{\rm S}_\rho
[{\rm D}_\mu,[{\rm D}_\nu,[u_\lambda,v\cdot u]_+]_+]$
& F & E \\ \hline
55 & & $i\epsilon^{\mu\nu\rho\lambda}v_\rho{\rm S}_\lambda
[{\rm D}_\mu,[{\rm D}_\kappa,[u_\nu,u^\kappa]]]$ &
D[$d_{179}$] & E \\ \hline
56  & & $i\epsilon^{\mu\nu\rho\lambda}v_\rho{\rm S}_\lambda
[{\rm D}_\mu,[{\rm D}_\kappa,[u_\nu,u^\kappa]]_+]_+$ &
F & E \\ \hline
57  & & $i\epsilon^{\mu\nu\rho\lambda}v_\rho{\rm S}_\lambda
v\cdot{\rm D}[v\cdot u,u_\mu]\rm D_\nu+h.c.$ & D[$d_{176}$] & E \\ \hline
58 & & $i\epsilon^{\mu\nu\rho\lambda}v_\rho{\rm S}_\lambda
[{\rm D}_\mu,[v\cdot{\rm D},[u_\nu,v\cdot u]]]$ & F& E
\\ \hline
59 & & $[{\rm D}_\mu, u_\nu]^2$ & D[$d_{136}$]& ON \\ \hline
60 & & ${\rm D}_\mu u^2{\rm D}^\mu$ & D[$d_{150}$] & E \\
\hline
61 & & $i\epsilon^{\mu\nu\rho\lambda}
v_\rho{\rm S}_\lambda[{\rm D}_\mu,[{\rm D}_\nu,u^2]]_+$ & F & E \\
\hline
62 & & $i\epsilon^{\mu\nu\rho\lambda}
v_\rho{\rm S}_\lambda[{\rm D}_\mu,[{\rm D}_\nu,(v\cdot u)^2]]_+$ 
& F & E \\ 
\hline
63 & &   
$[v\cdot u,[[v\cdot{\rm D},v\cdot u],v\cdot{\rm D}]]_+$ & D[$d_{129}$] & ON 
\\ \hline
64 & & $[v\cdot{\rm D},[[v\cdot{\rm D},v\cdot u],v\cdot u]]_+$ & 
D[$d_{138}$] & E \\ \hline
65 & &  $[u_\mu,[[v\cdot{\rm D}, u^\mu], v\cdot{\rm D}]]_+$ & D[$d_{134}$]
& ON \\ \hline
\end{tabular}
\end{table}

\addtocounter{table}{-1}
\begin{table} [htbp]
\centering
\caption{continued}
\begin{tabular} {|c|c|c|c|c|} \hline
$i$ & $(m,n,p,q)$ & Terms &  F($\equiv$Finite) & E\\
& & &  D($\equiv$Divergent)[$d_i$] & ON \\ \hline
66 & & $[v\cdot{\rm D},[[v\cdot{\rm D},u_\mu], u^\mu]]_+$ 
& D[$d_{142}$] & E \\ \hline
67 & & $[v\cdot u,[[{\rm D}^\mu,v\cdot u],{\rm D}_\mu]]_+$ 
& D[$d_{131}$] & ON \\ \hline
68 & & $[{\rm D}_\mu,[[{\rm D}^\mu,v\cdot u],v\cdot u]]_+$ 
& D[$d_{141}$] & ON \\ \hline
69 & & $[u^\nu,[[{\rm D}_\mu, u_\nu]_+,{\rm D}^\mu]]$ & D[$d_{137}$] 
& E \\ \hline
70 & & $[{\rm D}_\mu,[[{\rm D}^\mu, u_\nu], u^\nu]]_+$ & D[$d_{143}$]
& E \\ \hline
71 & & $[{\rm D}_\mu,[[{\rm D}_\nu,u^\mu],u^\nu]_+]$ & F & E  \\ \hline
72 & & $[{\rm D}_\mu,[[{\rm D}_\nu,u^\mu]_+,u^\nu]_+]_+$ & F & E \\ \hline
73 & & $[{\rm D}_\mu,[[{\rm D}_\nu,u^\mu],u^\nu]]_+$ & D[$d_{145}$] 
& E \\ \hline
74 & & $[{\rm D}_\mu,[[{\rm D}_\nu,u^\mu]_+,u^\nu]]$ & F & E \\
\hline
75 & & $[u_\mu,[{\rm D}_\nu,[{\rm D}^\mu,u^\nu]]_+]$
& F & ON \\ \hline
76 & & $[u_\mu,[{\rm D}_\nu,[{\rm D}^\mu,u^\nu]]]_+$
& F & ON \\ \hline
77 & & $[u_\mu,[[{\rm D}^\mu,v\cdot u],v\cdot{\rm D}]]_+$ & 
D[$d_{130}$] & E \\ \hline
78 & & $[v\cdot{\rm D},[[{\rm D}_\nu,v\cdot u],u^\nu]]_+$ & 
D[$d_{144}$] & E \\ \hline
79 & & $[[{\rm D}_\mu,v\cdot u],[v\cdot{\rm D},u^\mu]]_+$ & 
D[$d_{132}$] & ON \\ \hline
80 & & $[{\rm D}_\mu,[[v\cdot{\rm D},u^\mu],v\cdot u]]_+$ & D[$d_{140}$] & E \\
\hline 
81& & $[{\rm D}_\mu,[[v\cdot{\rm D},u^\mu]_+,v\cdot u]]$ & F  & E \\
\hline 
82 & & $[{\rm D}_\mu,[[v\cdot{\rm D},u^\mu]_+,v\cdot u]_+]_+$ & F & E \\
\hline
83& & $[v\cdot{\rm D},[u^\mu,[{\rm D}_\mu,v\cdot u]]_+]$ & F & E
\\ \hline 
84& & $[v\cdot{\rm D},[u^\mu,[{\rm D}_\mu,v\cdot u]]]_+$ & F & E
\\ \hline 
85 & & $[{\rm D}_\mu,[[{\rm D}_\nu,u^\nu]_+,u^\mu]]$ & F &  E \\ \hline
86 & & $[{\rm D}_\mu,[[{\rm D}_\nu,u^\nu],u^\mu]]_+$ & D[$d_{123}$] & E \\
\hline
87
 & & $[{\rm D}_\mu,[[{\rm D}_\nu,u^\nu]_+,u^\mu]_+]_+$  & F & E \\ \hline 
88 & & $[{\rm D}_\mu,[[{\rm D}_\nu,u^\nu],u^\mu]_+]$ & F  & E \\ \hline
89 & & $[u_\mu,[[{\rm D}_\nu,u^\nu],{\rm D}^\mu]_+]$ & F  & ON \\ \hline
90& & $[u_\mu,[[{\rm D}_\nu,u^\nu],{\rm D}^\mu]]_+$ & F  & ON \\ \hline
91 & & $[{\rm D}_\mu,[[v\cdot{\rm D},v\cdot u]_+,u^\mu]]$ & F &  ON \\ \hline
92  & & $[{\rm D}_\mu,[[v\cdot{\rm D},v\cdot u], u^\mu]]_+$ & D[$d_{139}$] & ON \\
\hline
93 & & $[{\rm D}_\mu,[[v\cdot{\rm D},v\cdot u]_+, u^\mu]_+]_+$  & F  & ON \\ \hline
94 & & $[{\rm D}_\mu,[[v\cdot{\rm D},v\cdot u],u^\mu]_+]$ 
& F & ON \\ \hline
95& & $[u_\mu,[[v\cdot{\rm D},v\cdot u],{\rm D}^\mu]_+]$ 
& F & ON \\ \hline
96 & & $[u_\mu,[[v\cdot{\rm D},v\cdot u],{\rm D}^\mu]]_+$ 
& F & ON \\ \hline
97& & $[v\cdot{\rm D},[[u_\mu,{\rm D}^\mu]_+,v\cdot u]_+]_+$ & F & E 
\\ \hline
98 & & $[v\cdot{\rm D},[[u_\mu,{\rm D}^\mu],v\cdot u]]_+$ & F & E 
\\ \hline
\end{tabular}
\end{table}

\addtocounter{table}{-1}
\begin{table} [htbp]
\centering
\caption{continued}
\begin{tabular} {|c|c|c|c|c|} \hline
$i$ & $(m,n,p,q)$ & Terms &  F($\equiv$Finite) & E\\
& & &  D($\equiv$Divergent)[$d_i$] & ON \\ \hline
99 & & $i\epsilon^{\mu\nu\rho\lambda}
v_\rho[{\rm D}_\mu,[[{\rm D}_\nu,u_\lambda],{\rm S}\cdot u]]$ & F & E \\
\hline
100 & & $i\epsilon^{\mu\nu\rho\lambda}
v_\rho[{\rm D}_\mu,[[{\rm D}_\nu,u_\lambda]_+,{\rm S}\cdot u]_+]$ & F & E \\
\hline
101  & & $i\epsilon^{\mu\nu\rho\lambda}
v_\rho[{\rm D}_\mu,[[{\rm D}_\nu,u_\lambda]_+,{\rm S}\cdot u]]_+$ & F & E \\
\hline
102 & & $i\epsilon^{\mu\nu\rho\lambda}
v_\rho[{\rm D}_\mu,[[{\rm D}_\nu,u_\lambda],{\rm S}\cdot u]_+]_+$ & F & E \\
\hline
103 & & $i\epsilon^{\mu\nu\rho\lambda}
v_\rho[{\rm D}_\nu,[[{\rm D}_\mu,{\rm S}\cdot u],u_\lambda]_+]_+$ & F & E \\  
\hline
104& & $i\epsilon^{\mu\nu\rho\lambda}
v_\rho[{\rm D}_\nu,[[{\rm D}_\mu,{\rm S}\cdot u]_+,u_\lambda]_+]$ & F & E \\  
\hline
105& & $i\epsilon^{\mu\nu\rho\lambda}
v_\rho[{\rm D}_\nu,[[{\rm D}_\mu,{\rm S}\cdot u]_+,u_\lambda]_+]$ & F & E \\  
\hline
106
& & $i\epsilon^{\mu\nu\rho\lambda}
v_\rho[{\rm D}_\nu,[[{\rm D}_\mu,{\rm S}\cdot u],u_\lambda]]$ & F & E \\  
\hline
107 & & $i\epsilon^{\mu\nu\rho\lambda}
{\rm S}_\rho[{\rm D}_\mu,[[{\rm D}_\nu,u_\lambda],v\cdot u]]$ & F & E \\
\hline
108 & & $i\epsilon^{\mu\nu\rho\lambda}
{\rm S}_\rho[{\rm D}_\mu,[[{\rm D}_\nu,u_\lambda]_+,v\cdot u]_+]$ & F & E \\
\hline
109 & & $i\epsilon^{\mu\nu\rho\lambda}
{\rm S}_\rho[{\rm D}_\mu,[[{\rm D}_\nu,u_\lambda]_+,v\cdot u]]_+$ & F & E \\
\hline
110 & & $i\epsilon^{\mu\nu\rho\lambda}
{\rm S}_\rho[{\rm D}_\mu,[[{\rm D}_\nu,u_\lambda],v\cdot u]_+]_+$ & F & E \\    
\hline
111& & $i\epsilon^{\mu\nu\rho\lambda}
{\rm S}_\rho[{\rm D}_\nu,[[{\rm D}_\mu,v\cdot u]_+,u_\lambda]]_+$ & F & E \\    
\hline
112 & & $i\epsilon^{\mu\nu\rho\lambda}
{\rm S}_\rho[{\rm D}_\nu,[[{\rm D}_\mu,v\cdot u]_+,u_\lambda]_+]$ & F & E \\    
\hline
113& & $i\epsilon^{\mu\nu\rho\lambda}
{\rm S}_\rho[{\rm D}_\nu,[[{\rm D}_\mu,v\cdot u],u_\lambda]]$ & F & E \\    
\hline
114 & & $i\epsilon^{\mu\nu\rho\lambda}
{\rm S}_\rho[{\rm D}_\nu,[[{\rm D}_\mu,v\cdot u],u_\lambda]_+]_+$ & F & E \\    
\hline
115 & & $i\epsilon^{\mu\nu\rho\lambda}
v_\rho[{\rm D}_\mu,[[{\rm S}\cdot{\rm D},u_\nu],u_\lambda]]$ & F  & E \\
\hline
116 & & $i\epsilon^{\mu\nu\rho\lambda}
v_\rho[{\rm D}_\mu,[[{\rm S}\cdot{\rm D},u_\nu]_+,u_\lambda]_+]$ & F & E \\
\hline
117 & & $i\epsilon^{\mu\nu\rho\lambda}
v_\rho[{\rm D}_\mu,[[{\rm S}\cdot{\rm D},u_\nu],u_\lambda]_+]_+$ & F & E  \\
\hline
118 & & $i\epsilon^{\mu\nu\rho\lambda}
v_\rho[{\rm D}_\mu,[[{\rm S}\cdot{\rm D},u_\nu]_+,u_\lambda]]_+$ & F & E \\
\hline
119 & & $i\epsilon^{\mu\nu\rho\lambda}
{\rm S}_\rho[{\rm D}_\mu,[[v\cdot{\rm D},u_\nu],u_\lambda]]$ & F & E \\
\hline
120 & & $i\epsilon^{\mu\nu\rho\lambda}
v_\rho[{\rm S}\cdot{\rm D},[[{\rm D}_\mu,u_\lambda]_+,u_\nu]]_+$ & F & E \\
\hline
121 & & $i\epsilon^{\mu\nu\rho\lambda}
v_\rho[{\rm S}\cdot{\rm D},[[{\rm D}_\mu,u_\lambda]_+,u_\nu]]_+$ & F & E \\
\hline
122 & & $i\epsilon^{\mu\nu\rho\lambda}
v_\rho[{\rm S}\cdot{\rm D},[[{\rm D}_\mu,u_\lambda]_+,u_\nu]_+]$ & F & E \\
\hline
123 & & $i\epsilon^{\mu\nu\rho\lambda}
v_\rho[{\rm S}\cdot{\rm D},[[{\rm D}_\mu,u_\lambda],u_\nu]]$ & F & E \\
\hline
124 & & $i\epsilon^{\mu\nu\rho\lambda}
{\rm S}_\rho[{\rm D}_\mu,[[v\cdot{\rm D},u_\nu]_+,u_\lambda]]_+$ & F & E \\
\hline
125& & $i\epsilon^{\mu\nu\rho\lambda}
{\rm S}_\rho[{\rm D}_\mu,[[v\cdot{\rm D},u_\nu],u_\lambda]_+]_+$ & F & E \\
\hline
126
& & $i\epsilon^{\mu\nu\rho\lambda}
{\rm S}_\rho[[{\rm D}_\mu,u_\lambda],[v\cdot{\rm D},u_\nu]_+]_+$ & 
D[$d_{188}$] & E \\
\hline
127 & & $i\epsilon^{\mu\nu\rho\lambda}
{\rm S}_\rho[v\cdot{\rm D},[[{\rm D}_\mu,u_\lambda],u_\nu]_+]_+$ & F & E \\
\hline
128 & & $i\epsilon^{\mu\nu\rho\lambda}
{\rm S}_\rho[v\cdot{\rm D},[[{\rm D}_\mu,u_\lambda]_+,u_\nu]]_+$ & F & E \\
\hline
129& & $i\epsilon^{\mu\nu\rho\lambda}
{\rm S}_\rho[v\cdot{\rm D},[[{\rm D}_\mu,u_\lambda]_+,u_\nu]]_+$ & F & E \\
\hline
130& & $i\epsilon^{\mu\nu\rho\lambda}
{\rm S}_\rho[v\cdot{\rm D},[[{\rm D}_\mu,u_\lambda],u_\nu]]$ & F & E \\
\hline
131& & $i\epsilon^{\mu\nu\rho\lambda}
v_\rho{\rm S}_\lambda[{\rm D}_\mu,[[{\rm D}_\nu,u^\kappa],u_\kappa]]$ & F & E \\
\hline
\end{tabular}
\end{table}

\addtocounter{table}{-1}
\begin{table} [htbp]
\centering
\caption{continued}
\begin{tabular} {|c|c|c|c|c|} \hline
$i$ & $(m,n,p,q)$ & Terms &  F($\equiv$Finite) & E\\
& & &  D($\equiv$Divergent)[$d_i$] & ON \\ \hline
132 & & $i\epsilon^{\mu\nu\rho\lambda}
v_\rho{\rm S}_\lambda[{\rm D}_\mu,[[{\rm D}_\nu,u^\kappa]_+,u_\kappa]_+]$ & F& E \\
\hline
133 & & $i\epsilon^{\mu\nu\rho\lambda}
v_\rho{\rm S}_\lambda[{\rm D}_\mu,[[{\rm D}_\nu,u^\kappa],u_\kappa]_+]_+$ & F & E \\
\hline
134 & & $i\epsilon^{\mu\nu\rho\lambda}
v_\rho{\rm S}_\lambda[{\rm D}_\mu,[[{\rm D}_\nu,u^\kappa]_+,u_\kappa]]_+$ & F & E \\
\hline
135& & $i\epsilon^{\mu\nu\rho\lambda}
v_\rho{\rm S}_\lambda[u_\kappa,[[{\rm D}_\nu,u^\kappa]_+,{\rm D}_\mu]]_+$ & F & E \\
\hline
136 & & $i\epsilon^{\mu\nu\rho\lambda}
v_\rho{\rm S}_\lambda[u_\kappa,[[{\rm D}_\nu,u^\kappa]_+,{\rm D}_\mu]_+]$ & F & E \\
\hline
137 & & $i\epsilon^{\mu\nu\rho\lambda}
v_\rho{\rm S}_\lambda[{\rm D}_\mu,[[{\rm D}_\nu,v\cdot u],v\cdot u]]$ & F & E \\
\hline
138 & & $i\epsilon^{\mu\nu\rho\lambda}
v_\rho{\rm S}_\lambda[{\rm D}_\mu,[[{\rm D}_\nu,v\cdot u]_+,v\cdot u]_+]$ & F & E\\
\hline
139 & & $i\epsilon^{\mu\nu\rho\lambda}
v_\rho{\rm S}_\lambda[{\rm D}_\mu,[[{\rm D}_\nu,v\cdot u],v\cdot u]_+]_+$ & F & E \\
\hline
140& & $i\epsilon^{\mu\nu\rho\lambda}
v_\mu{\rm S}_\lambda[{\rm D}_\mu,[[{\rm D}_\nu,v\cdot u]_+,v\cdot u]]_+$ & F & E \\
\hline
141 & & $i\epsilon^{\mu\nu\rho\lambda}
v_\mu{\rm S}_\lambda[v\cdot u,[[{\rm D}_\nu,v\cdot u]_+,u_\mu]]_+$ & F & E \\
\hline
142 & & $i\epsilon^{\mu\nu\rho\lambda}
v_\mu{\rm S}_\lambda[v\cdot u,[[{\rm D}_\nu,v\cdot u]_+,u_\mu]_+]$ & F & E \\
\hline
143& & $i\epsilon^{\mu\nu\rho\lambda}
v_\rho{\rm S}_\lambda[{\rm D}_\mu,[[{\rm D}_\kappa,u_\nu],u^\kappa]]$ & F & E \\
\hline
144 & & $i\epsilon^{\mu\nu\rho\lambda}
v_\rho{\rm S}_\lambda[{\rm D}_\mu,[[{\rm D}_\kappa,u_\nu]_+,u^\kappa]_+]$ & F & E \\
\hline
145 & & $i\epsilon^{\mu\nu\rho\lambda}
v_\rho{\rm S}_\lambda[{\rm D}_\mu,[[{\rm D}_\kappa,u_\nu],u^\kappa]_+]_+$ & F 
& E \\
\hline
146 &  & $i\epsilon^{\mu\nu\rho\lambda}
v_\rho{\rm S}_\lambda[{\rm D}_\mu,[[{\rm D}_\kappa,u_\nu]_+,u^\kappa]]_+$ & F & E\\
\hline
147 &  & $i\epsilon^{\mu\nu\rho\lambda}
v_\rho{\rm S}_\lambda[{\rm D}_\kappa,[[{\rm D}_\mu,u_\kappa],u^\nu]_+]_+$ & F & E\\
\hline
148&  & $i\epsilon^{\mu\nu\rho\lambda}
v_\rho{\rm S}_\lambda[{\rm D}_\kappa,[[{\rm D}_\mu,u_\kappa]_+,u^\nu]]_+$ & F & E\\
\hline
149 &  & $i\epsilon^{\mu\nu\rho\lambda}
v_\rho{\rm S}_\lambda[{\rm D}_\kappa,[[{\rm D}_\mu,u_\kappa]_+,u^\nu]_+]$ & F & E\\
\hline
150&  & $i\epsilon^{\mu\nu\rho\lambda}
v_\rho{\rm S}_\lambda[{\rm D}_\kappa,[[{\rm D}_\mu,u_\kappa],u^\nu]]$ & F & E\\
\hline
151 & & $i\epsilon^{\mu\nu\rho\lambda}
v_\rho{\rm S}_\lambda[{\rm D}_\mu,[[v\cdot{\rm D},u_\nu],v\cdot u]]$ & 
D[$d_{166}$] & E \\
\hline
152 & & $i\epsilon^{\mu\nu\rho\lambda}
v_\rho{\rm S}_\lambda[[{\rm D}_\mu,v\cdot u],[v\cdot{\rm D},u_\nu]]$ 
& D[$d_{163}$] & ON \\
\hline
153 & & $i\epsilon^{\mu\nu\rho\lambda}
v_\rho{\rm S}_\lambda[{\rm D}_\mu,[[v\cdot{\rm D},u_\nu], v\cdot u]_+]_+$ 
& D[$d_{172}$] & E \\
\hline
154
 & & $i\epsilon^{\mu\nu\rho\lambda}
v_\rho{\rm S}_\lambda[v\cdot{\rm D},[[{\rm D}_\mu,v\cdot u],u_\nu]]$ &  
D$d_{164}$] & E \\
\hline
155& & $i\epsilon^{\mu\nu\rho\lambda}
v_\rho{\rm S}_\lambda[{\rm D}_\mu,[[v\cdot{\rm D},u_\nu]_+, v\cdot u]]_+$ & F & E
\\ \hline
156& & $i\epsilon^{\mu\nu\rho\lambda}
v_\rho{\rm S}_\lambda[{\rm D}_\mu,[[v\cdot{\rm D},u_\nu]_+, v\cdot u]_+]$ 
& F & E \\ \hline
157& & $i\epsilon^{\mu\nu\rho\lambda}
v_\rho{\rm S}_\lambda[v\cdot{\rm D},[[{\rm D}_\mu,v\cdot u]_+,u_\nu]_+]$ & F & E
\\ \hline
158& & $i\epsilon^{\mu\nu\rho\lambda}
v_\rho{\rm S}_\lambda[v\cdot{\rm D},[[{\rm D}_\mu,v\cdot u]_+,u_\nu]]_+$ & F & E
\\ \hline  
159 & & $i\epsilon^{\mu\nu\rho\lambda}
v_\rho{\rm S}_\lambda[{\rm D}_\kappa,[[{\rm D}_\mu,u_\nu], u^\kappa]]$ & F & E \\
\hline
160 & & $i\epsilon^{\mu\nu\rho\lambda}
v_\rho{\rm S}_\lambda[{\rm D}_\kappa,[[{\rm D}_\mu,u_\nu]_+,u^\kappa]_+]$ & F & E
\\ \hline
161 & & $i\epsilon^{\mu\nu\rho\lambda}
v_\rho{\rm S}_\lambda[{\rm D}_\kappa,[[{\rm D}_\mu,u_\nu],u^\kappa]_+]_+$ & F & E \\
\hline
162 & & $i\epsilon^{\mu\nu\rho\lambda}
v_\rho{\rm S}_\lambda[{\rm D}_\kappa,[[{\rm D}_\mu,u_\nu]_+,u^\kappa]]_+$ & F & E \\
\hline
\end{tabular}
\end{table}

\addtocounter{table}{-1}
\begin{table} [htbp]
\centering
\caption{continued}
\begin{tabular} {|c|c|c|c|c|} \hline
$i$ & $(m,n,p,q)$ & Terms &  F($\equiv$Finite) & E\\
& & &  D($\equiv$Divergent)[$d_i$] & 0N \\ \hline
163 & & $i\epsilon^{\mu\nu\rho\lambda}
v_\rho{\rm S}_\lambda[{\rm D}_\mu,[[{\rm D}_\kappa,u^\kappa],u_\nu]]$ & F & E \\
\hline
164& & $i\epsilon^{\mu\nu\rho\lambda}
v_\rho{\rm S}_\lambda[{\rm D}_\mu,[[{\rm D}_\kappa,u^\kappa]_+,u_\nu]_+]$ & F & E \\
\hline
165& & $i\epsilon^{\mu\nu\rho\lambda}
v_\rho{\rm S}_\lambda[{\rm D}_\mu,[[{\rm D}_\kappa,u^\kappa]_+,u_\nu]]_+$ & F & E \\
\hline
166& & $i\epsilon^{\mu\nu\rho\lambda}
v_\rho{\rm S}_\lambda[{\rm D}_\mu,[[{\rm D}_\kappa,u^\kappa],u_\nu]_+]_+$ & F & E \\
\hline
167 & & $i\epsilon^{\mu\nu\rho\lambda}
v_\rho{\rm S}_\lambda[v\cdot{\rm D},[[{\rm D}_\mu,u_\nu],v\cdot u]]$ & 
D[$d_{165}$]
& E \\ \hline
168 & & $i\epsilon^{\mu\nu\rho\lambda}
v_\rho{\rm S}_\lambda[[v\cdot{\rm D},v\cdot u],[{\rm D}_\mu,u_\nu]]$ & 
D[$d_{162}$]&  ON \\ \hline
169 & & $i\epsilon^{\mu\nu\rho\lambda}
v_\rho{\rm S}_\lambda[v\cdot{\rm D},[[{\rm D}_\mu,u_\nu],v\cdot u]_+]_+$ 
& D[$d_{173}$] & E\\
\hline
170 & & $i\epsilon^{\mu\nu\rho\lambda}
v_\rho{\rm S}_\lambda[v\cdot{\rm D},[[{\rm D}_\mu,u_\nu]_+,v\cdot u]]_+$ & 
F & E \\ \hline
171 & & $i\epsilon^{\mu\nu\rho\lambda}
v_\rho{\rm S}_\lambda[v\cdot{\rm D},[[{\rm D}_\mu,u_\nu]_+,v\cdot u]_+]$ & 
F & E \\ \hline
172 & & $i\epsilon^{\mu\nu\rho\lambda}
v_\rho{\rm S}_\lambda[{\rm D}_\mu,[[v\cdot{\rm D},v\cdot u],u_\nu]]$ 
& D[$d_{171}$]
& E \\ \hline
173
& & $i\epsilon^{\mu\nu\rho\lambda}
v_\rho{\rm S}_\lambda[{\rm D}_\mu,[[v\cdot{\rm D},v\cdot u]_+,u_\nu]]_+$ 
& F & E \\ \hline
174& & $i\epsilon^{\mu\nu\rho\lambda}
v_\rho{\rm S}_\lambda[{\rm D}_\mu,[[v\cdot{\rm D},v\cdot u]_+,u_\nu]_+]$ 
& F & E \\ \hline
175 & & $i\epsilon^{\mu\nu\rho\lambda}
v_\rho{\rm S}_\lambda[u_\mu,[[{\rm D}^\kappa,u_\nu],{\rm D}_\kappa]]$ & 
D[$d_{170}$] 
& ON \\ \hline
176 & & $i\epsilon^{\mu\nu\rho\lambda}
v_\rho{\rm S}_\lambda[[{\rm D}_\kappa,u_\mu],[{\rm D}^\kappa,u_\nu]]$ & D[$d_{169}$] &
ON \\ \hline
177 & & $i\epsilon^{\mu\nu\rho\lambda}
v_\rho{\rm S}_\lambda[{\rm D}_\kappa,[[{\rm D}^\kappa,u_\mu],u_\nu]_+]_+$ & D[$d_{175}$] 
& E\\ \hline
178 & & $i\epsilon^{\mu\nu\rho\lambda}
v_\rho{\rm S}_\lambda[{\rm D}_\kappa,[[{\rm D}^\kappa,u_\mu]_+,u_\nu]]_+$ & F 
& E \\ \hline
179& & $i\epsilon^{\mu\nu\rho\lambda}
v_\rho{\rm S}_\lambda[u_\mu,[[{\rm D}^\kappa,u_\nu]_+,{\rm D}_\kappa]]_+$ & 
F & E \\ \hline
180& & $i\epsilon^{\mu\nu\rho\lambda}
v_\rho{\rm S}_\lambda[{\rm D}_\kappa,[[{\rm D}^\kappa,u_\mu]_+,u_\nu]_+]$ & F 
& E \\ \hline
181 & & $i\epsilon^{\mu\nu\rho\lambda}
v_\rho{\rm S}_\lambda[u_\mu,[[v\cdot{\rm D},u_\nu],v\cdot{\rm D}]]$ & 
D[$d_{168}$]
& ON \\ \hline
182 & & $i\epsilon^{\mu\nu\rho\lambda}
v_\rho{\rm S}_\lambda[[v\cdot{\rm D},u_\mu],[v\cdot{\rm D},u_\mu]]$ & D[$d_{167}$]
& ON\\ \hline
183 & & $i\epsilon^{\mu\nu\rho\lambda}
v_\rho{\rm S}_\lambda[v\cdot{\rm D},[[v\cdot{\rm D},u_\mu],u_\nu]_+]_+$ & D[$d_{174}$]
& E \\ \hline
184 & & $i\epsilon^{\mu\nu\rho\lambda}
v_\rho {\rm S}_\lambda[v\cdot{\rm D},[[v\cdot{\rm D},u_\mu]_+,u_\nu]]_+$ & F 
& E \\ \hline
185& & $i\epsilon^{\mu\nu\rho\lambda}
v_\rho{\rm S}_\lambda[u_\mu,[[v\cdot{\rm D},u_\nu]_+,v\cdot{\rm D}]_+]$ & 
F & E \\ \hline
186& & $i\epsilon^{\mu\nu\rho\lambda}
v_\rho {\rm S}_\lambda[v\cdot{\rm D},[[v\cdot{\rm D},u_\mu]_+,u_\nu]_+]$ & F 
& E \\ \hline
\end{tabular}
\end{table}

\addtocounter{table}{-1}
\begin{table} [htbp]
\centering
\caption{continued}
\begin{tabular} {|c|c|c|c|c|} \hline
$i$ & $(m,n,p,q)$ & Terms &  F($\equiv$Finite) & E\\
& & &  D($\equiv$Divergent)[$d_i$] & ON\\ \hline
187 & (2,0,1,0)  &  
$[\rm D_\mu,[\rm D^\mu,\chi_+]]$ & D[$d_{158}$] & E \\ \hline 
188 & & $\rm D_\mu\chi_+\rm D^\mu$ & D[$d_{160}$] & E\\ \hline
189 & & $v\cdot\rm D\chi_+v\cdot\rm D$  & D[$d_{159}$] & E \\ \hline
190
& & $[v\cdot\rm D,[v\cdot\rm D,\chi_+]]$ &  D[$d_{157}$] & E \\ \hline
191 & &  
$i\epsilon^{\mu\nu\rho\lambda}
v_\rho\rm S_\lambda[\rm D_\mu,[\rm D_\nu,\chi_+]_+]$ & F & E \\ \hline
192 & (1,1,0,1) & 
$i[v\cdot{\rm D},[v\cdot u,\chi_-]]_+$ & F & E \\ \hline 
193 & & $i[v\cdot{\rm D},[v\cdot u,\chi_-]_+]$ & F & E
\\ \hline 
194& & $i[v\cdot u,[v\cdot {\rm D},\chi_-]]_+$ & F & ON
\\ \hline 
195 & &  $i[{\rm D}_\mu,[u^\mu,\chi_-]]_+$ & F & E  \\ \hline
196& &  $i[u_\mu,[{\rm D}^\mu,\chi_-]]_+$ & F & ON  \\ \hline
197 & &  $i[{\rm D}_\mu,[u^\mu,\chi_-]_+]$
&  F & E \\ \hline 
198 & & $\epsilon^{\mu\nu\rho\lambda}
v_\rho{\rm S}_\lambda[{\rm D}_\mu,[u_\nu,\chi_-]],$ &  F & E \\ \hline
199& & $\epsilon^{\mu\nu\rho\lambda}
v_\rho{\rm S}_\lambda[u_\mu,[{\rm D}_\nu,\chi_-]],$ &  F & ON \\ \hline
200 & &  $\epsilon^{\mu\nu\rho\lambda}
v_\rho{\rm S}_\lambda[{\rm D}_\mu,[u_\nu,\chi_-]_+]_+$ & F & E \\ \hline
201 & (0,2,1,0)  & $u_\mu\chi_+u^\mu$ & F & ON \\ \hline
202 & & $u^2\chi_+$ &  D[$d_{10}$] & ON \\ \hline
203 & & $v\cdot u\chi_+v\cdot u$ & F & ON \\ \hline
204 & & $(v\cdot u)^2\chi_+$ & D[$d_{13}$] & ON   \\ \hline 
205 & &  
$i\epsilon^{\mu\nu\rho\lambda}v_\rho{\rm S}_\lambda[[u_\mu,u_\nu],\chi_+]_+$
&  D[$d_{51}$] & ON \\ \hline
206 & (0,0,2,0) & $\chi_+^2$       &  D[$d_{21}$] & ON \\ \hline
207 & (0,0,0,2) & $\chi_-^2$       &  F & ON \\ \hline
\end{tabular}
\end{table}

\begin{table} [htbp]
\centering
\caption{The Allowed O$(q^4,\phi^{2n+1})$
Terms}
\begin{tabular} {|c|c|c|c|c|} \hline
$i$ & $(m,n,p,q)$ & Terms &  F($\equiv$Finite) & E\\
& & &  D($\equiv$Divergent)[$d_i$] & ON \\ \hline
208 & (3,1,0,0) &   
$i[{\rm D}_\mu,[v\cdot{\rm D},[{\rm D}^\mu,{\rm S}\cdot u]]]_+$ & D[$d_{191}$] 
& E \\ \hline
209& & $i[v\cdot{\rm D},[{\rm D}_\mu,[{\rm D}^\mu,{\rm S}\cdot u]_+]]$
& F & E \\ \hline
210& & $i[{\rm D}_\mu,[{\rm D}^\mu,[v\cdot{\rm D},{\rm S}\cdot u]]]_+$
& F & E \\ \hline
211& & $i[{\rm S}\cdot{\rm D},[v\cdot{\rm D},[{\rm D}_\mu,u^\mu]]]_+$
& F & ON$^\prime$ \\ \hline
212& & $i[v\cdot{\rm D},[{\rm S}\cdot{\rm D},[{\rm D}_\mu,u^\mu]_+]]$
& F & E \\ \hline
213& & $i[{\rm D}_\mu,[v\cdot{\rm D},[{\rm S}\cdot{\rm D},u^\mu]]]_+$
& F & E \\ \hline
214& & $i[v\cdot{\rm D},[{\rm D}_\mu[{\rm S}\cdot{\rm D},u^\mu]_+]]$
& F & E \\ \hline
215 & & $i[{\rm D}_\mu,[{\rm S}\cdot{\rm D},[v\cdot{\rm D},u^\mu]_+]_+]_+$ & F & E\\
\hline
216& & $i[{\rm D}_\mu,[{\rm S}\cdot{\rm D},[v\cdot{\rm D},u^\mu]_+]]$ & F & E\\
\hline
217& & $i[{\rm D}_\mu,[{\rm S}\cdot{\rm D},[{\rm D}^\mu,v\cdot u]_+]]$ & F & E\\
\hline
218& & $i[{\rm D}_\mu,[{\rm S}\cdot{\rm D},[{\rm D}^\mu,v\cdot u]_+]_+]_+$ & F & E\\
\hline
219& & $i[{\rm S}\cdot{\rm D},[{\rm D}_\mu,[v\cdot{\rm D},u^\mu]]]_+$
& F & E \\ \hline
220& & $i[{\rm S}\cdot{\rm D},[{\rm D}_\mu,[{\rm D}^\mu,v\cdot u]]]_+$
& F & E \\ \hline
221 & & $i[v\cdot{\rm D},[{\rm S}\cdot{\rm D},[{\rm D}_\mu,u^\mu]_+]_+]_+$ & F & E \\
\hline
222 & & $i[{\rm S}\cdot{\rm D},[v\cdot{\rm D},[v\cdot{\rm D}, v\cdot u]]]_+$ & 
D[$d_{189}$]& E \\ \hline
223 & & $i[v\cdot{\rm D},[{\rm S}\cdot{\rm D},[v\cdot{\rm D}, v\cdot u]_+]]$ & F & ON
 \\ \hline
224 & & $i[v\cdot{\rm D},[{\rm S}\cdot{\rm D},[v\cdot{\rm D}, v\cdot u]_+]_+]_+$ & F & ON
\\ \hline
225 & 
& $i[v\cdot{\rm D},[v\cdot{\rm D},[{\rm S}\cdot{\rm D},v\cdot u]_+]]$ & D[$d_{194}$] & E \\
\hline
226 & & $i{\rm D}_\mu[v\cdot{\rm D},{\rm S}\cdot u]_+{\rm D}^\mu$ & D[$d_{196}$] & E  \\
\hline
227 & & $i[{\rm D}_\mu,[{\rm D}^\mu,[{\rm S}\cdot{\rm D},v\cdot u]]]_+$ & F & E \\
\hline
228 & &  $i[v\cdot{\rm D},[v\cdot{\rm D},[v\cdot{\rm D},{\rm S}\cdot u]]]_+$ 
& D[$d_{190}$] & E \\ \hline
229 & & $i[{\rm  S}\cdot u,(v\cdot{\rm D})^3]_+$ &  D[$d_{193}$] & 
ON$^\prime$ \\ \hline
230 & & $iv\cdot{\rm D}[v\cdot{\rm D},{\rm S}\cdot u]_+v\cdot{\rm D}$
& D[$d_{195}$] & E \\ \hline
231 & & $\epsilon^{\mu\nu\rho\lambda} 
[{\rm D}_\mu,[{\rm D}_\nu,[{\rm D}_\rho,u_\lambda]_+]_+]$
& F & E \\ \hline
232 & & $\epsilon^{\mu\nu\rho\lambda} 
[{\rm D}_\mu,[{\rm D}_\nu,[{\rm D}_\rho,u_\lambda]]_+]_+$
& F & E \\ \hline
233 & & $\epsilon^{\mu\nu\rho\lambda} 
[{\rm D}_\mu,[{\rm D}_\nu,[{\rm D}_\rho,u_\lambda]_+]]_+$
& F & E \\ \hline
234 & & $\epsilon^{\mu\nu\rho\lambda}v_\rho
[{\rm D}_\mu,[{\rm D}_\nu,[v\cdot{\rm D},u_\lambda]]_+]_+$
& D[$d_{72}$] & ON \\
\hline
235
 & & $\epsilon^{\mu\nu\rho\lambda}v_\rho
[{\rm D}_\mu,[{\rm D}_\nu,[v\cdot{\rm D},u_\lambda]_+]]_+$
& F & E \\ \hline
236 & & $\epsilon^{\mu\nu\rho\lambda}v_\rho
[{\rm D}_\mu,[{\rm D}_\nu,[v\cdot{\rm D},u_\lambda]_+]_+]$
& F & E \\ \hline
237 & & $\epsilon^{\mu\nu\rho\lambda}v_\rho
[{\rm D}_\mu,[{\rm D}_\nu,[{\rm D}_\lambda,v\cdot u]]_+]_+$
& F & ON \\ \hline
238& & $\epsilon^{\mu\nu\rho\lambda}v_\rho
[{\rm D}_\mu,[{\rm D}_\nu,[{\rm D}_\lambda,v\cdot u]_+]]_+$
& F & ON \\ \hline
239& & $\epsilon^{\mu\nu\rho\lambda}v_\rho
[{\rm D}_\mu,[{\rm D}_\nu,[{\rm D}_\lambda,v\cdot u]_+]_+]$
& F & ON \\ \hline
240
& & $\epsilon^{\mu\nu\rho\lambda}v_\rho
[v\cdot{\rm D},[{\rm D}_\mu,[{\rm D}_\nu,u_\lambda]_+]]_+$
& F & E \\ \hline
241& & $\epsilon^{\mu\nu\rho\lambda}v_\rho
[v\cdot{\rm D},[{\rm D}_\mu,[{\rm D}_\nu,u_\lambda]_+]_+]$
& F & E \\ \hline
242 & & $\epsilon^{\mu\nu\rho\lambda}v_\rho
[[v\cdot{\rm D},{\rm D}_\mu]_+,[{\rm D}_\nu,u_\lambda]_+]$
& F & E \\ \hline
\end{tabular}
\end{table}

\addtocounter{table}{-1}
\begin{table} [htbp]
\centering
\caption{continued}
\begin{tabular} {|c|c|c|c|c|} \hline
$i$ & $(m,n,p,q)$ & Terms &  F($\equiv$Finite) & E\\
& & &  D($\equiv$Divergent)[$d_i$] & ON \\ \hline
243& & $\epsilon^{\mu\nu\rho\lambda}v_\rho
[u_\mu,[{\rm D}_\nu,[v\cdot{\rm D},{\rm D}_\lambda]_+]_+]$ 
& F & E \\ \hline
244 & (1,3,0,0) &  
$i[v\cdot u,u_\mu]_+,[u^\mu,{\rm S}\cdot{\rm D}]_+]_+$ & D[$d_{84}$] 
& ON$^\prime$ \\ \hline
245& & $i[[{\rm D}_\mu,{\rm S}\cdot u],[v\cdot u, u^\mu]]_+$ & D[$d_{81}$]
& ON \\ \hline
246& & $i[[{\rm D}_\mu,{\rm S}\cdot u],[v\cdot u, u^\mu]_+]$ & F 
& ON \\ \hline
247& & $i[{\rm S}\cdot u,[{\rm D}_\mu,[u^\mu,v\cdot u]]]_+$ & F
& ON \\ \hline
248& & $i[{\rm D}_\mu,[{\rm S}\cdot u,[u^\mu,v\cdot u]]]_+$ & F & E \\
\hline
249& & $i[v\cdot u,[u_\mu,[{\rm D}^\mu,{\rm S}\cdot u]]_+]$
& F & ON \\ \hline
250& & $i[u_\mu,[v\cdot u,[{\rm D}^\mu,{\rm S}\cdot u]]_+]$ & F & ON \\
\hline
251& & $i[v\cdot u,[u_\mu,[{\rm D}^\mu,{\rm S}\cdot u]]]_+$
& F & ON \\ \hline
252& & $i[u_\mu,[v\cdot u,[{\rm D}^\mu,{\rm S}\cdot u]]]_+$ & F & ON \\
\hline
253& & $i[[v\cdot u,{\rm S}\cdot u],[{\rm D}_\mu,u^\mu]_+]$ & F & ON$^\prime$ \\
\hline
254& & $i[u_\mu,[{\rm D}^\mu,[{\rm S}\cdot u,v\cdot u]]]_+$
& F & ON \\
\hline
255& & $i[{\rm D}_\mu,[u^\mu,[{\rm S}\cdot u,v\cdot u]]]_+$  & E & ON \\
\hline
256& & $i[{\rm S}\cdot u,[v\cdot u,[{\rm D}^\mu,u_\mu]]_+]$ & F & ON
\\ \hline
257& & $i[{\rm S}\cdot u,[v\cdot u,[{\rm D}^\mu,u_\mu]]]_+$ & F & ON
\\ \hline
258& & $i[v\cdot u,[{\rm S}\cdot u,[{\rm D}^\mu,u_\mu]]_+]$ & F & ON
\\ \hline
259& & $i[v\cdot u,[{\rm S}\cdot u,[{\rm D}^\mu,u_\mu]]]_+$ & F & ON
\\ \hline
260& & $i[{\rm D}_\mu,[u^\mu,[v\cdot u,{\rm S}\cdot u]_+]]$  & F & E \\
\hline
261& & $i[[v\cdot u,u_\mu],[{\rm S}\cdot{\rm D},u^\mu]_+]$ & F &
ON$^\prime$ \\ \hline
262& & $i[u^\mu,[{\rm S}\cdot{\rm D},[u_\mu,v\cdot u]]]_+$ & F & ON 
\\ \hline
263& & $i[[{\rm S}\cdot{\rm D},[u^\mu,[u_\mu,v\cdot u]]]_+$ & F & E 
\\ \hline
264& & $i[v\cdot u,[u^\mu,[{\rm S}\cdot{\rm D},u_\mu]]_+]$ & F & ON
\\ \hline
265&  & $i[v\cdot u,[u^\mu,[{\rm S}\cdot{\rm D},u_\mu]]]_+$ & F & ON
\\ \hline
266& & $i[u^\mu,[v\cdot u,[{\rm S}\cdot{\rm D},u_\mu]]_+]$ & F & ON
\\ \hline
267& & $i[u^\mu,[v\cdot u,[{\rm S}\cdot{\rm D},u_\mu]]]_+$ & F & ON
\\ \hline
268 & & $i[u^2,[v\cdot u,{\rm S}\cdot{\rm D}]_+]_+$ & D[$d_{83}$]
& ON$^\prime$ \\ \hline
269& & $i[[u^\mu,{\rm S}\cdot u],[v\cdot{\rm D},u_\mu]_+]$ & F & ON$^\prime$
\\ \hline
270& & $i[v\cdot{\rm D},[u_\mu,[{\rm S}\cdot u,u^\mu]]]_+$ & F & ON \\
\hline
271& & $i[u_\mu,[v\cdot{\rm D},[{\rm S}\cdot u,u^\mu]]]_+$ & F & ON \\
\hline
272& & $i[{\rm S}\cdot u,[u_\mu,[v\cdot{\rm D},u^\mu]]_+]$ & F & ON \\
\hline
273& & $i[{\rm S}\cdot u,[u_\mu,[v\cdot{\rm D},u^\mu]]]_+$ & F & ON \\
\hline
274& & $i[u_\mu,[{\rm S}\cdot u,[v\cdot{\rm D},u^\mu]]_+]$ & F & ON \\
\hline
275& & $i[u_\mu,[{\rm S}\cdot u,[v\cdot{\rm D},u^\mu]]]_+$ & F & ON \\
\hline
\end{tabular}
\end{table}
\addtocounter{table}{-1}
\begin{table} [htbp]
\centering
\caption{continued}
\begin{tabular} {|c|c|c|c|c|} \hline
$i$ & $(m,n,p,q)$ & Terms &  F($\equiv$Finite) & E\\
& & &  D($\equiv$Divergent)[$d_i$] & ON \\ \hline
276  &
& $i[u^2,[{\rm S}\cdot u,v\cdot{\rm D}]_+]_+$ & D[$d_{85}$]
& ON$^\prime$ \\ \hline
277& & $i[[v\cdot u,{\rm S}\cdot u],[v\cdot{\rm D},v\cdot u]]_+$
& D[$d_{82}$] & ON$^\prime$
\\ \hline
278 & & $i[[v\cdot u,{\rm S}\cdot u]_+,[v\cdot{\rm D},v\cdot u]_+]_+$
& F & ON$^\prime$ \\ \hline
279& & $i[v\cdot{\rm D},[v\cdot u,[{\rm S}\cdot u,v\cdot u]]]_+$
& F 
& E \\ \hline
280& & $i[v\cdot u,[v\cdot{\rm D},[{\rm S}\cdot u,v\cdot u]]]_+$
& F & ON \\ \hline

281& & $i[{\rm S}\cdot u,[v\cdot u,[v\cdot{\rm D},v\cdot u]]_+]$ & F
& ON \\ \hline
282& & $i[{\rm S}\cdot u,[v\cdot u,[v\cdot{\rm D},v\cdot u]]]_+$ & F
& ON \\ \hline
283& & $i[v\cdot u,[{\rm S}\cdot u,[v\cdot{\rm D},v\cdot u]]_+]$ & F
& ON \\ \hline
284& & $i[v\cdot u,[{\rm S}\cdot u,[v\cdot{\rm D},v\cdot u]]]_+$ & F
& ON \\ \hline
285& & $\epsilon^{\mu\nu\rho\lambda} 
[u_\mu,[u_\nu,[{\rm D}_\rho,u_\lambda]]_+]_+$ & F & ON \\ \hline
286& & $\epsilon^{\mu\nu\rho\lambda} 
[{\rm D}_\mu,[u_\nu,[u_\rho,u_\lambda]]_+]_+$ & F & E
\\ \hline
287& & $\epsilon^{\mu\nu\rho\lambda}v_\rho
[v\cdot{\rm D},[u_\mu,[u_\nu,u_\lambda]]_+]_+$ & 
D[$d_6$] & E \\ \hline
288 & & $\epsilon^{\mu\nu\rho\lambda}v_\rho
[u_\mu,[u_\nu,[v\cdot{\rm D},u_\lambda]]_+]_+$ & F & E \\
\hline
289& & $\epsilon^{\mu\nu\rho\lambda}v_\rho
[u_\mu,[u_\nu,[{\rm D}_\lambda,v\cdot u]]_+]_+$ & F & ON \\
\hline
290
& & $\epsilon^{\mu\nu\rho\lambda}v_\rho
[{\rm D}_\mu,[v\cdot u,[u_\nu,u_\lambda]]_+]_+$ & 
D[$d_{67}$] & ON \\ 
\hline
291 & & $\epsilon^{\mu\nu\rho\lambda}v_\rho
[v\cdot u,[u_\mu,[u_\nu,{\rm D}_\lambda]]_+]_+$ &   
D[$d_{71}$] & ON\\ \hline
292& & $\epsilon^{\mu\nu\rho\lambda}v_\rho
[[{\rm D}_\mu,u_\nu],[u_\lambda,v\cdot u]_+]_+$ & D[$d_{69}]$ & ON
\\ \hline
293& & $\epsilon^{\mu\nu\rho\lambda}v_\rho
[u_\mu,[{\rm D}_\nu,[v\cdot u,u_\lambda]]_+]_+$ & F & ON \\
\hline 
294 & & $\epsilon^{\mu\nu\rho\lambda}v_\rho
[u_\mu,[{\rm D}_\nu,[v\cdot u,u_\lambda]]]$ & F & ON \\
\hline 
295 & & $\epsilon^{\mu\nu\rho\lambda}v_\rho
[u_\mu,[v\cdot u,[u_\nu,{\rm D}_\lambda]]_+]_+$ & D[$d_{70}$] & ON
\\ \hline
296
& & $\epsilon^{\mu\nu\rho\lambda}v_\rho
[u_\mu,[v\cdot u,[u_\nu,{\rm D}_\lambda]]]$ & F & ON \\
\hline
297& & $\epsilon^{\mu\nu\rho\lambda}v_\rho
[u_\mu,[v\cdot u,[u_\nu,{\rm D}_\lambda]_+]]_+$ & F & ON \\
\hline
298
 & & $\epsilon^{\mu\nu\rho\lambda}v_\rho
[{\rm D}_\mu,[u_\nu,[v\cdot u,u_\lambda]_+]_+]$ & F & ON \\ \hline
299& & $\epsilon^{\mu\nu\rho\lambda}v_\rho
[{\rm D}_\mu,[u_\nu,[v\cdot u,u_\lambda]_+]]_+$ & F & ON \\ \hline
300 & & $i[v\cdot u,u_\mu]_+{\rm S}\cdot u{\rm D}^\mu+$h.c. 
& D[$d_{87}$] & E \\ \hline
301 & & $i[v\cdot u,{\rm S}\cdot u]_+u^\mu{\rm D}_\mu$+h.c. & D[$d_{88}$]
 & E \\ \hline
302 & & $i[u_\mu,{\rm S}\cdot u]_+v\cdot u{\rm D}^\mu$+h.c.
& D[$d_{89}$] & E \\ \hline
303 & & $i[u_\mu,{\rm S}\cdot u]_+u^\mu v\cdot{\rm D}+$h.c. & D[$d_{86}$]
& E \\ \hline
304 & & $i[v\cdot u,{\rm S}\cdot u]_+v\cdot u v\cdot{\rm D}+$ h.c.
& D[$d_{91}$] & E \\ \hline
305 & & $i[(v\cdot u)^2,[v\cdot u,{\rm S}\cdot{\rm D}]_+]_+$ & D[$d_{92}$]
& ON$^\prime$ \\ \hline
\end{tabular}
\end{table}

\addtocounter{table}{-1}
\begin{table} [htbp]
\centering
\caption{continued}
\begin{tabular} {|c|c|c|c|c|} \hline
$i$ & $(m,n,p,q)$ & Terms &  F($\equiv$Finite) & E\\
& & &  D($\equiv$Divergent)[$d_i$] & ON \\ \hline
306& & $i[{\rm S}\cdot{\rm D},(v\cdot u)^3]_+$ & F & E \\ \hline
307& & $i[{\rm D}_\mu,[v\cdot u,[{\rm S}\cdot u,u^\mu]]]_+$
& F & E \\ \hline 
308& & $i[v\cdot u,[{\rm D}_\mu,[{\rm S}\cdot u,u^\mu]]]_+$
& F & E \\ \hline 
309  & & $i[{\rm S}\cdot u,[u_\mu, [{\rm D}^\mu,v\cdot u]]_+]$ & F & ON \\ \hline
310& & $i[{\rm S}\cdot u,[u_\mu, [{\rm D}^\mu,v\cdot u]]]_+$ & F & ON \\ \hline
311& & $i[u_\mu,[{\rm S}\cdot u,[{\rm D}^\mu,v\cdot u]]]_+$ & F & ON \\ \hline
312& & $i[u_\mu,[{\rm S}\cdot u,[{\rm D}^\mu,v\cdot u]]_+]$ & F & ON \\ \hline
313  & & $i[v\cdot u,[v\cdot u,[v\cdot u,{\rm S}\cdot{\rm D}]]]_+$ & F & ON \\  
\hline
314 & & $i[(v\cdot u)^2,[{\rm S}\cdot u,v\cdot{\rm D}]_+]_+$ & D[$d_{90}$]
& ON \\ \hline
315
 & & $i[[u^\mu,v\cdot u],[{\rm S}\cdot{\rm D},u_\mu]]_+$ &  D[$d_{80}$] & ON \\
\hline
316 & & $i[u_\mu,{\rm S}\cdot u],[{\rm D}^\mu,v\cdot u]]_+$&  D[$d_{79}$]& ON
\\ \hline
317
& (2,0,0,1) &  
$[{\rm S}\cdot{\rm D},[v\cdot{\rm D},\chi_-]]_+$
&  F & E\\ \hline
318 & (0,2,0,1) &  $[{\rm S}\cdot u,[v\cdot u,\chi_-]]_+$
&  F & ON \\ \hline
319  & (1,1,1,0) &  
$i[[v\cdot{\rm D},{\rm S}\cdot u]_+,\chi_+]_+$ & D[$d_{117}$] & E \\
\hline
320 & & $i[[{\rm S}\cdot{\rm D},v\cdot u]_+,\chi_+]_+$
&  D[$d_{118}$] & E \\ \hline
\end{tabular}
\end{table}

\begin{table}[htbp]
\centering
\caption{BChPT counterparts of L.C.-independent HBChPT terms}
\begin{tabular} {|c|c|c|}  \hline
$i$ & HBChPT term & BChPT counterpart \\ \hline
1 & $({\cal V}^\mu{\cal A}_\mu)^{m_3}\equiv(v\cdot u)^{j_1}{\rm S}\cdot{\rm D}
(u_\rho{\rm D}^\rho)^{j_2}$ &
$({i\over{\rm m}}u\cdot{\rm D})^{j_1}
\gamma^5\rlap/{\rm D}(u^\rho{\rm D}_\rho)^{j_2}$ \\ \hline
2 & $({\cal V}^\nu{\cal V}_\nu)^{m_1}\equiv (v\cdot\rm D)^{l_1}(\rm D_\nu\rm D^\nu
)^{l_2}$ & $\Biggl((\rm D^2+{\rm m}^2)$ \\
& & $-\biggl(-(i\rlap/{\rm D}-{\rm m})^2+{i\over8}\sigma^{\mu\nu}[u_\mu,u_\nu]\biggr)
\Biggr)^{l_1}$ \\
& & $\times\Biggl[({\rm D}_\nu{\rm D}^\nu)^{l_2}$ \\
& & $-{1\over 2}\sum\Biggl[\Biggl(({\rm D}^2+{\rm m}^2)-\biggl(
-(i\rlap/{\rm D}-{\rm m})^2$ \\ 
& & $+{i\over 8}\sigma^{\mu\nu}[u_\mu,u_\nu]\biggr)
\Biggr)+{\rm m}^2\Biggr]$ \\
& & $\times({\rm D}^\nu{\rm D}_\nu)^{l_2-1}\Biggr]$ \\ \hline
3 & $({\cal A}^\alpha{\cal A}_\alpha)^{m_2}\equiv(u^\mu u_\mu)^{m_2-1}
{\rm S}\cdot u$ & $(u^\mu u_\mu)^{m_2-1}\gamma^5\rlap/u$ \\ \hline
\end{tabular}
\end{table}

\begin{table} [htbp]
\centering
\caption{The 2  L.C.-independent O$(q^4,\phi^{2n})$ terms 
whose LECs are fixed relative
to an O$(q^2,\phi^{2n})$ and  O$(q^3)$ terms}
\begin{tabular} {|c|c|c|} \hline
& (2,2,0,0) HBChPT Term & BChPT Counterpart \\ \hline
1 & $[{\rm D}_\mu,[[{\rm D}_\nu,u^\mu]_+,u^\nu]_+]_+$ &
$[{\rm D}_\mu,[[{\rm D}_\nu,u^\mu]_+,u^\nu]_+]_+$ \\ 
& & $+2i{\rm m}[[{\rm D}_\mu,\rlap/u]_+,u^\mu]_+ + 2i{\rm m}
[{\rm D}_\mu,[u^\mu,\rlap/u]_+]_+\equiv$ O$(q^2)$ \\ 
& & or  $[{\rm D}_\mu,[[{\rm D}_\nu,u^\mu]_+,u^\nu]_+]_+$ \\
& & $+2i{\rm m}[[{\rm D}_\mu,\rlap/u]_+,u^\mu]_+\equiv$ O$(q^3)$ \\
\hline
2 & $[{\rm D}_\mu,[[{\rm D}_\nu, u^\nu]_+,u^\mu]_+]_+$ & 
 $[{\rm D}_\mu,[[{\rm D}_\nu, u^\nu]_+,u^\mu]_+$ \\ 
& & $+2i{\rm m}[{\rm D}_\mu,[\rlap/u,u^\mu]_+]_+
+2i{\rm m}[[{\rm D}_\mu,u^\mu]_+,\rlap/u]_+\equiv$ O$(q^2)$ \\ 
& & or $[{\rm D}_\mu,[[{\rm D}_\nu, u^\nu]_+,u^\mu]_+]_+$ \\
& & $+2i{\rm m}[{\rm D}_\mu,[\rlap/u,u^\mu]_+]_+\equiv$ O$(q^3)$ \\
\hline
3 & $[{\rm D}_\mu,[{\rm D}_\nu,[u^\mu,u^\nu]_+]_+]_+$ &
$[{\rm D}_\mu,[{\rm D}_\nu,[u^\mu,u^\nu]_+]_+]_+ + 4i{\rm m^0}
[{\rm D}_\mu,[\rlap/u,u^\mu]_+]_+\equiv$ O$(q^2)$ \\
& & or  $[{\rm D}_\mu,[{\rm D}_\nu,[u^\mu,u^\nu]_+]_+]_+$ \\
& & $+2i{\rm m^0}[{\rm D}_\mu,[\rlap/u,u^\mu]_+]_+\equiv$ O$(q^3)$ \\
\hline
\end{tabular}
\end{table}

\begin{table}[htbp]
\centering
\caption{BChPT counterparts of L.C.-dependent HBChPT terms}
\begin{tabular} {|c|c|c|}  \hline
$i$ & HBChPT term & BChPT counterpart \\ \hline
1 & $\epsilon^{\lambda_1\lambda_2\lambda_3\lambda_4}
\quad \prod_{i=1}^{M_1}{\cal V}_{\mu_i}\prod_{j=1}^{M_2}{\cal A}_{\nu_j}:$
& \\ \cline{2-3} 
& $(a)\ \epsilon^{\lambda_1\lambda_2\lambda_3\lambda_4}\ v_{\mu_k}
{\rm S}_{\nu_l}\prod_{i\neq k}^{N_1}{\rm D}_{\mu_i}\prod_{j\neq l}^{N_2}
u_{\nu_j} $
& $\epsilon^{\lambda_1\lambda_2\lambda_3\lambda_4}
\gamma^5\gamma_{\nu_l}$ \\
& & $\times {i\over{\rm m}}{\rm D}_{\mu_k}
\prod_{i\neq k}^{N_1}{\rm D}_{\mu_i}\prod_{j\neq l}^{N_2}
u_{\nu_j}$ \\ \cline{2-3}
& (b)\ ${\epsilon}^{\lambda_1\lambda_2\lambda_3\lambda_4}\ v_{\mu_k}
\prod_{i\neq k}^{M_1}{\rm D}_{\mu_i}\prod_{j=1}^{M_2}u_{\nu_j}$ & 
${\epsilon}^{\lambda_1\lambda_2\lambda_3\lambda_4}$
 \\
& & $\times{i\over{\rm m}}{\rm D}_{\mu_k}
\prod_{i\neq k}^{M_1}{\rm D}_{\mu_i}\prod_{j=1}^{M_2}u_{\nu_j}$ \\ \cline{2-3} 
& (c)\ ${\epsilon}^{\lambda_1\lambda_2\lambda_3\lambda_4}\ 
\prod_{i=1}^{M_1}{\rm D}_{\mu_i} \prod_{j=1}^{M_2} u_{\nu_j}$ 
& \\ 
& $ {\rm e.g.}$ $\epsilon^{\mu\nu\rho\lambda}\biggl(
{\rm D}_\mu u_\nu[{\rm D}_\rho,
{\rm D}_\lambda]$, 
&  $\epsilon^{\mu\nu\rho\lambda}$  \\
& $u_\mu{\rm D}_\nu[u_\rho,u_\lambda]\biggr)$ & 
$\times\biggl({\rm D}_\mu u_\nu[{\rm D}_\rho,
{\rm D}_\lambda]+i{\rm m}\gamma_\mu u_\nu[{\rm D}_\rho,{\rm D}_\lambda]$, \\ 
& & $u_\mu{\rm D}_\nu[u_\rho,u_\lambda]
+i{\rm m}\gamma_\nu u_\mu[u_\rho,u_\lambda]\biggr)$ \\ \cline{2-3} 
& (d)\ ${\epsilon}^{\lambda_1\lambda_2\lambda_3\lambda_4}\ {\rm S}_{\nu_l}\prod_{i=1}^{M_1}
{\rm D}_{\mu_i}\prod_{j\neq l}^{M_2} u_{\nu_j}$ & \\
& ${\rm e.g.}$ $\epsilon^{\mu\nu\rho\lambda}{\rm S}_\lambda\biggl({\rm D}_\mu
[{\rm D}_\rho,{\rm D}_\nu]$, 
& $\epsilon^{\mu\nu\rho\lambda}\gamma^5\gamma_\lambda
{\rm D}_\mu[{\rm D}_\rho,{\rm D}_\nu]$ \\
& $ u_\mu{\rm D}_\nu u_\rho\biggr)$ 
& $-i{\rm m}\sigma^{\nu\rho}
[{\rm D}_\rho,{\rm D}_\nu],$ \\
& &  $\epsilon^{\mu\nu\rho\lambda}
\gamma^5\gamma_\lambda u_\mu
{\rm D}_\nu u_\rho$ \\
& & $-i{\rm m}\sigma^{\rho\mu}[u_\mu,u_\rho]$ \\ \hline
\end{tabular}
\end{table} 

\clearpage
\addtocounter{table}{-1}
\begin{table}[htbp]
\centering
\caption{continued}
\begin{tabular}{|c|c|c|} \hline 
$i$ & HBChPT term & BChPT counterpart \\ \hline
2 & $\epsilon^{\lambda_1\lambda_2\lambda_3\lambda_4}
\prod_{i=1}^{M_1}{\cal V}_{\mu_i}\prod_{j=1}^{M_2}{\cal A}_{\nu_j}(v\cdot u)^{j_1}$
 & \\ \cline{2-3}
& $(a)\ \epsilon^{\mu\nu\rho\lambda}{\rm S}_{\nu_l}
\prod_{i=1}{\rm D}_{\mu_i}\prod_{j\neq l} u_{\nu_j}(v\cdot u):$ & \\
& ${\rm e.g.}:\ i\epsilon^{\mu\nu\rho\lambda}
{\rm S}_\lambda\biggl({\rm D}_\nu u_\mu{\rm D}_\rho v\cdot u+{\rm h.c.},$ &
$\epsilon^{\mu\nu\rho\lambda}\gamma^5\gamma_\lambda\biggl(
{\rm D}_\nu u_\mu{\rm D}_\rho[{\rm D}^\kappa,u_\kappa]_+,$ \\
& $[u_\nu,u_\rho]u_\mu v\cdot u+{\rm h.c.},\ $ ${\rm D}_\nu u_\mu{\rm D}_\rho 
v\cdot u +{\rm h.c.}$ & 
$[u_\nu,u_\rho]u_\mu[{\rm D}^\kappa,u_\kappa]_+\biggr),$ \\
& & ${1\over 2}\epsilon^{\mu\nu\rho\lambda}\gamma_\lambda\gamma^5{\rm D}_\nu u_\mu
{\rm D}_\rho[{\rm D}^\kappa,u_\kappa]_+$ \\
& & $-i{{\rm m}\over 2}
\sigma^{\lambda\rho}[u_\lambda,{\rm D}_\rho]
[{\rm D}_\kappa,u^\kappa]_+$ \\ \cline{2-3}
& $(b)\ {\epsilon}^{\lambda_1\lambda_2\lambda_3\lambda_4}\ 
v_{\mu_k}\prod_{i\neq k}^{M_1}
{\rm D}_{\mu_i} \prod_{j=1}^{M_2} u_{\nu_j}(v\cdot u)^{j_1}$ & \\
& ${\rm up\ to\ O}(q^4)\ {\rm sufficient\ to\ consider}\ j_1=1$ &
${\epsilon}^{\lambda_1\lambda_2\lambda_3\lambda_4}$ \\ 
& & $\times{{{\rm D}_{\mu_k}}\over{\rm m^0}}
\prod_{i\neq k}^{M_1}
{\rm D}_{\mu_i} \prod_{j=1}^{M_2} u_{\nu_j}[{\rm D}_\kappa,u^\kappa]_+$  
\\ \cline{2-3}
& $(c)\ \epsilon^{\lambda_1\lambda_2\lambda_3\lambda_4}
 \prod_{\mu_i}^{M_1}{\rm D}_{\mu_i}\prod_{j=1}^{M_2}u_{\nu_j}(v\cdot u)^{j_1}$
& \\
& $\equiv$ O$(q^5)\ {\rm at\ least}$ & \\ \cline{2-3}
& $(d)\ {\epsilon}^{\lambda_1\lambda_2\lambda_3\lambda_4}\ v_{\mu_k}{\rm S}_{\nu_l}
\prod_{i\neq k}^{M_1} {\rm D}_{\mu_i}\prod_{j\neq l}^{M_2} u_{\nu_j}(v\cdot u)^{j_1}$
& \\
& ${\rm for}$\ O$(q^4)\ $ ${\rm terms}\ j_1=2:$ & \\
& $\epsilon^{\mu\nu\rho\lambda}v_\mu{\rm S}_\nu[{\rm D}_\rho,{\rm D}_\lambda](v\cdot u)^2$
& $\sigma^{\nu\rho}[{\rm D}_\nu,{\rm D}_\rho]
[{\rm D}_\kappa,u^\kappa]_+^2$ \\ \hline
3 & $\epsilon^{\lambda_1\lambda_2\lambda_3\lambda_4}
\prod_{i=1}^{M_1}{\cal V}_{\mu_i}\prod_{j=1}^{M_2}u_{\nu_j}(\rm S\cdot\rm D)$ 
& \\ \cline{2-3}
&  
$(a)\ \epsilon^{\lambda_1\lambda_2\lambda_3\lambda_4}
 \prod_{\mu_i}^{M_1}{\rm D}_{\mu_i}\prod_{j=1}^{M_2}u_{\nu_j}(\rm S\cdot D)$ & \\
& $\equiv$ O$(q^5)\ {\rm at\ least}$ & \\ \cline{2-3}
& $(b)\ {\epsilon}^{\lambda_1\lambda_2\lambda_3\lambda_4}\ 
v_{\mu_k}\prod_{i\neq k}^{M_1}
{\rm D}_{\mu_i} \prod_{j=1}^{M_2} u_{\nu_j}(\rm S\cdot\rm D)$ & 
${\epsilon}^{\lambda_1\lambda_2\lambda_3\lambda_4}$ \\ 
& & $\times{{{\rm D}_{\mu_k}}\over{\rm m}}
\prod_{i\neq k}^{M_1}
{\rm D}_{\mu_i} \prod_{j=1}^{M_2} u_{\nu_j}\gamma^5\rlap/\rm D$
\\ \hline  
4 & $\epsilon^{\lambda_1\lambda_2\lambda_3\lambda_4}
\prod_{i=1}^{M_1}{\cal V}_{\mu_i}
\prod_{j=1}^{M_2}{\cal A}_{\nu_j}(u^\rho{\rm D}_\rho)^{j_2}
(v\cdot u)^{j_1}$ & \\
& $\equiv{\rm at\ least\ of\ O}(q^5)$ & \\ \hline
\end{tabular}
\end{table}

\addtocounter{table}{-1}
\begin{table}[htbp]
\centering
\caption{continued}
\begin{tabular}{|c|c|c|} \hline 
$i$ & HBChPT term & BChPT counterpart \\ \hline
5 & $\epsilon^{\lambda_1\lambda_2\lambda_3\lambda_4}\prod_{i=1}^{M_1}{\cal V}_{\mu_i}
\prod_{j=1}^{M_2}u_{\nu_j}(u^\rho{\rm D}_\rho)^{j_2}
(\rm S\cdot\rm D)$ & \\
& $\equiv{\rm at\ least\ of\ O}(q^6)$ & \\ \hline
6 & $\epsilon^{\lambda_1\lambda_2\lambda_3\lambda_4}\prod_{i=1}^{M_1}{\cal V}_{\mu_i}
\prod_{j=1}^{M_2}u_{\nu_j}({\rm S}\cdot{\rm D})
(v\cdot u)^{j_1}$ & \\
& $\equiv{\rm at\ least\ of\ O}(q^5)$ & \\ \hline
7 & $\epsilon^{\lambda_1\lambda_2\lambda_3\lambda_4}\prod_{i=1}^{M_1}{\cal V}_{\mu_i}
\prod_{j=1}^{M_2}u_{\nu_j}({\rm S}\cdot{\rm D})
(u_\rho{\rm D}^\rho)^{j_2}(v\cdot u)^{j_1}$ & \\
& $\equiv{\rm at\ least\ of\ O}(q^7)$ & \\ \hline
8 & $\epsilon^{\lambda_1\lambda_2\lambda_3\lambda_4}\prod_{i=1}^{M_1}
{\cal V}_{\mu_i}\prod_{j=1}^{M_2}{\cal A}_{\nu_j}
(u^\rho{\rm D}_\rho)^{j_2}:$ & \\ \cline{2-3}
& $(a)\ \epsilon^{\lambda_1\lambda_2\lambda_3\lambda_4}\prod_{i=1}^{M_1}{\rm D}_{\mu_i}
\prod_{j=1}^{M_2}u_{\nu_j}(u^\rho{\rm D}_\rho)^{j_2}$ & \\
& $\equiv{\rm O}(q^6)\ {\rm at\ least}$ & \\ \cline{2-3}
& $(b)\ \epsilon^{\lambda_1\lambda_2\lambda_3\lambda_4}v_{\mu_k}\prod_{i\neq k}^{M_1}
{\rm D}_{\mu_i}\prod_{j=1}^{M_2} u_{\nu_j}(u_\rho{\rm D}^\rho)^{j_2}$ & \\
& $\equiv{\rm O}(q^5)\ {\rm at\ least}$ & \\ \cline{2-3}
& $(c)\ \epsilon^{\lambda_1\lambda_2\lambda_3\lambda_4}\ {\rm S}_{\nu_l}\prod_{i=1}^{M_1}
{\rm D}_{\mu_i}\prod_{j\neq l}^{M_2}u_{\nu_j}(u_\rho{\rm D}^\rho)^{j_2}$
 & \\
& $\equiv{\rm O}(q^5)\ {\rm at\ least}$ & \\ \cline{2-3}
& $(d)\ \epsilon^{\lambda_1\lambda_2\lambda_3\lambda_4}{\rm S}_{\nu_l}v_{\mu_k}
\prod_{i\neq k}^{M_1}{\rm D}_{\mu_i}
\prod_{j\neq l}u_{\nu_j}(u^\rho{\rm D}_\rho)^{j_2}$ & \\
& ${\rm up\ to\ O}(q^4)\ j_2=1:$ & \\
& ${\rm e.g}\ i\epsilon^{\mu\nu\rho\lambda}v_\nu{\rm S}_\mu u^\kappa u_\rho
{\rm D}_\lambda {\rm D}_\kappa+{\rm h.c.}$ & 
${\epsilon^{\mu\nu\rho\lambda}\over{\rm m}}
\gamma^5\gamma_\mu{\rm D}_\nu u^\kappa u_\rho
{\rm D}_\lambda {\rm D}_\kappa$ \\
& & $+2i\sigma^{\rho\lambda}[u^\kappa u_\rho,{\rm D}_\lambda]
{\rm D}_\kappa$ \\ \hline
\end{tabular}
\end{table}
\clearpage

\begin{figure}[p]
\centerline{\mbox{\psfig{file=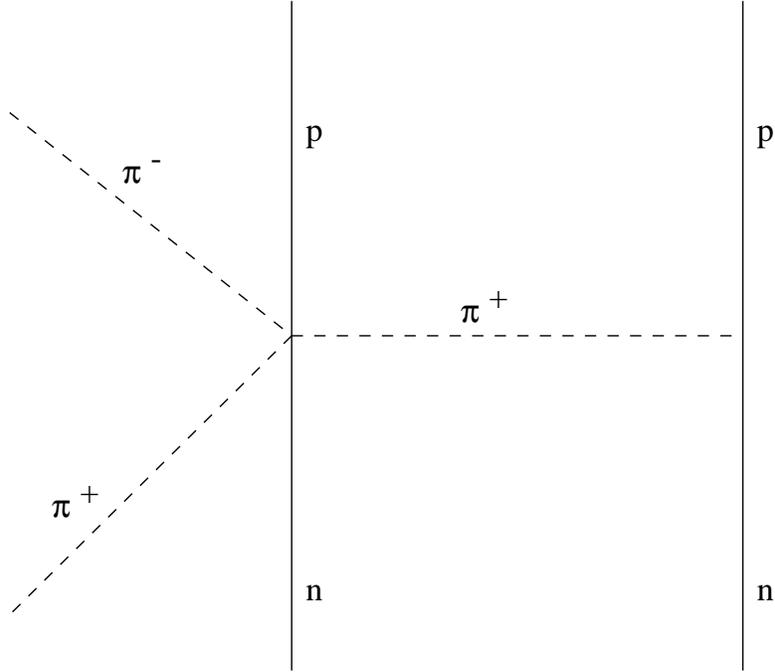,width=0.8\textwidth}}}
\caption{``Contact graph" of pion DCX}
\end{figure}

\begin{figure}[p]
\centerline{\mbox{\psfig{file=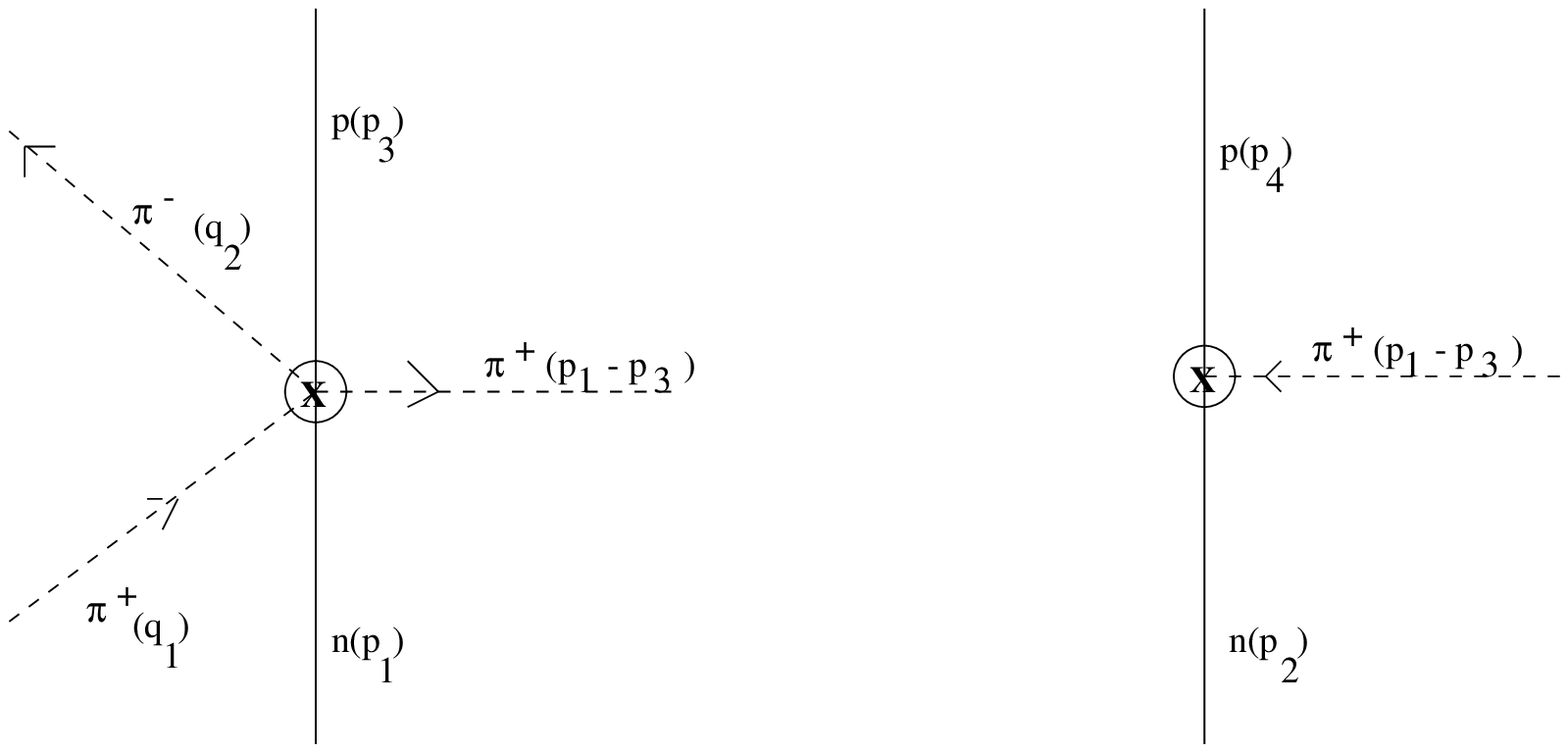,width=0.8\textwidth}}}
\caption{O($q^4$) operator insertion (indicated by a cross in a circle)
for ${\bar p}(\pi^+)^2\pi^-n$- and ${\bar p}\pi^+n$-vertices
that figure in the ``contact graph" of pion DCX}
\end{figure}


\begin{thebibliography}{99}
\bibitem{1n} A.Misra, D.S.Koltun, Nucl. Phys. A ${\bf 646}$, 343 (1999). 
\bibitem{jm} E.Jenkins and
A.V.Manohar, Phys. Lett. B {\bf 255} (1991) 558.
\bibitem{bkm} V.Bernard, N.Kaiser, J.Kambor and Ulf-G Meissner,
Nucl.Phys.B {\bf 388} (1992) 315.
\bibitem{bkm1} V.Bernard,
N.Kaiser and Ulf-G.Meissner, Int. J. Mod. Phys. ${\bf E4}$, 193 (1995).
\bibitem{em} G.Ecker and M.Mojzis, Phys. Lett. B {\bf 365}, 312 (1996). 
\bibitem{bkm2} V.Bernard, N.Kaiser  and Ulf-G Meissner,
Nucl.Phys.B ${\bf 457}$, 147 (1995).
\bibitem{mms}Ulf-G.Meissner, G.Muller, S.Steininger, hep-ph/9809446. 
\bibitem{mnnl} T.Mannel, W.Roberts. W.Ryzak, Nucl. Phys. B {\bf 368}, 204 (1992).
\bibitem{kj} M.F.Jiang, D.S.Koltun, Phys. Rev C {\bf 42}, 2662 (1990).
\bibitem{mk} A.Misra, D.S.Koltun, nucl-th/9810075, submitted to Phys Rev C.
\bibitem{krause} A.Krause, Helv. Phys. Acta 63, 3 (1990) .
\bibitem{lm} M.Luke and A.V.Manohar Phys. Lett.B ${\bf 286}$, 348 (1992).
\end{thebibliography}
\end{document}